\documentclass{pinchcr_ll}
\usepackage{amsmath,amssymb}

\usepackage{diff-commands,hyperref}
\usepackage[stable]{footmisc}

\usepackage{feynmf}
\renewcommand{\subsubsection}[1]{\medskip
\noindent
{\bf #1}
\smallskip}

\def\be{\begin{equation}}
\def\ee{\end{equation}}
\def\bea{\begin{eqnarray}}
\def\eea{\end{eqnarray}}

\def\reff#1{\ref{#1}}
\def\labell#1{\label{#1}}

\def\eq#1{Eq.~(\reff{#1})}
\def\eqs#1#2{Eqs.~(\reff{#1})--(\reff{#2})}
\def\eqss#1#2#3{Eqs.~(\reff{#1}), (\reff{#2}) and (\reff{#3})}
\def\fig#1{Fig.~\reff{#1}}

\def\sec#1{Sec.~\reff{#1}}
\def\tab#1{Table~\reff{#1}}

\def\WLO#1{W_{#1}^{\mbox{\tiny(0)}}}
\def\kLO#1{{k_{#1}^{\mbox{\tiny(0)}}}}
\def\qLO#1{{q_{#1}^{\mbox{\tiny(0)}}}}
\def\WNLO#1{W_{#1}}
\def\kNLO#1{k_{#1}}
\def\qNLO#1{q_{#1}}

\newcommand{\sumprime}{\vbox{\vspace{-0.3cm}\hbox{$\prime$}\vspace{0.3cm}}}
\def\sumngtq{\sum_{\vec{n}^2>\Lambda}}
\def\sumnleq{\sum_{\vec{n}^2\le \Lambda}}
\def\sumn{\sum_{\vec{n}\in\mathbb{Z}^3}}

\newcommand{\Ckpipi}{C_{K\to\pi\pi}}
\newcommand{\Cpipi}{C_{\pi\pi\to\pi\pi}}

\def\Rlam#1#2{R_\Lambda\left(#1,#2\right)}

\newcommand{\lsim}{ {\
\lower-1.2pt\vbox{\hbox{\rlap{$<$}\lower5pt\vbox{\hbox{$\sim$}}}}\ } }
\newcommand{\gsim}{ {\
\lower-1.2pt\vbox{\hbox{\rlap{$>$}\lower5pt\vbox{\hbox{$\sim$}}}}\ } }

\def\er#1#2{\relax\ifmmode{}^{+#1}_{-#2}\else$^{+#1}_{-#2}$\fi}
\def\erparen#1#2{\relax\ifmmode{}(^{#1}_{#2})\else$(^{#1}_{#2})$\fi}
\def~{\ifmmode\phantom{0}\else\penalty10000\ \fi}

\def\hc{\mathrm{h.c.}}

\def\d#1#2{d\mskip 1.5mu^{#1}\mkern-1mu{#2}\,}
\def\D#1#2{{\d#1{#2} \over (2\pi)^{#1}}\,}

\def\det{\mathrm{det}}

\def\fm{\mathrm{fm}}
\def\ev{\mathrm{e\kern-0.1em V}}
\def\kev{\mathrm{ke\kern-0.1em V}}
\def\mev{\mathrm{Me\kern-0.1em V}}
\def\gev{\mathrm{Ge\kern-0.1em V}}
\def\tev{\mathrm{Te\kern-0.1em V}}

\def\re{\mathrm{Re}}
\def\im{\mathrm{Im}}

\def\n#1e#2n{{#1}\times 10^{#2}}

\def\ra{\rangle}
\def\la{\langle}
\def\l{\left}
\def\r{\right}
\def\ord#1{\mathcal{O}(#1)}
\def\nn{\nonumber}

\def\cO{\mathcal{O}}

\def\cH{\mathcal{H}}
\def\cL{\mathcal{L}}
\def\cM{\mathcal{M}}
\def\cD{\mathcal{D}}
\def\cE{\mathcal{E}}

\def\ods2{\mathcal{O}_{\Delta S=2}}
\def\zds2{Z_{\Delta S=2}}

\def\msbar{{\overline{\mathrm{MS}}}}

\def\RGI{\mathrm{RGI}}

\def\lqcd{\Lambda_\mathrm{QCD}}
\def\lat{\mathrm{lat}}

\def\langleV{\;_V\hspace{-0.05cm}\langle}
\def\rangleV{\rangle_V}

\makeatletter
\def\slash#1{{\mathpalette\c@ncel{#1}}} 
\def\big#1{{\hbox{$\left#1\vbox to1.012\ht\strutbox{}\right.\n@space$}}}
\def\Big#1{{\hbox{$\left#1\vbox to1.369\ht\strutbox{}\right.\n@space$}}}
\def\bigg#1{{\hbox{$\left#1\vbox to1.726\ht\strutbox{}\right.\n@space$}}}
\def\Bigg#1{{\hbox{$\left#1\vbox
to2.083\ht\strutbox{}\right.\n@space$}}}
\makeatother

\def\sectiondum{\chapter}

\def\subsectiondum{\section}
\def\subsubsectiondum{\subsection}

\title{Flavor physics and lattice quantum chromodynamics}

\author{Laurent {\sc Lellouch}}

\affiliation{\ \\[0.5cm]\vbox{Centre de Physique
    Th\'eorique~\thanks{CPT is research unit UMR 7332 of the CNRS, of
      Aix-Marseille U. and of U. Sud Toulon-Var; it is also affiliated
      with the CNRS' research federation FRUMAM (FR 2291).}\\ CNRS UMR
    7332\\Aix-Marseille U. and U. Sud Toulon-Var\\ F-13288 Marseille
    Cedex 9\\ France\\[1cm]
\includegraphics[width=0.3\textwidth]{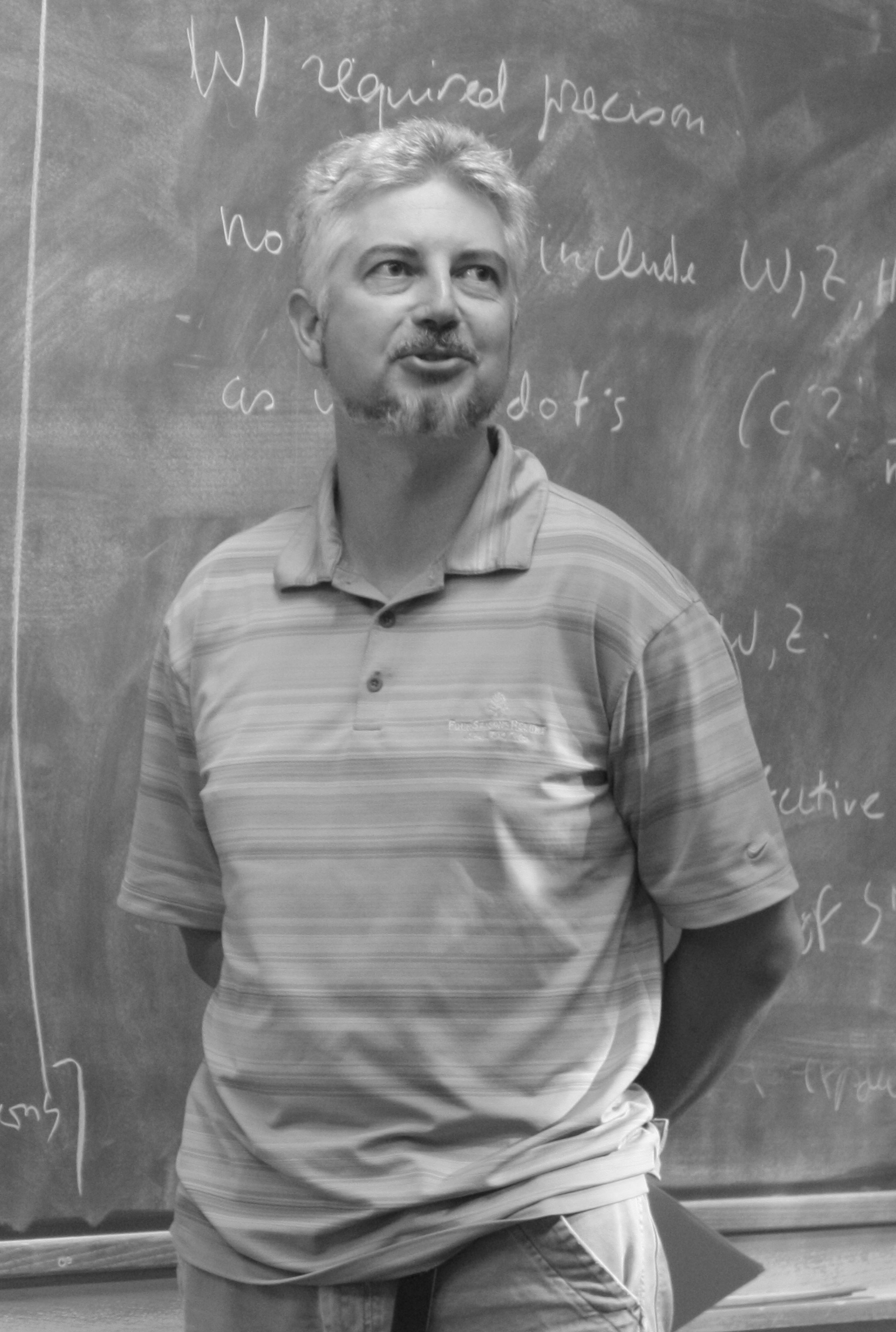}
\\[1cm]
    \large Summer school on ``Modern perspectives in lattice
    QCD''\\ \large \'Ecole de Physique des Houches, August 3--28,
    2009\\[1cm]}}

\begin{document}

\maketitle










\dedication{To Annemarie, Benjamin and Niels}

\acknowledgements

I am indebted to my fellow organizers for an enjoyable collaboration
in preparing this Summer School and for unanimously designating me to
give the traditional public lecture! A school is only as good as the
students who attend it are, and I would like to thank them for their
strong motivation, unrelenting questioning and ability to put together
great parties seven nights a week. I am also grateful to the other
teachers for preparing excellent lectures which were profitable, not
only for the students. Moreover, I wish to thank Leticia Cugliandolo
for her masterful direction of the \'Ecole des Houches, and to thank
Brigitte Rousset and Murielle Gardette for their seamless running of
the program. Finally, the help of Antonin Portelli and Alberto Ramos
in preparing many of the Feynman diagrams in these notes, as well as
the careful reading of the manuscript by J\'erome Charles, Marc
Knecht, Thorsten Kurth, Eduardo de Rafael and Alberto Ramos, are
gratefully acknowledged. This work is supported in part by EU grant
MRTN-CT-2006-035482 (FLAVIAnet), CNRS grant GDR 2921 and a CNRS
``formation permanente'' grant.

\preface{
Quark flavor physics and lattice quantum chromodynamics (QCD) met many
years ago, and together have given rise to a vast number of very
fruitful studies. All of these studies certainly cannot be reviewed
within the course of these lectures. Instead of attempting to do so, I
discuss in some detail the fascinating theoretical and
phenomenological context and background behind them, and use the rich
phenomenology of nonleptonic weak kaon decays as a template to present
some key techniques and to show how lattice QCD can effectively help
shed light on these important phenomena. Even though the lattice study
of $K\to\pi\pi$ decays originated in the mid-eighties, it is still
highly relevant. In particular, testing the consistency of the
Standard Model with the beautiful experimental measurements of direct
CP violation in these decays remains an important goal for the
upcoming generation of lattice QCD practitioners.

The course begins with an introduction to the Standard Model, viewed
as an effective field theory. Experimental and theoretical limits on
the energy scales at which New Physics can appear, as well as current
constraints on quark flavor parameters, are reviewed. The role of
lattice QCD in obtaining these constraints is described. A second section
is devoted to explaining the Cabibbo-Kobayashi-Maskawa mechanism for
quark flavor mixing and CP violation, and to detailing its most
salient features. The third section is dedicated to the study of
$K\to\pi\pi$ decays. It comprises discussions of indirect CP violation
through $K^0$-$\bar K^0$ mixing, of the $\Delta I=1/2$ rule and of
direct CP violation. It presents some of the lattice QCD tools required to
describe these phenomena {\em ab initio.} 
}

\tableofcontents

\maintext

\sectiondum{Introduction and motivation}

\subsectiondum{The Standard Model as a low-energy effective 
field theory}

If elementary particles were massless, their fundamental interactions
would be well described by the most general perturbatively
renormalizable~\footnote{Beyond fixed-order perturbation theory, the
$U(1)_Y$ of hypercharge is {\em trivial:} the renormalized coupling
constant vanishes when the cutoff of the regularized theory is taken
to infinity, a notion first suggested in~\shortcite{Wilson:1973jj}.}
relativistic quantum field theory based on:
\begin{itemize}
\item the gauge group
\be
\labell{eq:SMGG}
SU(3)_c\times SU(2)_L\times U(1)_Y\ ,
\ee
where the subscript $c$ stands for ``color'', $L$ for left-handed weak
isospin and $Y$ for hypercharge;

\item three families of quarks and leptons
\be
\labell{eq:families}
\left\{
\begin{array}{l}
u\ d\ e^-\ \nu_e\\
c\ s\ \mu^-\ \nu_\mu\\
t\ b\ \tau^-\ \nu_\tau
\end{array}\ ,
\right.
\ee
with prescribed couplings to the gauge fields (i.e. in specific
representations of the gauge groups);

\item and the absence of anomalies.

\end{itemize}
In the presence of masses for the weak gauge bosons $W^\pm$ and $Z^0$,
for the quarks and for the leptons, the most economical way known to
keep this construction perturbatively renormalizable is to implement
the Higgs mechanism~\shortcite{Englert:1964et,Higgs:1964pj}, as done
in the Standard Model (SM). However, this results in adding a yet
unobserved degree of freedom to the model, the Higgs boson.

By calling a theory renormalizable we mean that it can be used to make
{\em predictions of arbitrarily high accuracy} over a very large
interval of energies, ranging from zero to possibly infinite energy,
with only a finite number of coupling constants.~\footnote{If one
sticks to perturbation theory, the precision reached is actually
limited by the fact that perturbative expansions in field theory are
typically asymptotic expansions. Moreover, the triviality of the Higgs
and $U(1)_Y$ sectors means that the cutoff, which we generically call
$\Lambda$ here, has to be kept finite. This limits the accuracy of
predictions through the presence of regularization dependent
corrections which are proportional to powers of $E/\Lambda$, where $E$
is an energy typical of the process studied. In that sense, only
asymptotically free theories can be fundamental since they are the
only ones that can be used to describe phenomena up to arbitrarily
high energies.
} These
couplings are associated with operators of mass dimension less or
equal to four in $3+1$ dimensions.

Renormalizable field theories are remarkable in many ways. Consider an
arbitrary high-energy theory described by a Lagrangian $\cL_\mathrm{UV}$
(e.g. a GUT, a string theory, \ldots) with given low-energy spectrum
and symmetries. At sufficiently low energies this theory is described
by the unique renormalizable theory with the given spectrum and
symmetries, whose Lagrangian we will denote $\cL_\mathrm{ren}$. Moreover, the
deviations between the predictions of the two theories can be
parametrized through a local low-energy effective field theory (EFT)
\be \cL_\mathrm{UV} = \cL_\mathrm{ren} + \sum_{d\ge
  4}\sum_i\frac{C_{d,i}}{\Lambda_i^{d-4}}O_i^{(d)}
\ ,\ee
where the $O_i^{(d)}$ are operators of mass dimension $d\ge 4$ built
up from fields of $\cL_\mathrm{ren}$. The $\Lambda_i$ are mass scales
which are much larger than the masses in the spectrum of
$\cL_\mathrm{ren}$--there may be one or many of them depending on the number
of distinct scales in $\cL_\mathrm{UV}$. The $C_{d,i}$ are dimensionless
coefficients whose sizes depend on how the corresponding operators are
generated in the UV theory, e.g. at tree or loop level.

Thus, very generally, we can write down the Lagrangian of particle
physics as a low-energy EFT with the gauge group of \eq{eq:SMGG}, the
matter content of \eq{eq:families} and a Higgs mechanism:
\be 
\labell{eq:LPP}
\cL_\mathrm{SM}^\mathrm{eff} = \cL_\mathrm{SM} +\frac1M O_\mathrm{Maj}^{(5)} + 
\sum_{d\ge 6}\sum_i\frac{C_{d,i}}{\Lambda_i^{d-4}}O_i^{(d)}
\ ,\ee
where the left-handed neutrino Majorana mass term, $O_\mathrm{Maj}^{(5)}$,
and the $O_i^{(d)}$ must be invariant under the Standard Model gauge
group (\reff{eq:SMGG}). In \eq{eq:LPP}, $\cL_\mathrm{SM}$ is the
renormalizable Standard Model Lagrangian
\be
\labell{eq:LSM}
\cL_\mathrm{SM} = \cL_\mathrm{g+f}+\cL_\mathrm{flavor}+
\cL_\mathrm{EWSB}+\cL_\nu
\ .
\ee
where $\cL_\mathrm{g+f}$ contains the gauge and fermion kinetic and
coupling terms, $\cL_\mathrm{flavor}$, the Higgs-Yukawa terms,
$\cL_\mathrm{EWSB}$, the Higgs terms and $\cL_\nu$, the possible
renormalizable neutrino mass and right-handed neutrino kinetic terms.
In that sense, the renormalizable Standard Model is a low-energy
approximation of a more complete high-energy theory involving scales
of New Physics much larger than $M_W$.

Schematically, the gauge and fermion Lagrangian reads
\be
\cL_\mathrm{g+f} = \frac14 F_{\mu\nu}^a F^{\mu\nu}_a+\bar\psi\slash{D}\psi
\ .\ee
It has 3 parameters, the gauge couplings $(g_1,g_2,g_3)$, and is
very well tested through experiments conducted at LEP, SLC, the
Tevatron, etc. Its parameters are known to better than per mil
accuracy.

The Higgs-Yukawa terms are given by
\be
\cL_\mathrm{flavor} = -\bar\psi_R^{(-1/2)}Y_{(-1/2)}\phi^\dagger\psi_L
-\bar\psi_R^{(1/2)}Y_{(1/2)}\tilde\phi^\dagger\psi_L+\hc
\ ,\ee
with $\psi_L$ corresponding to the left-handed $SU(2)_L$ doublets and
$\psi_R^{(\pm1/2)}$ the right-handed $SU(2)_L$ singlets, associated with
the $I_3=\pm\frac12$ component of the doublets. In this equation, $\phi$
is the Higgs field and $\tilde\phi$ its conjugate,
$(\phi^0,-\phi^{+*})$.  The flavor component 
of the Standard Model Lagrangian has
many more couplings, 13 in fact. It gives rise to the 3 charged lepton
masses, 6 quark masses and the quark flavor mixing matrix which has 3
mixing angles and 1 phase.~\footnote{Remember that we have separated
  out into $\cL_\nu$ possible renormalizable neutrino mass terms.} 
The understanding of this quark mixing and
its associated CP violation will be the main focus of the present
course.

There is also the electroweak symmetry
breaking (EWSB) contribution
\be
\cL_\mathrm{EWSB}=(D_\mu\phi)^\dagger(D^\mu\phi)-\mu^2\phi^\dagger\phi - \lambda(\phi^\dagger\phi)^2
\ .\ee
It has only 2 couplings, the Higgs mass and self-coupling
$(\mu,\lambda)$, and is very poorly tested so far, a situation which
will change radically with the LHC.

As for the neutrino Lagrangian, little is known from experiment about its
form. There are theoretically two possible, nonexclusive scenarios:
\begin{enumerate}

\item

There are no right-handed neutrinos in sight. Thus, we give our
left-handed neutrinos a mass without introducing a right-handed
partner. In that case, $\cL_\nu=0$ in \eq{eq:LSM} and we have a
Majorana mass term for the left-handed neutrinos in \eq{eq:LPP}, with
\be O_\mathrm{Maj}^{(5)}=-\frac12\nu_L^TC\tilde\phi^TA^L_\nu\tilde\phi
\nu_L+\hc\ , \ee
where $C$ is the charge conjugation matrix (see \eq{eq:Cfermion}). 
That is, after EWSB the
neutrino acquire a Majorana mass through the introduction of a
nonrenormalizable dimension-5 operator. This implies that the
Standard Model is an EFT and that we already have a signal for a new mass
scale. Indeed, with $m_\nu\sim 0.1$~eV (a plausible value),
eigenvalues of the coupling matrix $A^L_\nu$ of order 1 and
$\langle\phi\rangle\sim 246\,\gev$, one finds for the mass scale $M$
of \eq{eq:LPP}
\be
M\sim\frac{\langle\phi\rangle^2}{m_\nu}\sim 10^{15}\,\gev
\ee
which is tantalizingly close to a possible unification scale.

\item

We choose to allow right-handed neutrinos, $N_R$. These neutrinos must
be singlets under the Standard Model group. Thus, they themselves may
have a Majorana mass, but this time a renormalizable one, in addition
to allowing the presence of a Dirac mass term:
\bea
\cL_\nu&=&N_Ri\slash\partial N_R-
(\bar L_L Y_\nu^\dagger\tilde\phi N_R+\frac12 N_R^TCM_\nu^R N_R + \hc)\\
&=& N_Ri\slash\partial N_R-\frac12(\nu_L^T,N_R^{cT})C\l(
\begin{array}{cc}
0 & Y_\nu^\dagger\phi^0\\
Y_\nu^*\phi^0 & M_\nu^R
\end{array}
\r)
\l(\begin{array}{c}
\nu_L\\ N_R^c
\end{array}
\r)
+\cdots
\ ,\eea
where $L_L$ stands for the left-handed lepton doublets, and $N_R^c$
for the charge conjugate of $N_R$ (see \eq{eq:Cfermion}).

There are here three more possibilities:
\begin{itemize}

\item[a)] $M_\nu^R=0$

In that case, the three neutrinos have Dirac masses and lepton
number is conserved.

\item[b)] $M_\nu^R\gg Y_\nu\la\phi\ra$

Here, the see-saw mechanism comes into play:
there are no right-handed neutrinos in sight and all three left-handed
neutrinos acquire a mass through $d=5$ operators:
\be
\cL_\mathrm{eff}^{m_\nu}=-\frac12
\nu_L^TC\tilde\phi^T\l(Y_\nu^\dagger(M_\nu^R)^{-1}Y_\nu^*\r)\tilde\phi \nu_L + O\l((M_\nu^R)^{-2}\r)
\ .\ee
Thus, we have an explicit realization of scenario 1). 

Taking $\re Y_\nu\la\phi\ra\sim 1\,\gev$ in rough analogy with the $\tau$
and again, $m_\nu\sim 0.1$~eV, we obtain for the mass scale $M$ of \eq{eq:LPP}
\be
M\sim \frac{(\re Y_\nu\la\phi\ra)^2}{m_\nu}\sim 10^{10}\,\gev
\ .\ee

\item[c)] Some eigenvalues of $M$ $\sim$ some eigenvalues of $Y_\nu\la\phi\ra$

Then, the sea-saw neutrino mass matrix will have more than three small
distinct eigenvalues (actually up to six), leading to more than three
light neutrinos. Such a possibility is constrained by phenomenology,
but is not excluded.

\end{itemize}
\end{enumerate}
Though the topic of neutrino masses and associated mixing and CP violation is
fascinating, it is not the flavor physics which is of interest to us
here. Thus, this is all that we will say about the subject and, for the
remainder of the course, we can safely take $m_\nu=0$, forgetting about
$O_\mathrm{Maj}^{(5)}$ and $\cL_\nu$ altogether.

Having explored the neutrino mass Lagrangian and some of the
constraints which neutrinos place on the scale of New Physics, we now
do the same for the other components of $\cL_\mathrm{SM}^\mathrm{eff}$,
generically denoting the scale of New Physics by $\Lambda$:
\begin{enumerate}

\item {\em EWSB and naturalness:} besides possible right-handed
  neutrinos, the Higgs boson is the only Standard Model particle whose
  mass is not protected by a symmetry from the physics at energy
  scales much larger than $M_W$. To get a very rough estimate of what
  the contributions of New Physics to the Higgs mass could be, we
  assume that the effect of the new degrees of freedom can be
  approximated by computing Standard Model loop corrections to this
  mass, cutoff at a scale $\Lambda\gg M_W$ that is characteristic of
  the new phenomena. Then, the contributions to the Higgs mass at one
  loop are given by the diagrams in \fig{fig:dmh2} with a cutoff
  $\Lambda$. They yield
\be
\labell{eq:deltamh2}
\delta M_H^2 = \frac{3\Lambda^2}{16\pi^2\la\phi\ra^2}(4m_t^2-2M_W^2-M_Z^2-M_H^2)
\ee
which is dominated by the top contribution for $M_H\ll
350\,\mev$. If, for naturalness reasons, we require that the physical 
squared Higgs mass is no less than a fraction $f$ of the correction of
\eq{eq:deltamh2}, then we find that
\bea
\Lambda_\mathrm{nat}&\le&\frac{4\pi\la\phi\ra}{\sqrt{3f}}\frac{M_H}{\sqrt{4m_t^2-
2M_W^2-M_Z^2-M_H^2}}\nonumber\\
&\sim& \frac{700\,\gev}{\sqrt
  f}\times\frac{(M_H/115\,\gev)}{\sqrt{1-\left(\frac{M_H-115\,\gev}{310\,\gev}
\right)^2}}
\labell{eq:Lambdanat}
\ .\eea
Thus, if we allow at most 1\% of fine tuning on the Higgs mass
squared, \eq{eq:Lambdanat} says that new physics must appear below
$\Lambda_\mathrm{nat}\sim 7\,\tev$.

\begin{figure}[t]
  \centering
  \includegraphics[width=0.9\textwidth]{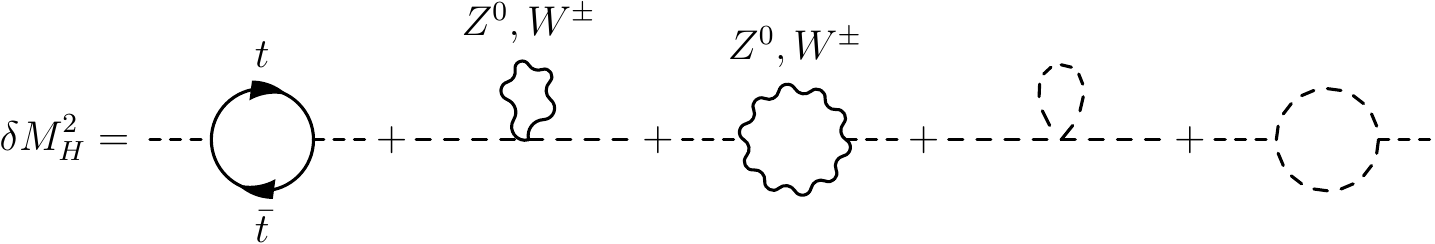}
  \caption{Diagrams which contribute radiative 
corrections, $\delta M_{H}^2$, to the Higgs mass squared at one loop}
\labell{fig:dmh2} 
\end{figure}

\item
{\em Gauge sector and flavor conserving $d=6$ operators:} consider, for
instance, $O_{WB}=g_1g_2(\phi^\dagger\sigma^a\phi)W_{\mu\nu}^a
  B_{\mu\nu}$, which couples the $W$ bosons to the $U(1)_Y$ gauge boson
  $B$. Precision electroweak data, assuming that the Wilson
  coefficient of this operator is of order one, impose the following
  constraint on the scale of New Physics~\shortcite{Barbieri:2004qk}:
\be
\Lambda \gsim 5\,\tev \qquad 95\%\, \mathrm{CL}
\ .
\ee

\item
{\em Flavor physics and, in particular, flavor changing neutral currents
(FCNC):} consider, for instance, $K^0$-$\bar K^0$ mixing. In the
absence of electroweak interactions, the long-lived $K^0_L$ is a CP
odd state, whereas the short-lived $K^0_S$ is a CP even state. When
these interactions are turned on, these two degenerate particles
acquire a minuscule mass difference, which is measured experimentally
to be:
\be
\Delta M_K\equiv M_{K_L}-M_{K_S}\simeq 3.5\times 10^{-12}\,\mev
\ .\ee
Consider now the contribution to $\Delta M_K$ of an arbitrary $d=6$,
$\Delta S=2$ operator schematically written as $(\bar ds)(\bar ds)$:
\be
\Delta M_K = 2\times \frac1{2M_K}\frac{\re C^*_{\Delta
    S=2}}{\Lambda^2}\la \bar K^0|(\bar sd)(\bar sd)|K^0\ra
\ .\ee
Now, assuming that $\re C^*_{\Delta S=2}\sim 1$ and that the matrix
element is of order the fourth power of a typical QCD scale,
e.g. $\sim M_\rho^4$, we get
\be
\Lambda > \frac{M_\rho^2}{\sqrt{M_K\,\Delta M_K}}\sim 10^3\,\tev
\, \ee
which is orders of magnitudes larger than the lower bound imposed by
gauge sector and flavor conserving transitions, as well as than the
upper bound imposed by naturalness.
\end{enumerate}
Thus, if we do not make any assumptions about how the New Physics
breaks flavor symmetries, we are forced to push this physics to very
high scales. Said differently, flavor physics is sensitive to very
high energy scales if the New Physics is allowed to have a flavor
structure which differs from that of the Standard Model. Therefore,
one assumption commonly made is that the New Physics breaks the flavor
symmetries with the same Yukawa couplings as in the Standard
Model. This assumption is called Minimal Flavor Violation (MFV). For
instance, we might have, in the case of $K^0$-$\bar K^0$ mixing, the
following operators contributing: $\frac1{\Lambda^2}(\bar s_RY_{sd}
d_L)^2$, $\frac1{\Lambda^2}(\bar s_LY_{sd}^*Y_{sd}\gamma_\mu d_L)^2$,
etc. I will leave you work out the corresponding scales, $\Lambda$,
but they are certainly much lower and in line with those obtained from
flavor-conserving physics.

\subsectiondum{Flavor physics phenomenology}

As we shall see shortly in more detail, the Standard Model has a 
very rich and constrained flavor structure, which includes:
\begin{itemize}
\item mixing of quark flavors;
\item CP violation by a unique invariant $J$
  \shortcite{Jarlskog:1985ht}, discussed in \sec{sec:jarlskog};
\item the absence of tree-level flavor changing neutral currents (FCNC).
\end{itemize}
All of these features are encapsulated in:
\begin{itemize}

\item the Cabibbo-Kobayashi-Maskawa (CKM) 
matrix~\shortcite{Cabibbo:1963yz,Kobayashi:1973fv}
\be
V=\left(\begin{array}{ccc}
V_{ud} & V_{us} & V_{ub}\\
V_{cd} & V_{cs} & V_{cb}\\
V_{td} & V_{ts} & V_{tb}
\end{array}\right)
\ ,\ee
which is unitary. It has 3 mixing angles and a single phase,
which is responsible for CP violation.

\item
the quark masses: $m_q$, with $q=u,d,s,c,b,t$.

\end{itemize}
For their discovery, in 1973, that Nature's CP violation and rich flavor
structure can be well described when a third generation is added to the
$SU(2)_L\times U(1)_Y$ electroweak model~\shortcite{Kobayashi:1973fv},
Kobayashi and Maskawa were awarded part of the 2008 Physics Nobel Prize.

Because this flavor structure is so intriguing and most probably
contains important information about physics at much higher energies
than currently explored,
particle physicists have invested a considerable amount of effort in
exploring it theoretically and experimentally over the past five
decades. This exploration has multiple goals:

\medskip
\noindent
1) To determine from experiment the matrix elements of the CKM
matrix $V$, which are important parameters of our fundamental theory.

\smallskip
\noindent
2) To verify that the CKM description of quark flavor mixing and CP violation
is correct, e.g.:
\begin{itemize}
\item Can all of the observed CP violation in the quark sector be explained in
  terms of a single phase?
\item Is the measured matrix $V$ unitary?
\end{itemize}

The latter can be tested by verifying whether
\be
\sum_{D=d,s,b}|V_{UD}|^2=1 \qquad\mbox{and}\qquad \sum_{U=u,c,t}|V_{UD}|^2=1
\ .\ee
If either of these sums turn out to be less than
1, that would signal an additional generation or family. On the other hand,
if either one is larger than 1, completely new physics would have to
be invoked. The unitarity of $V$ also implies that the scalar product
of any two distinct columns or rows of the matrix must vanish, i.e.
\bea
\labell{eq:D1D2triangle}
\sum_{U=u,c,t}V_{UD_1}V_{UD_2}^* &=& 0,  \qquad\mbox{for } D_1\ne D_2\ ,\\
\labell{eq:U1U2triangle}
\sum_{D=d,s,b}V_{U_1D}V_{U_2D}^* &=& 0,  \qquad\mbox{for } U_1\ne U_2
\ .
\eea
These relations can be represented as triangles in the complex
plane, which are traditionally labeled by their unsummed flavors, 
i.e. $(D_1,D_2)$ for those of 
\eq{eq:D1D2triangle} and $(U_1,U_2)$ for those of \eq{eq:U1U2triangle}. 
In the absence of CP violation, unitarity triangles would
become degenerate. The (db) triangle is shown in
\fig{fig:dbtriangle}. In \fig{fig:dbsbdsUTs}, it is drawn to scale with two
other triangles to give you a sense of the variety of unitarity
triangles and the difficulties there may be in measuring some of their
sides and angles.

\begin{figure}
\centering
\includegraphics[width=0.7\textwidth]{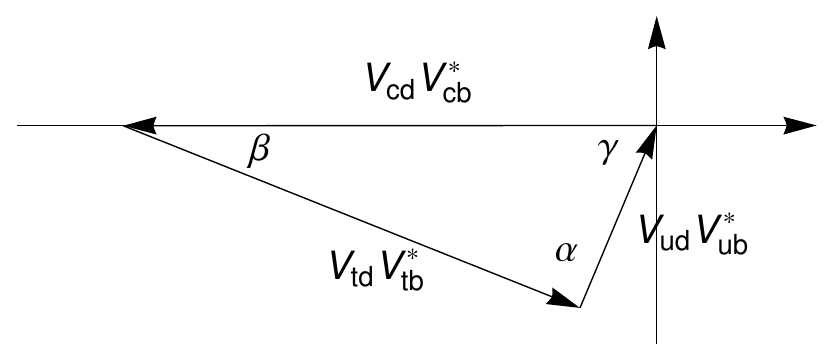}
\caption{The (db) unitarity triangle.}
\labell{fig:dbtriangle}
\end{figure}

\begin{figure}
\centering
\includegraphics[width=0.9\textwidth]{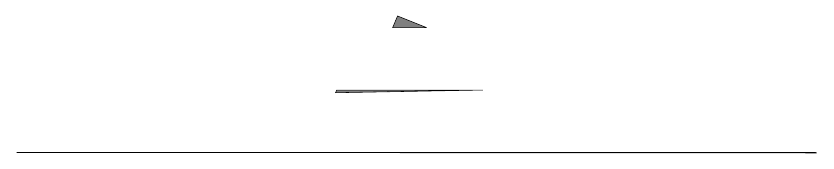}
\caption{From top to bottom, the (db), (sb) and (ds) unitary
  triangles, normalized by $V_{cD_1}V_{cD_2}^*/|V_{cD_1}V_{cD_2}^*|$ with $(D_1,D_2)=(d,b)$, $(s,b)$ and $(d,s)$ 
respectively, and drawn to a common scale.}
\labell{fig:dbsbdsUTs}
\end{figure}

The (db) triangle has been the focus of considerable experimental
(LEP, $B$-factories, Tevatron, $\ldots$) and theoretical (QCD
factorization, Soft Collinear Effective Theory (SCET), lattice QCD
(LQCD), $\ldots$) effort in the last ten to fifteen years. With the
arrival of the LHC and, in particular, the experiment LHCb, the focus
is shifting from the study of the $B_d$ towards the study of the $B_s$
meson and thus toward investigations of the (sb) triangle. Here
too, lattice QCD has a considerable role to play, most notably in the
study of the $B_s$-$\bar B_s$ mass and width differences, of the leptonic
decay $B_s\to\mu^+\mu^-$ or of the semileptonic decay 
$B_s\to\phi\mu^+\mu^-$.

The strategy here is to verify the unitarity of the CKM matrix
by performing redundant
measurements of: 
\begin{itemize}
\item triangle sides with CP conserving decays,
\item angles with CP violating processes,
\end{itemize}
and checking that the triangles indeed close.

\smallskip
\noindent
3) To determine in what processes there is still room for significant
New Physics contributions. For instance, $O(40\%)$ effects are
still possible in $B^0$-$\bar B^0$ mixing from New Physics with a
generic weak phase~\shortcite{Lenz:2010gu}.

\smallskip
\noindent
4) To constrain the flavor sectors of beyond the Standard Model (BSM)
candidates. As we saw above, it is difficult to add new physics to the
Standard Model without running into serious problems in the flavor
sector.

\smallskip
\noindent
5) To actually find evidence for beyond the Standard Model
physics. This is most likely to be found in processes which are highly
suppressed in the Standard Model, such as FCNC.

\smallskip
\noindent
6) If new particles and interactions are discovered, it is important
to investigate their quark and flavor structure.

\medskip

All of these goals require being able to compute reliably and
precisely flavor observables in the Standard Model or beyond. A
high level of precision has been reached already on the magnitudes of
individual CKM matrix elements~\cite{Charles:2004jd}:
\be
\begin{array}{rcl}
& & \hspace{1.8cm} d \hspace{2.3cm} s \hspace{2.3cm} b\\
|V|&=&\begin{array}{c}u \\[0.2cm] c \\[0.2cm] t\end{array}
\left(
\begin{array}{ccc}
0.97425^{+0.00018}_{-0.00018} &
0.22543^{+0.00077}_{-0.00077} &
0.00354^{+0.00016}_{-0.00014} \\[0.2cm]
0.22529^{+0.00077}_{-0.00077} &
0.97342^{+0.00021}_{-0.00019} &
0.04128^{+0.00058}_{-0.00129} \\[0.2cm]
0.00858^{+0.00030}_{-0.00034} &
0.04054^{+0.00057}_{-0.00129} &
0.999141^{+0.000053}_{-0.000024}
\end{array}
\right)
\end{array}
\ ,\ee
assuming the correctness of the Standard Model and, in particular,
CKM unitarity.  The most poorly known CKM matrix elements are
$|V_{ub}|$ and $|V_{td}|$, both with an uncertainty around 4\%. Then
come $|V_{ts}|$ and $|V_{cb}|$, with an uncertainty of about 2\%. Thus,
to have an impact in testing the CKM paradigm of quark flavor mixing
and CP violation, and to take full advantage of LHCb results, the
precision of theoretical predictions must be of order a few percent
(better in many cases). This is no small challenge when
nonperturbative QCD dynamics is involved.

\subsectiondum{Flavor physics and lattice QCD}

Lattice QCD plays and will continue to play a very important role in
flavor physics, by providing reliable calculations of nonperturbative
strong interaction corrections to weak processes involving quarks.

The processes for which LQCD gives the most reliable predictions are
those which involve a single hadron that is stable against strong
interaction decay in the initial state and, at most, one stable hadron
in the final state. Resonances (i.e. unstable hadrons) are much more
difficult to contend with, especially if many decay channels are
open. Similarly, final states with more than a single stable hadron
are much more difficult, especially if these hadrons can rescatter
inelastically. This will be discussed in much more detail in
\sec{sec:kpipiinfv}.

Thus, the processes typically considered for determining the absolute
values of the CKM matrix elements are the following
%
\be
\begin{array}{ccc}

|V_{ud}| & |V_{us}| & |V_{ub}|\\
\quad\includegraphics[width=0.25\textwidth]{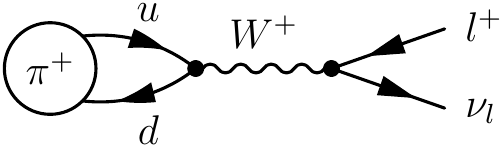}\quad
&
\quad\includegraphics[width=0.25\textwidth]{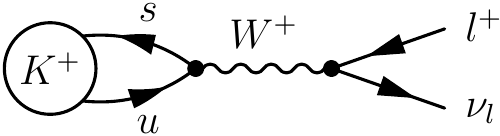}\quad
&
\quad\includegraphics[width=0.25\textwidth]{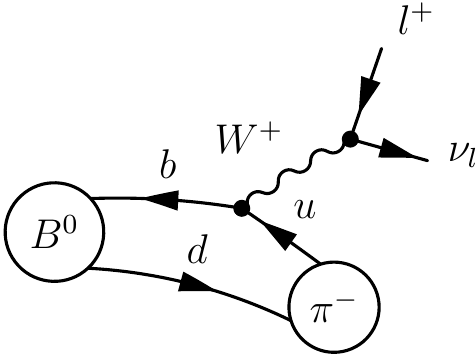}\quad\\

&
\includegraphics[width=0.25\textwidth]{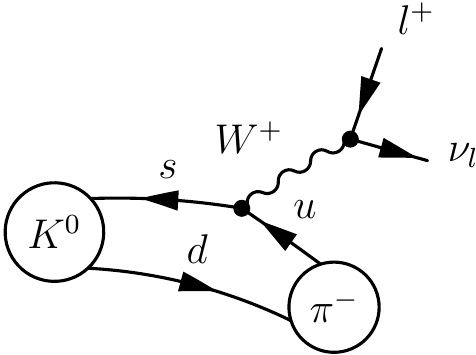}
&\\[0.3cm]

\end{array}
\nonumber
\ee

\be
\begin{array}{ccc}

|V_{cd}| & |V_{cs}| & |V_{cb}|\\
\quad\includegraphics[width=0.25\textwidth]{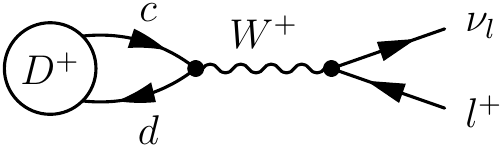}\quad
&
\quad\includegraphics[width=0.25\textwidth]{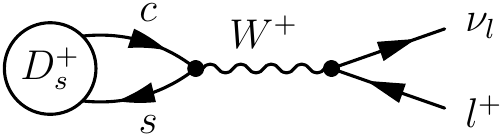}\quad
&
\quad\includegraphics[width=0.25\textwidth]{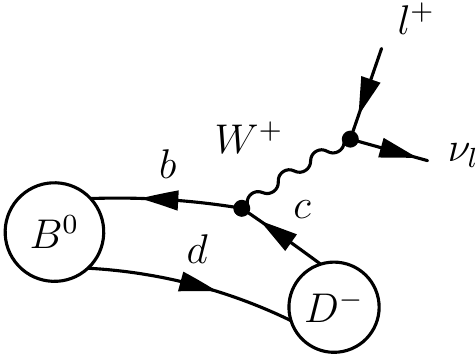}\quad\\

\includegraphics[width=0.25\textwidth]{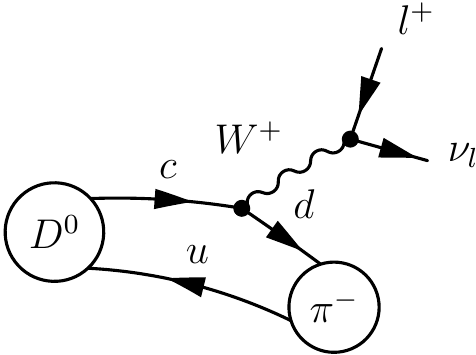}
&
\includegraphics[width=0.25\textwidth]{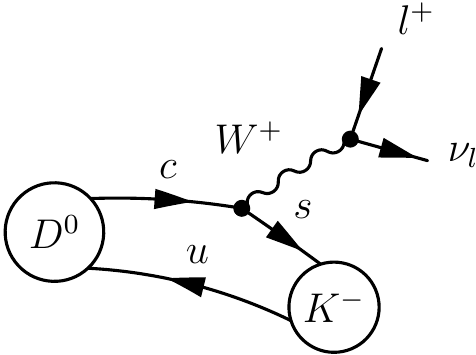}
&\\\\[0.3cm]
\end{array}
\nonumber
\ee

\be
\begin{array}{c@{\hspace{1.2cm}}c@{\hspace{1.2cm}}c@{\hspace{1.2cm}}}

|V_{ct}| & |V_{ts}| & |V_{tb}|\\[0.3cm]
\quad\includegraphics[width=0.25\textwidth]{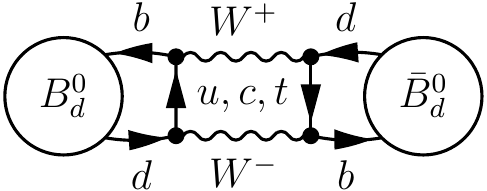}\quad
&
\quad\includegraphics[width=0.25\textwidth]{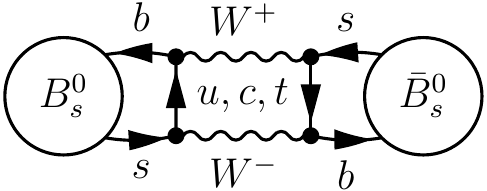}\quad
&
\end{array}
\nonumber\ee

Now, to determine the unique CKM matrix phase or, more precisely, the
CP violating parameter $J$, lattice QCD can have an important impact
through the
following processes:
\begin{itemize}
\item Indirect CP violation in $K\to\pi\pi$ decays. This occurs through the
  process of $K^0$-$\bar K^0$ mixing, which is given by the imaginary part
$$
\im\;\;\left[\parbox{0.30\textwidth}{\includegraphics[width=0.30\textwidth]{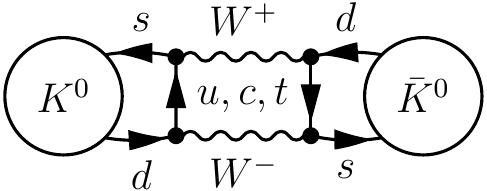}}+\;\;\cdots\right]\ ,
$$
where the ellipsis stands for the other box contribution. 
It is a $|\Delta S|=2$ FCNC that will be discussed in detail below.

\item Direct CP violation in $K\to\pi\pi$ decays. The processes which
  contribute are given by the following $|\Delta S|=1$ amplitudes
$$
\im\left\{\parbox{0.30\textwidth}{\includegraphics[width=0.30\textwidth]{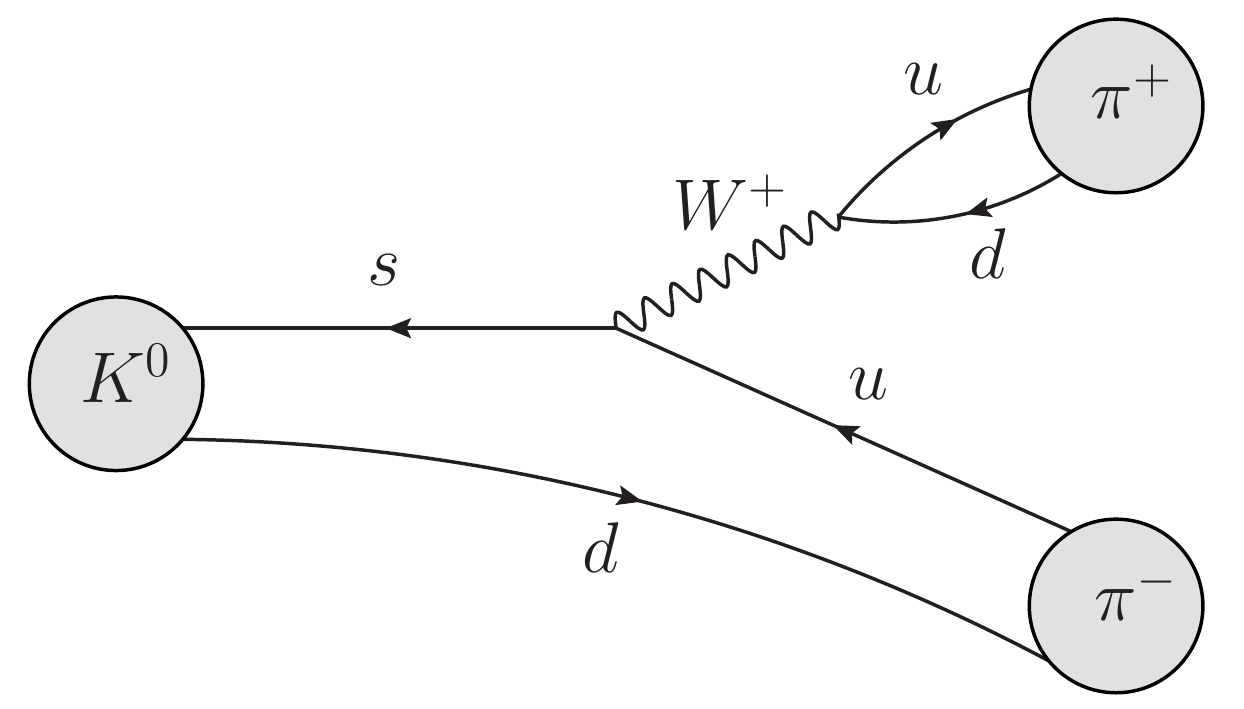}}\qquad+\qquad\parbox{0.30\textwidth}{\includegraphics[width=0.30\textwidth]{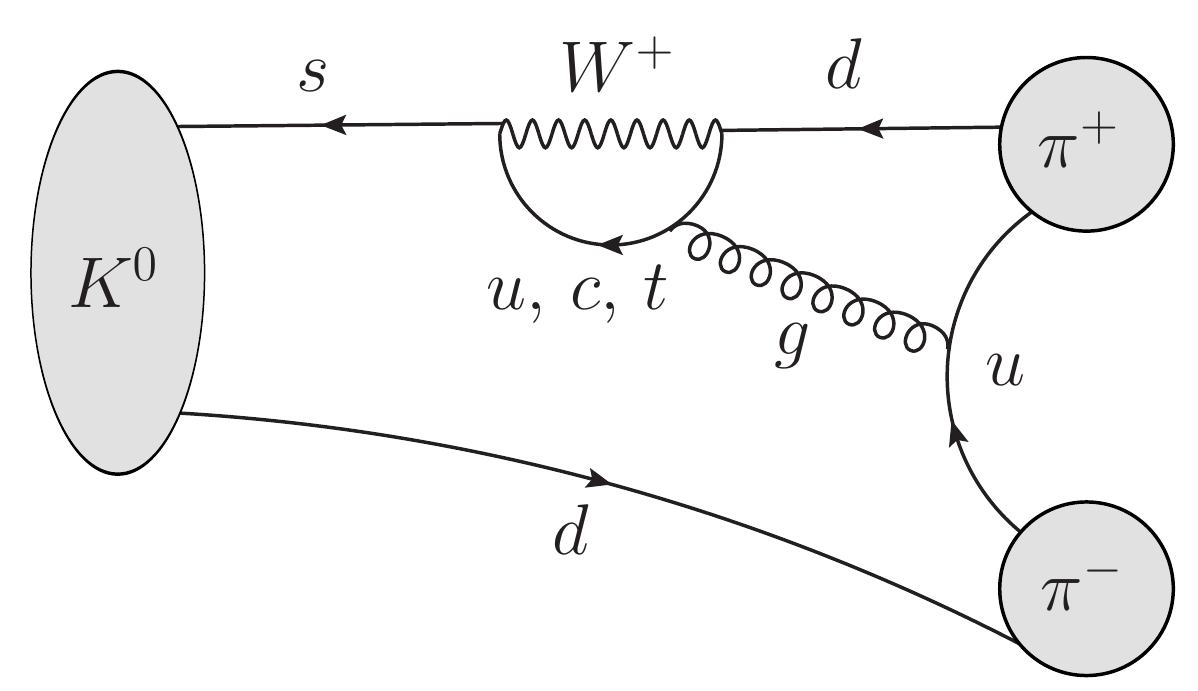}}\r.
$$
$$
+\qquad\parbox{0.30\textwidth}{\includegraphics[width=0.30\textwidth]{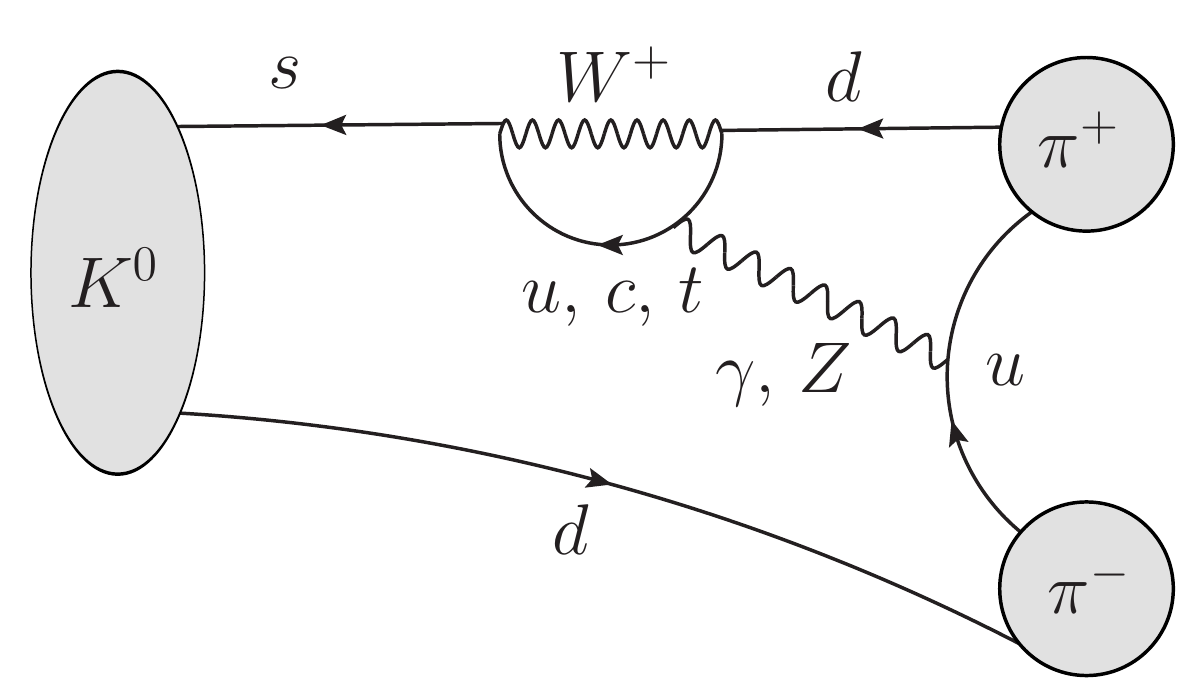}}+\qquad\parbox{0.30\textwidth}{\includegraphics[width=0.30\textwidth]{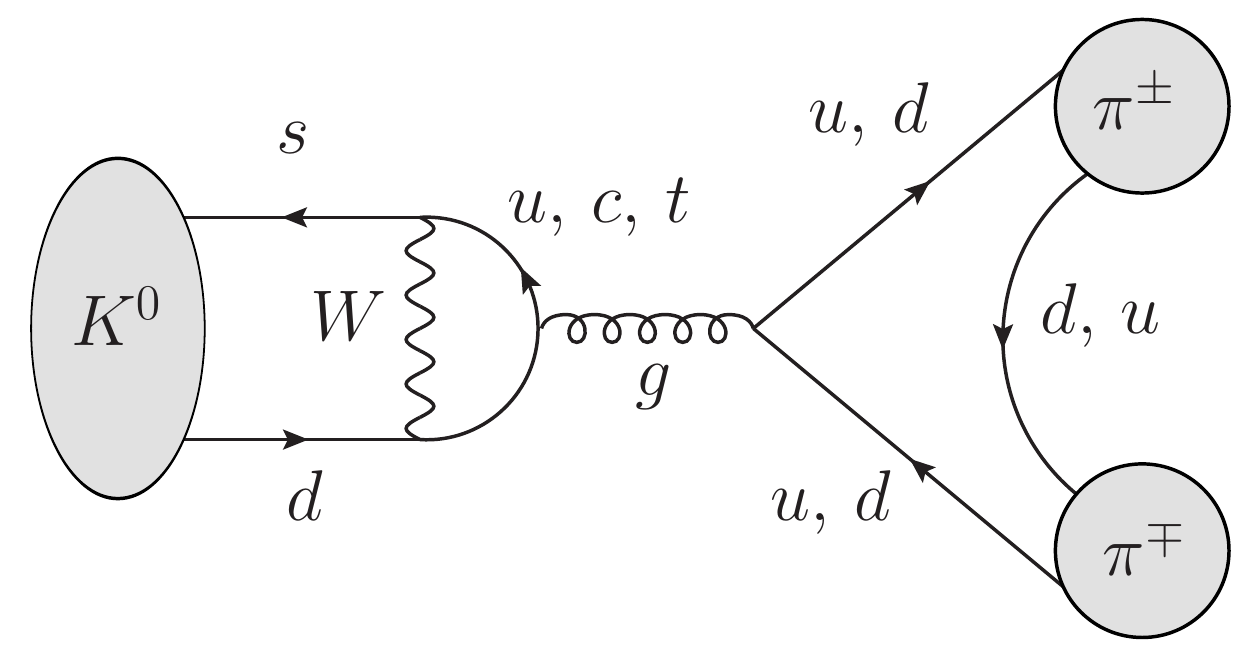}}
$$
$$
\left.+\qquad\parbox{0.30\textwidth}{\includegraphics[width=0.30\textwidth]{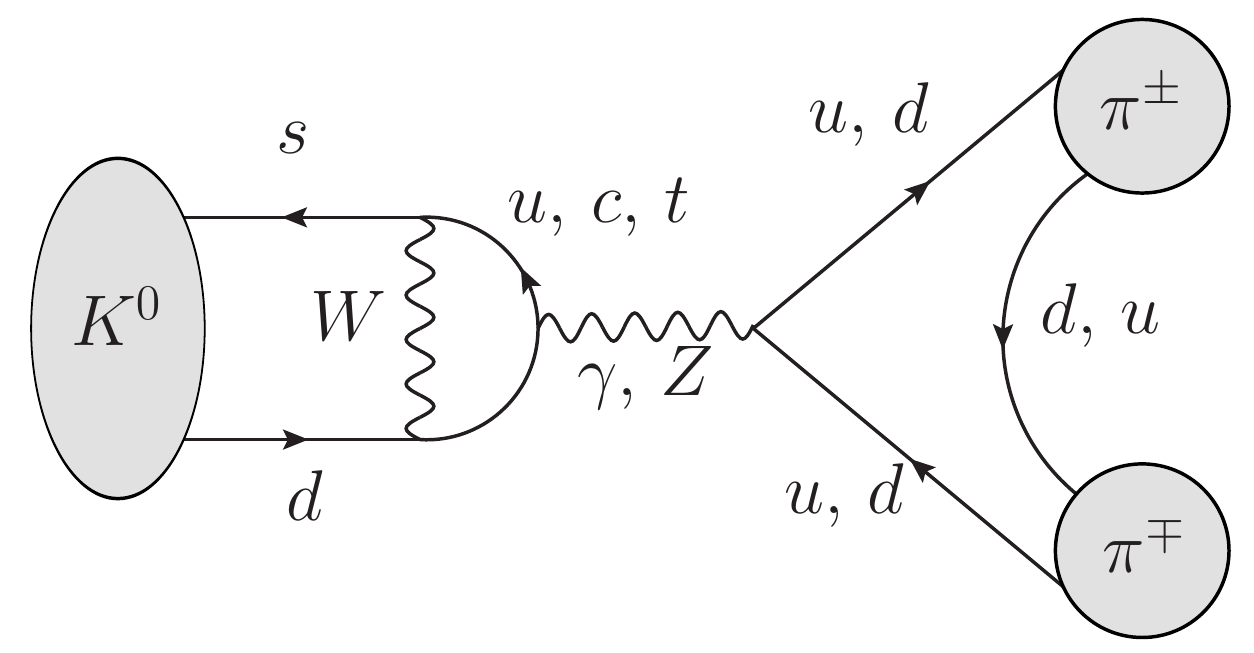}}+\cdots\right\}
$$
in the case of $K^0\to\pi^+\pi^-$, where the ellipsis stands for missing 
diagrams similar to those drawn. The diagrams for
$K^0\to\pi^0\pi^0$ are analogous. Again, this particular phenomenon will be
discussed in detail in the sequel.
\end{itemize}

\subsectiondum{Low-energy effective field theories of the Standard Model}

With present knowledge and present computer resources, we cannot
simulate the full Standard Model in lattice field theory calculations.
In particular various degrees of freedom must be ``eliminated'' from
the calculations for the following reasons:
\begin{itemize}

\item
$W$, $Z$ and $t$: there is no hope to be able to simulate these
  degrees of freedom whose masses are $M_{W,Z}\sim 80\div 90\,\gev$
  and $m_t\sim 175\,\gev$ on lattices which must be large enough to
  accommodate $135~\mev$ pions, i.e. with sizes $L\gsim 4/M_\pi\sim
  6\,\fm$.~\footnote{The factor of 4 in $4/M_\pi$ is a conservative
    rule-of-thumb estimate which guarantees that finite-volume
    corrections to stable hadron masses, proportional
    to $e^{-M_\pi L}$, are typically below the percent level.} Since we
  would also have to have $am_t\ll 1$, with $a$ the lattice spacing,
  to guarantee controlled discretization errors, the number of points
  on the lattice would have to be $L/a\gg 4m_t/M_\pi\sim 5.2\times
  10^3$, which is beyond any foreseeable computing capabilities. Perhaps even
  more important, however, is the fact that we just do not know how to
  discretize nonabelian, chiral gauge theories (please see David's
  lecture notes in this volume~\shortcite{david}).

\item $b$: even the $b$ quark, with $m_b\sim 4.2\,\gev$, would require
  lattices with $L/a\gg 120$, which is already too much for present
  technology.

\item $c$: with $m_c\sim 1.3\,\gev$, the charm is a borderline case,
  both on the lattice and in terms of QCD. On the lattice it can be
  included in simulations, but with $am_c\sim 0.35$ at best,
  discretization errors remain an important preoccupation. From the
  point of view of QCD, the charm is not quite a heavy quark--heavy
  quark effective theory is only marginally applicable since $m_c$ is
  not much larger than typical QCD scales--and it is clearly not
  light--it is not in the regime of chiral perturbation theory. So its
  inclusion should be considered with care. In addition, its inclusion
  in weak processes can mean significantly more complicated
  correlation functions to compute. For instance, \fig{fig:kkbarcharm}
  illustrates the type of correlation functions required to determine
  the amplitude for $K^0$-$\bar K^0$ mixing, in the absence of charm
  (diagram on left) and in the presence of a dynamical charm quark
  (diagram on right). In the absence of charm, it is a rather standard
  three-point function which must be computed while in its presence,
  it is a four-point function, with two four-quark operator
  insertions. However, in certain circumstances, the inclusion of a
  dynamical charm quark significantly simplifies the renormalization
  of the weak effective theory, as briefly discussed in
  \sec{sec:DI12inSM} for the $\Delta I=1/2$ rule.

\begin{figure}[t]
\centering
\includegraphics[width=0.4\textwidth]{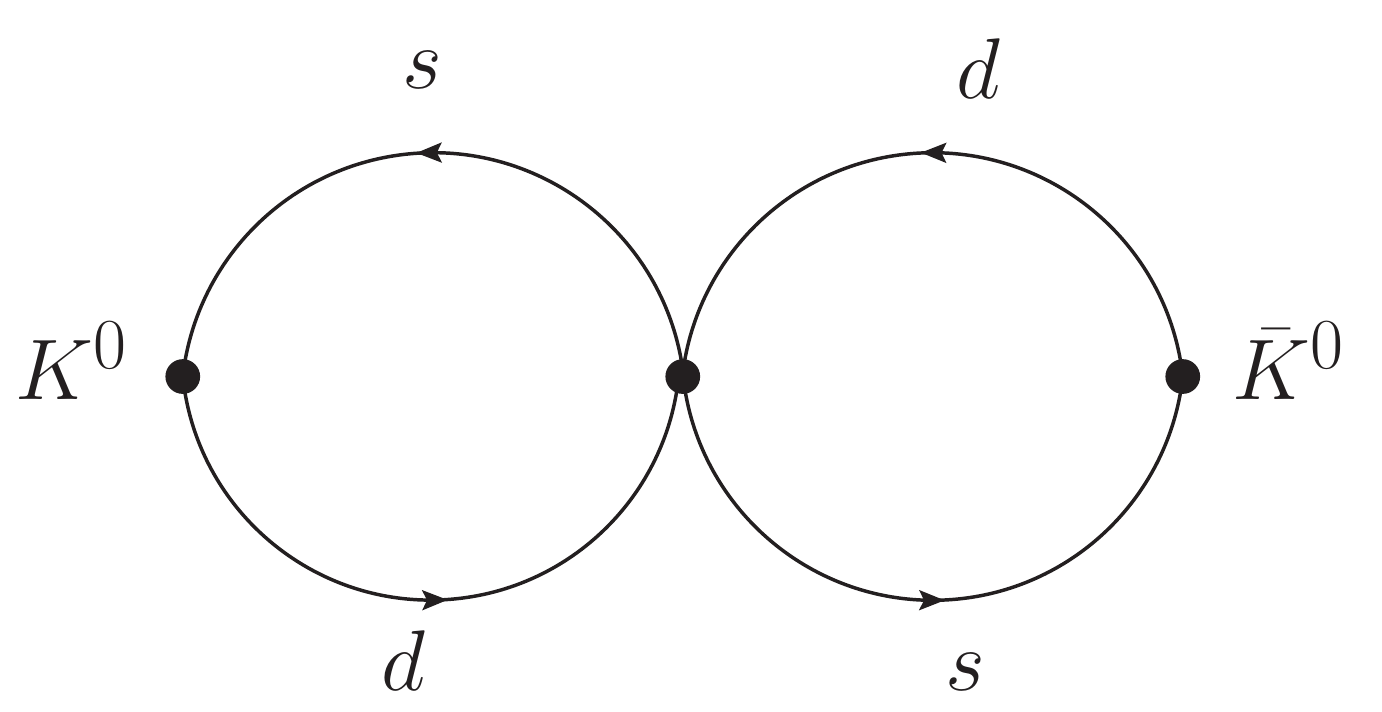}
\qquad
\includegraphics[width=0.4\textwidth]{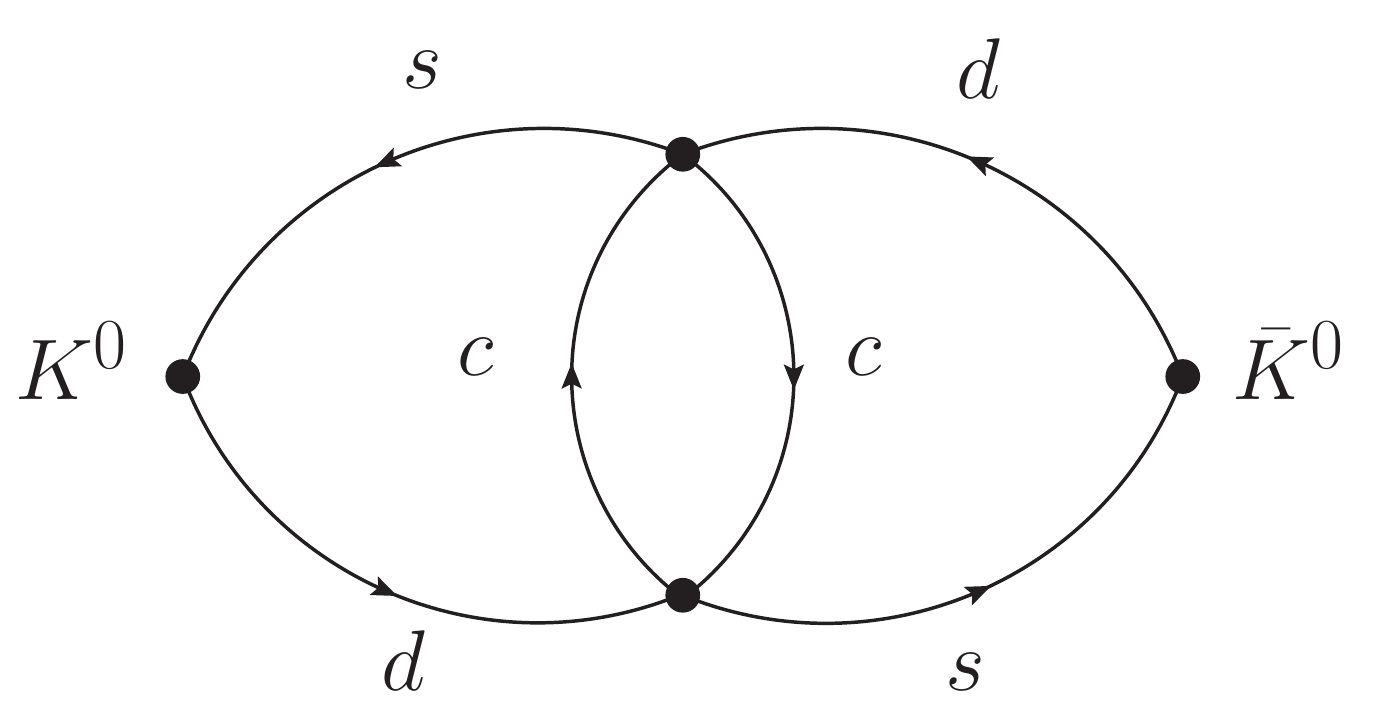}
\caption{Examples of correlation functions required for the lattice
  computation of the $K^0$-$\bar K^0$ mixing amplitude. The diagram on the left
  exhibits the type of three-point function with a four-quark operator
  insertion, required to obtain the $K^0$-$\bar K^0$ amplitude when
  the charm quark is integrated out. The diagram on the right shows a
  four-point function which is required when the charm quark is kept
  active. } \labell{fig:kkbarcharm}
\end{figure}

\end{itemize}

Fortunately, with the precisions required at present and in any
foreseeable future, it is not necessary to include virtual $W$, $Z$,
$t$ and $b$ contributions. The situation with the charm is less clear,
as $(M_\rho/m_c)^2$ sea corrections can, in principle, play a role
when percent precisions are reached. So in considering processes
involving these massive particles, we can turn to effective field
theories (EFTs) in which the $W$, $Z$, $t$, $b$ and possibly $c$ and
$\tau$ are no longer dynamical degrees of freedom. Thus, we are left
with an $SU(3)_c\times U(1)_\mathrm{EM}$ gauge theory of color and
electromagnetism. This theory includes the virtual effects of the following
degrees of freedom:
\begin{itemize}
\item 2 massless gauge bosons: the qluon, $g$, and the photon, $\gamma$
\item 3 to 4 quarks: the $u$, $d$, $s$ and possibly the $c$
\item 2 to 3 leptons: the $e$, $\mu$ and possibly the $\tau$
\item 3 neutrinos: the $\nu_e$, $\nu_\mu$ and the $\nu_\tau$.
\end{itemize}
It also includes local operators of naive mass-dimension $d\ge 5$,
which result from integrating out the heavy $W$, $Z$ and
$t$. Moreover, in this theory, the $b$--and possibly the $c$ and the
$\tau$--are described by heavy fermion effective theories, in which
the antiparticle of the fermion considered is integrated out. It is at
this level that LQCD enters, to describe the nonperturbative effects
of the strong interaction.

\sectiondum{Standard Model and quark flavor mixing}

As already discussed, the Standard Model has a highly constrained
quark flavor structure, parametrized by the CKM matrix and the quark
masses. We will now see more precisely how it arises and what
its basic implications are.

\subsectiondum{On the origin of quark flavor mixing in the Standard Model}
\labell{sec:CKMorigin}

Let us look in more detail at the quark and gauge sectors of the
Standard Model. With the notation
$(\mbox{dim. rep.}\,SU(3)_c,\,\mbox{dim. rep.}\,SU(2)_L)_{Y}$,
where ``rep.'' stands for representation, the quark content of the
Standard Model is given by
\bea
Q_L=\left(\begin{array}{c}U_L\\D_L\end{array}\right)\sim
  (3,2)_\frac12\nonumber\\
U_R\sim (3,1)_{2/3}\\
D_R\sim (3,1)_{-1/3}\nonumber
\ ,
\eea
with $U=(u,c,t)^T$ and $D=(d,s,b)^T$. The coupling of these quarks
with the gauge bosons is given by
\be
\cL_\mathrm{g+q}= \cdots + \bar Q_L\Dslash Q_L + \bar U_R\Dslash U_R
+  D_R\Dslash D_R
\ ,
\ee
where the ellipsis stands for the gauge kinetic and self-interaction
terms. This Lagrangian has the following global symmetries: $U(3)_L$ on
$Q_L$, $U(3)_{U_R}$ on $U_R$ and $U(3)_{D_R}$ on $D_R$.

Now let us look at the quark Yukawa terms. After spontaneous symmetry
breaking by the Higgs field, keeping only the terms proportional to
the Higgs v.e.v. $\la \phi\ra$, we have:
\be
\cL_m^q \stackrel{\la\phi\ra\ne 0}{\longrightarrow }  - \bar
U_R M_U U_L - \bar D_R M_D D_L + \hc
\ ,\ee
where $M_U$ and $M_D$ are arbitrary $3\times 3$ complex matrices. On
these terms we perform the following set of flavor transformations
which leave $\cL_{g+q}$ invariant:
\begin{itemize} 
\item the $SU(3)_{U_R}$ transformation: $U_R\to V_R^U U_R$,
\item the $SU(3)_L$ transformation: $U_L\to V_L^U U_L$ and $D_L\to V_L^U D_L$,
\item and the $SU(3)_{D_R}$ transformation: $D_R\to V_R^D D_R$,
\end{itemize}
with $V_R^U$, $V_L^U$ and $V_R^D$ such that:
\be
M_U^d=V_R^{U\dagger} M_U V_L^U
\ ,\ee
and
\be
V_R^{D\dagger}M_D=M_D^dV_L^{D\dagger}
\ee
 where $M_{U,D}^d$ are diagonal matrices with real positive entries
 which are the quark masses.  The second equation
 defines a fourth unitary matrix $V_L^D$. Under these
 rotations, the quark mass Lagrangian transforms as
\be
\labell{eq:Lmqdiag}
\cL_m^q\to -\bar U_R M_U^d U_L - \bar D_R M_D^d[V_L^{D\dagger}V_L^U D_L]+\hc
\ .
\ee
The up quark mass matrix is diagonal in this basis, but not the down
quark matrix. In addition, we have exhausted the flavor transformations
allowed by $\cL_{g+q}$. Thus, if we want to work in a mass basis
(i.e. a basis in which all quark masses are diagonal), we have to
perform the additional transformation:
\begin{itemize}
\item $D_L\to V_L^{U\dagger}V_L^D D_L$,
\end{itemize}
which is not a symmetry of $\cL_{g+q}$. Clearly, the only terms
in $\cL_{g+q}$ which are affected are those which couple
$U_L$ and $D_L$. They are transformed in the following way:
\bea
\cL_{CC}^q&=&\frac{g_2}{\sqrt 2} \bar U_L \slash W^{(+)} D_L +
\hc
\nonumber\\
&\to& \frac{g_2}{\sqrt 2} \bar U_L \slash W^{(+)} V D_L+
\hc
\ ,
\eea
where $CC$ stands for charged current and $V\equiv V_L^{U\dagger}V_L^D$ is
the CKM matrix. All other terms are left unchanged. In particular, the
neutral currents:
\bea
\cL_{NC}^q &=& \frac{g_2}{2} (\bar U_L,\bar D_L) \slash
W^3
\left(\begin{array}{cc}1&0\\0&-1\end{array}\right)\left(\begin{array}{c}
U_L\\D_L\end{array}\right)\nonumber\\
&& +\frac{g_1}{6} (\bar U_L,\bar D_L)\slash B \left(\begin{array}{c}
U_L\\D_L\end{array}\right)\\
&& +\frac{2}{3} g_1 \bar U_R\slash B U_R - \frac{1}{3} g_1 \bar D_R
\slash B D_R\nonumber
\ ,\eea
which are flavor diagonal, remain diagonal in the mass
basis. Moreover, the Higgs couples to fermions through their masses
and therefore has a diagonal coupling to the quarks in the mass
basis. The fact that all uncharged couplings remain diagonal in the
mass basis implies that there are no tree level FCNC transitions in
the Standard Model.

\subsectiondum{Properties of the CKM matrix}

In this section, we look in more detail at what are the key properties
of the CKM matrix.

\subsubsectiondum{Degrees of freedom of the CKM matrix}
\labell{sec:CKMdof}

The flavor eigenstates, $(d',s',b')$, are related to the mass
eigenstates, $(d,s,b)$, through:
\be
\left(\begin{array}{c}d'\\s'\\b'\end{array}\right)=V\left(\begin{array}{c}d\\s\\b\end{array}\right)
\ ,\ee
where the CKM matrix is unitary, i.e. $V^\dagger
V=VV^\dagger=1$. Since $V$
is a $3\times 3$ complex matrix, it has 9 phases and 9
moduli. Unitarity imposes 3 real and 3 complex constraints. Thus, we
are left with 6 phases and 3 moduli which, because of the
normalization of the rows and columns to one, can be written as
angles.

Now, aside from the CC interactions, all other terms involving quarks
are diagonal in flavor, and have LL, LR, RL and RR chiral
structures. Thus, we can perform vector (not axial) phase rotations on
each flavor and leave all of these other terms invariant. However,
under the phase rotations
\be
U_{L,R}\to e^{i\theta_U} U_{L,R}\qquad\mbox{and}\qquad D_{L,R}\to e^{i\theta_D} 
D_{L,R}
\ ,
\labell{eq:rephase}\ee
with $U=u$, $c$ or $t$ and $D=d$, $s$ or $b$, the $UD$ component of
the CKM matrix transforms as
\be
V_{UD}\to V_{UD} \, e^{i(\theta_D-\theta_U)}
\ .\ee
These transformations can be used to eliminate phases
in $V$. Although there are 6 phases $\theta_U$ and $\theta_D$, only 5
phase differences $\theta_D-\theta_U$ are independent. Thus, only 5 of
the 6 phase can be eliminated.

We have now exhausted the field transformations that can be used
to reduce the CKM matrix' degrees of freedom. Thus, $V$ has 3
angles and 1 phase.

\subsubsectiondum{Standard parametrization of the CKM matrix}

The idea behind this parametrization is to write $V$ as a product of
3 rotations between pairs of generations, throwing the phase into the
$1\to 3$ rotation, so that it multiplies the smallest mixing
coefficients. Thus,
\be
V=R_{32}\,\mathrm{diag}\{1,1,e^{i\delta}\}R_{31}\{1,1,e^{-i\delta}\}R_{21}
\ ,\ee
with the rotations
\be
R_{21}=\left(\begin{array}{ccc}c_{12}&s_{12}&0\\-s_{12}&c_{12}&0\\0&0&1\end{array}
\right)
\ ,\ee
$c_{12}=\cos\theta_{12}$ and $s_{12}=\sin\theta_{12}$, and similarly
for the other rotations. This yields the following expression for the
CKM matrix:
\be
V=\left(\begin{array}{ccc}
c_{12}c_{13} & c_{13}s_{12} & s_{13}e^{-i\delta}\\
-s_{12}c_{23}-c_{12}s_{23}s_{13}e^{i\delta} & c_{13}c_{23}-s_{12}s_{23}s_{13}e^{i\delta} & c_{13}s_{23}\\
s_{12}s_{23}-c_{12}c_{23}c_{13}e^{i\delta} & -c_{12}s_{23}-c_{23}s_{13}s_{13}e^{i\delta} & c_{13}c_{23}\\
\end{array}\right)
\ ,
\labell{eq:standardVCKM}\ee
and the angles are chosen to lie in the first quadrant. Note that this 
parametrization is not rephasing invariant.

\subsubsectiondum{Wolfenstein parametrization}

Experimentally, it is found that $1\gg s_{12}\gg s_{23}\gg s_{13}$,
i.e. mixing gets smaller as one moves off the diagonal. It is
convenient to exhibit this hierarchy by expanding $V$ in powers of
$s_{12}$, i.e. in the sine of the Cabibbo angle $\theta_{12}$
\shortcite{Cabibbo:1963yz,GellMann:1960np}. 
This yields the Wolfenstein parametrization
\shortcite{Wolfenstein:1983yz}. To implement this expansion,
we define~\shortcite{Buras:1994ec}
\bea
\lambda &\equiv&
s_{12}=\frac{|V_{us}|}{\sqrt{|V_{ud}|^2+|V_{us}|^2}}\nonumber\\
A\lambda^2 &\equiv& s_{23} =\frac{|V_{cb}|}{\sqrt{|V_{ud}|^2+|V_{us}|^2}}\\
A\lambda^3(\rho+i\eta) &\equiv& s_{13}e^{i\delta}=V_{ub}^*\nonumber
\ .\eea
and make the appropriate replacement in the standard parametrization of 
\eq{eq:standardVCKM}.Then,
\be
V = \left(\begin{array}{ccc}1-\lambda^2/2 & \lambda &
  A\lambda^3(\rho-i\eta)\\
-\lambda & 1-\lambda^2/2 & A\lambda^2\\
A\lambda^3(1-\rho-i\eta) & -A\lambda^2 & 1\end{array}\right)
+O(\lambda^4)\ ,
\ee
which clearly exhibits the hierarchy of mixing.

\subsubsectiondum{CP violation}

Let us now see how CP violation arises in the Standard Model. Under
parity, the charged $W$-boson fields transform as
\be
W_{\mu}^{(\pm)}(x)\stackrel{P}{\longrightarrow}W^{(\pm)\mu}(x_P)
\ ,\ee
with $x_P=(x^0,-\vec{x})$. Similarly, under charge conjugation,
\be
W_{\mu}^{(\pm)}(x)\stackrel{C}{\longrightarrow}-W_{\mu}^{(\mp)}(x)
\ .\ee
Thus, under CP, these bosons transform as
\be
W_{\mu}^{(\pm)}(x)\stackrel{CP}{\longrightarrow}-W^{(\mp)\mu}(x_P)
\ .\ee
In my favorite Dirac spinor basis, the parity transform of a fermion
field is given by:
\be
\left(\begin{array}{c}\psi_L\\ \psi_R\end{array}\right)(x)
\stackrel{P}{\longrightarrow}\left(\begin{array}{c}\psi_R\\ \psi_L
\end{array}\right)(x_P)
\ .\ee
In this basis,
\be
\gamma^\mu=\left(\begin{array}{cc} 0 & \sigma^\mu \\
\bar \sigma^\mu & 0 \end{array}\right)\qquad\mbox{and}\qquad
\gamma^5=\left(\begin{array}{cc} -I & 0 \\
 0 & I \end{array}\right)
\ ,\ee
where $I$ is the two-by-two unit matrix,
$\sigma^\mu=(I,\vec{\sigma})$,
$\bar\sigma^\mu=(I,-\vec{\sigma})$ and $\vec{\sigma}$ are the Pauli
matrices. Clearly, parity is not a symmetry
of the Standard Model since left and right fermions belong to
different representations of the Standard Model group. Under charge
conjugation, we have
\be
\psi(x)=\left(\begin{array}{c}\psi_L\\ \psi_R\end{array}\right)(x)
\stackrel{C}{\longrightarrow}\psi^c(x)=\left(\begin{array}{c}i\sigma^2
\psi_R^*\\ 
-i\sigma^2\psi_L^*
\end{array}\right)(x)=i\gamma^2\gamma^0\bar\psi^T(x)=C\bar\psi^T(x)
\labell{eq:Cfermion}\ .\ee
Again, charge conjugation
is clearly not a Standard Model symmetry. However, the CP operation,
\be
\left(\begin{array}{c}\psi_L\\ \psi_R\end{array}\right)(x)
\stackrel{CP}{\longrightarrow}\left(\begin{array}{c}-i\sigma^2\psi_L^*\\ 
i\sigma^2\psi_R^*
\end{array}\right)(x_P)=\gamma^0C\bar\psi^T(x_P)
\labell{eq:CPfermion}\ ,\ee
has a chance of being a symmetry transformation as it does not mix
left and right-handed fields. Using the well known Pauli matrix
identity, $\sigma^2\sigma^i\sigma^2=-\sigma^{i*}$ and the
anticommutation of fermion fields, the CC quark term of
the Standard Model Lagrangian transforms, under CP, as:
\be
\frac{g_2}{\sqrt2}\left\{\bar U_L\slash W^{(+)}VD_L+\bar D_L\slash
W^{(-)}V^\dagger U_L\right\}
\stackrel{CP}{\longrightarrow}
\frac{g_2}{\sqrt2}\left\{\bar U_L\slash W^{(+)}V^*D_L+\bar D_L\slash
W^{(-)}V^T U_L\right\}
\ .\ee
Since $V^*\ne V$ in the presence of a nonvanishing phase, $\delta$,
CP is potentially violated in the Standard Model. We will see below
what the necessary conditions for CP to be violated are.

\subsectiondum{CP violation and rephasing invariants}

The standard parametrization of the CKM matrix $V$, given in 
\eq{eq:standardVCKM}, corresponds to a particular choice of quark
field phases. Observables cannot depend on such
choices. Therefore, it is important to find rephasing invariant
combinations of CKM matrix elements.

\subsubsectiondum{Quadratic invariants}

The moduli
\be
I_{UD}^{(2)}\equiv \vert V_{UD}\vert^2
\ ,\ee
with $U=u,c,t$ and $D=d,s,b$, are clearly rephasing invariant. There are 9 of
these.

Now, unitarity requires that:
\be
\left\{
\begin{array}{l}
\sigma_U=\sum_{D=d,s,b} I_{UD}^{(2)}=1\\
\sigma_D=\sum_{U=u,c,t} I_{UD}^{(2)}=1
\end{array}\right.
\ ,\ee
which yields 6 constraints on the $I_{UD}^{(2)}$. However, we clearly
have $\sum_U\sigma_U=\sum_D\sigma_D$, which means that there are only
5 independent constraints. In turn, this means that there are 4
independent quadratic invariants $I_{UD}^{(2)}$, which are obviously
real.

\subsubsectiondum{Quartic invariants}
\labell{sec:CKMquartinv}

We now define
\be
\labell{eq:I4}
I_{U_1D_1U_2D_2}^{(4)}\equiv V_{U_1D_1}V_{U_2D_2}V_{U_1D_2}^*V_{U_2D_1}^*
\ .
\ee
These products of CKM matrix elements are also clearly rephasing
invariant, since for every field which one of the CKM factors in
(\reff{eq:I4}) multiplies,
another factor multiplies its Dirac conjugate. In \eq{eq:I4},
$U_1$, $U_2$ ($D_1$, $D_2$) are chosen cyclically amongst $u,c,t$
($d,s,b$) so as to avoid $I^{(4)}=(I^{(2)})^2$ as well as to avoid
obtaining complex conjugate invariants,
e.g. $I_{U_1D_2U_2D_1}^{(4)} = I_{U_1D_1U_2D_2}^{(4)*}$. With these
constraints, there are 9 invariants. 

However, not all of these invariants are independent. Indeed,
unitarity yields
\be
\left\{\begin{array}{rcll}
\sum_{D=d,s,b}V_{U_1D}V_{U_2D}^* & = & 0 &\qquad U_1\ne U_2\\
\sum_{U=u,c,t}V_{UD_1}V_{UD_2}^* & = & 0 &\qquad D_1\ne D_2
\end{array}\right.\ .\ee
This implies, in turn:
\be
\left\{\begin{array}{rcll}
 V_{U_1D_1}V_{U_2D_1}^* & = & \sum_{D\ne D_1}V_{U_1D}V_{U_2D}^* &\qquad \mbox{(I)}\\
 V_{U_1D_1}V_{U_1D_2}^* & = & \sum_{U\ne U_1}V_{UD_1}V_{UD_2}^* &\qquad \mbox{(II)}
\end{array}\right.\nonumber\ .\ee
Multiplying both sides of (I) by $V_{U_2D_2}V_{U_1D_2}^*$ while
maintaining the cyclicity of indices, yields
\be
I_{U_1D_1U_2D_2}^{(4)}=- \vert
V_{U_1D_1}V_{U_2D_2}\vert^2-I_{U_1D_2U_2D_3}^{(4)}
\qquad \mbox{(III)}
\nonumber\ .\ee
Similarly, multiplying both sides of (II) by
$V_{U_2D_2}V_{U_2D_1}^*$ gives
\be
I_{U_1D_1U_2D_2}^{(4)}=- \vert
V_{U_2D_1}V_{U_2D_2}\vert^2-I_{U_2D_1U_3D_2}^{(4)}
\qquad \mbox{(IV)}
\nonumber\ .\ee
Thus, the 9 $I^{(4)}$'s can all be written in terms of
$I^{(4)}_{udcs}$, for instance, and the 4 independent
$I^{(2)}_{UD}$. Moreover, (III) and (IV) imply that all 9 $I^{(4)}$
have the same imaginary part.

\subsubsectiondum{Higher-order invariants}

Higher-order invariants can, in fact, be written in terms of
$I^{(2)}$s and $I^{(4)}$s. For instance, the sextic rephasing
invariant
$V_{U_1D_1}V_{U_2D_2}V_{U_3D_3}V_{U_1D_2}^*V_{U_2D_3}^*V_{U_3D_1}^*$
is equal to
$I_{U_1D_1U_2D_2}^{(4)}I_{U_2D_1U_3D_3}^{(4)}/I_{U_2D_1}^{(2)}$. This
obviously fails in singular cases, but these will not be considered
here because they are irrelevant in practice.

\subsubsectiondum{Jarlskog's invariant}
\labell{sec:jarlskog}

This whole discussion of invariants implies that there is a unique,
imaginary rephasing invariant combination of CKM matrix
elements. This invariant must appear in all CP violating
observables, because it is the imaginary component of the CKM matrix
which is responsible for this violation. This invariant is known as
the Jarlskog invariant \shortcite{Jarlskog:1985ht}:
\bea
J\equiv \im\,I^{(4)} & = &
c_{13}^2c_{12}c_{23}s_{12}s_{23}s_{13}s_\delta\nonumber\\
& = & \lambda^6 A^2\bar\eta + O(\lambda^{10})
\ ,
\labell{eq:Jdef}
\eea
where $\bar\eta=\eta(1-\lambda^2/2)$. This, in turn, means that to
have CP violation in the Standard Model, $\theta_{12}$, $\theta_{23}$,
$\theta_{13}$ must not be 0 or $\pi/2$ and $\delta\ne
0,\pi$. Moreover, CP violation is maximal for $s_\delta=1$,
$\theta_{12}=\theta_{23}=\pi/4$ and $s_{13}=1/\sqrt3$. At that point,
the Jarlskog invariant takes the value
\be
J_{max}=\frac{1}{6\sqrt3}\simeq 0.1
\ .\ee
However, global CKM fits yield \shortcite{Charles:2004jd}
\be
J=2.96^{+18}_{-17}\times 10^{-5}\ll J_{max}
\ ,
\ee
i.e. CP violation in Nature is very far from being as large as it
could be.

Instead of looking at the CKM matrix for a rephasing invariant measure
of CP violation, we can go back to the mass matrices $M_U$ and $M_D$.
In \sec{sec:CKMorigin}, after performing only flavor transformations which are
symmetries of $\cL_{g+q}$, we reached the point where the
quark mass term was given by \eq{eq:Lmqdiag}
\be
\cL_m^q\to -\bar U_R M_U^d U_L - \bar D_R M_D^dV^{\dagger}D_L+\hc
\ ,
\ee
where $V$ is the CKM matrix. Now, by performing the flavor
symmetry transformation
\be
D_R\longrightarrow V^\dagger D_R
\ ,\ee
we obtain
\be
\cL_m^q\to -\bar U_R M_U^d U_L - \bar D_R VM_D^dV^{\dagger}D_L+\hc
\ .
\labell{eq:semidiagqmass}\ee
In this equation, both $M_U^d$ and $M_D^h\equiv VM_D^dV^{\dagger}$ are
hermitian matrices. This means that the commutator $C_J\equiv [M_U^d,M_D^h]$ is
pure imaginary. Since the only source of imaginary numbers in
\eq{eq:semidiagqmass} is the phase of the CKM matrix, $C_J$ carries information
about CP violation in the Standard Model. However, $C_J$ is not a
rephasing invariant. Defining the matrices
$P_U\equiv\mathrm{diag}\{e^{-i\theta_u}, e^{-i\theta_c},e^{-i\theta_t}
\}$ and $P_D\equiv\mathrm{diag}\{e^{-i\theta_d}, e^{-i\theta_s},e^{-i\theta_b}
\}$, under the rephasing operations of \eq{eq:rephase}, $C_J$ transforms as
\be
C_J\to
[M_U^d,P_DM_D^h P_D^{\dagger}]
\ ,\ee
where $P_U$ cancels against $P_U^{\dagger}$ in the first term,
because $M_U^d$ and $P_U$ are diagonal. Nonetheless, $\det \,C_J$ is
rephasing invariant, because $M_U^d$ and $P_D$ are diagonal. Thus,
following Jarlskog~\shortcite{Jarlskog:1985ht}, we consider
\be
\det\, C_J = 2i J\times \prod_{U_1>U_2}(m_{U_1}-m_{U_2})\prod_{D_1>D_2}(m_{D_1}-m_{D_2})
\ .\ee

In light of what was discussed in the section on rephasing invariants,
$\det\, C_J$ must be the only imaginary, rephasing-invariant quantity
that can be obtained from the mass term of \eq{eq:semidiagqmass},
which is the most general mass term that can be written for quarks in
the Standard Model. This means that the presence of CP violation is
the Standard Model is equivalent to $\det\, C_J\ne 0$. In turn this
implies that there will be CP violation if and only if the conditions
on the mixing angles and the phase of the CKM matrix given after
\eq{eq:Jdef} are obeyed, but also if and only if there are no mass
degeneracies in the up and down quark sectors.

\subsubsectiondum{Unitarity triangle areas}

As we saw earlier, the unitarity of the CKM matrix gives rise to 6
unitarity triangles, defined by
\eqs{eq:D1D2triangle}{eq:U1U2triangle}. The areas of the $(D_1,D_2)$
triangles are given by
\bea
A_{D_1D_2}&=&\frac12\vert V_{uD_1}V_{uD_2}^*\wedge
V_{cD_1}V_{cD_2}^*\vert\nonumber\\
&=& \frac12\vert -i\im\,(V_{uD_1}V_{uD_2}^*
V_{cD_1}^*V_{cD_2})\vert\\
&=&\frac12\im\,I_{uD_1cD_2}^{(4)}=\frac12 J\nonumber
\ ,\eea
where the last line follows from \eq{eq:Jdef}. Similarly the areas of
the $(U_1,U_2)$ triangles are
\bea
A_{U_1U_2}=\frac12\im\,I_{U_1dU_2s}^{(4)}=\frac12 J\nonumber
\ .\eea

Thus, all 6 triangles have the same area, which is given by the
Jarlskog invariant. Since CP violation can only arise if $J\ne 0$,
none of the unitarity triangles can be degenerate if Standard Model CP
violation is measured in Nature.

\sectiondum[A lattice case study: $K\to\pi\pi$,
CP violation and $\Delta I=1/2$ rule]{A lattice case study: {\huge
$K\to\pi\pi$}, CP violation and {\huge $\Delta I=1/2$} rule}

Having introduced the Standard Model and flavor physics, we now turn
to an important set of processes that have been nagging theorists for
over four decades: $K\to\pi\pi$ decays. These decays have been a rich
source of information and of constraints on the weak interaction. In
1964 they provided the first evidence in Nature for indirect CP violation,
which arises in the mixing of the neutral kaon with
its antiparticle before the
decay into two pion~\shortcite{Christenson:1964fg}. Then, in 1999, CP violation
which arises directly in the flavor-changing decay vertex was
discovered in these same decays, after more than 20
years of experimental
effort~\shortcite{Fanti:1999nm,AlaviHarati:1999xp}.

Currently these decays still give very important constraints on the
CKM paradigm, through the measurement of indirect CP violation,
parametrized by $\epsilon$. And as far as we presently know, direct CP
violation in these decays, parametrized by $\epsilon'$, may be
harboring New Physics. Moreover, $K\to\pi\pi$ decays display what we
believe are unusually large, and certainly poorly understood,
nonperturbative QCD corrections, which go under the name of $\Delta
I=1/2$ rule.

At first sight the study of these decays is a perfect problem for the
lattice. Only $u$, $d$ and $s$ (valence) quarks are involved, so that one
expects controllable discretization errors. Of course, pions are
light, which makes them difficult to simulate, and there are also two
hadrons in the final state. But $SU(3)$ chiral perturbation theory
($\chi$PT) at LO relates $K\to\pi\pi$ to $K\to\pi$ and $K\to 0$
amplitudes which are simpler to
compute~\shortcite{Bernard:1985wf}. Moreover, $\chi$PT at NLO relates
$K\to\pi\pi$ amplitudes obtained with heavier pions to the same
amplitudes with physically light
pions~\shortcite{Kambor:1989tz,Kambor:1991ah}. Given the typical size of
chiral corrections, one would expect that we could at least be able to
get an $O(20-30\%)$ estimate of the relevant amplitudes. In addition
to which we might expect good signals since only pseudoscalar mesons are
involved.

Despite all of the positive indications that the lattice should be
able to provide valuable information about these decays, all attempts
to account for nonperturbative strong interaction effects have failed,
except in the study of indirect CP violation.~\footnote{This was
certainly true at the time of the school, but the situation has been
moving quite fast since then [please
see~\shortcite{Christ:LGT10,Liu:LGT10,Sachrajda:2011tg} for an
update].} And though much progress has been made on many aspects of
these decays over the years, providing a fully quantitative
description still remains an open problem. Thus, I have chosen to
focus on this particular topic in my discussion of the application of
lattice methods to flavor phenomenology.

\subsectiondum[$K\to\pi\pi$ phenomenology]{{\large $K\to\pi\pi$} phenomenology~\footnote{Here and in the following, we will assume that CPT is
  conserved. We will also work in the strong isospin symmetry limit.}}

Kaons have strong isospin 1/2, while pions have isospin 1.  The weak
decays of a kaon into two pions can thus occur through two channels in
the isospin limit:
\begin{itemize}

\item the $\Delta I=3/2$ channel, where the final two pions are in a
  state of isospin $I=2$, a state which we label $(\pi\pi)_2$;

\item the $\Delta I=1/2$ channel, where the final two pions are in a
  state of isospin $I=0$, a state which we label $(\pi\pi)_0$.

\end{itemize}
Decay into a two-pion state with isospin $I=1$ is forbidden by Bose
symmetry.

We denote the amplitudes for $K\to\pi\pi$ decays by:
\be
T[K^0\to(\pi\pi)_I]=iA_Ie^{i\delta_I}
\ ,
\labell{eq:kpipiamp}\ee
where $\delta_I$ is the strong scattering phase of two pions in the
isospin $I$, angular momentum $J=0$ channel, defined through
\be
T[(\pi\pi)_I\to(\pi\pi)_I ]=2e^{i\delta_I}\sin\delta_I
\ .\ee
In \eq{eq:kpipiamp}, $K^0$ is the flavor eigenstate with $I_3=-1/2$
and strangeness $S=1$: it is composed of a $d$ and an $\bar s$ quark.

Using this notation, we have the following isospin decompositions for
the $K\to\pi\pi$ amplitudes:
\bea
-i T[K^0\to\pi^+\pi^-]&=&\frac1{\sqrt6} A_2e^{i\delta_2}+\frac1{\sqrt3}
A_0e^{i\delta_0}\nonumber\\
-i T[K^0\to\pi^0\pi^0]&=&\sqrt{\frac23} A_2e^{i\delta_2}-\frac1{\sqrt3}
A_0e^{i\delta_0}\labell{eq:isodecomp}\\
-i T[K^+\to\pi^+\pi^0]&=&\frac{\sqrt3}2 A_2e^{i\delta_2}\nonumber
\ ,\eea
where the coefficients of the various amplitudes are simply $SU(2)$
Clebsch-Gordan coefficients. If CP violation is present, then
$A_I^*\ne A_I$.

Now, in the absence of CP violation, the two physical neutral kaon
states $K_S$ and $K_L$ are also CP eigenstates:~\footnote{The CP
  transformation of the neutral kaon states is chosen here to be
  $CP\vert K^0(p)\rangle = -\vert \bar K^0(p_P)\rangle$ and $CP\vert
  \bar K^0(p)\rangle = -\vert K^0(p_P)\rangle$, where $p_P$ is the parity
  transformed four-momentum $p=(p^0,-\vec{p})$. In the following, we
  will ignore the momentum labels unless they play a relevant role. }
\be
\vert K_{S/L}\rangle\simeq \vert K_{\pm}\rangle\equiv
\frac1{\sqrt2}\left(\vert K^0\rangle\mp \vert\bar K^0\rangle\right)
\ ,\ee
with $CP\vert K_{\pm}\rangle=\pm \vert K_{\pm}\rangle $. The CP even
$K_S$ decays only into two pions, while the CP odd $K_L$ decays into
three. Because of the phase space available to the decay products, the
former is much shorter lived than the latter, with a lifetime
$\tau_S\sim 10^{-10}\,\mathrm{s}$ versus $\tau_L\sim 5\cdot
10^{-8}\,\mathrm{s}$. This explains the subscript $S$ for short and
$L$ for long.

In Nature, the weak interaction breaks CP, and $K_S$ and
$K_L$ are not pure CP eigenstates. As a result of $K^0$-$\bar K^0$
mixing through the weak interaction, we have
\bea
\vert K_{L/S}\rangle &=& \frac1{\sqrt{1+|\tilde\epsilon|^2}}\left(\vert K_{\mp}\rangle+\tilde\epsilon
\vert K_{\pm}\rangle\right)\labell{eq:KSLintermsK12}\\
&=&  \frac1{\sqrt2\sqrt{1+|\tilde\epsilon|^2}}\left((1+\tilde\epsilon)\vert K^0\rangle\pm(1-\tilde\epsilon)
\vert \bar K^0\rangle\right)\labell{eq:KSLintermsK0K0bar}
\ ,\eea
where $\tilde\epsilon$ is a small complex parameter.

The neutral kaons form a two state quantum mechanical system which can
be described by a nonhermitian, 2-by-2 Hamiltonian:
\be
H_{ij}=M_{ij}-\frac{i}2\Gamma_{ij}
\ ,
\labell{eq:Hijdef}\ee
where $i,j=1$ corresponds to $K^0$ and $i,j=2$ to $\bar K^0$. CPT
implies that $H_{11}=H_{22}$ and $H_{21}=H_{12}^*$. To determine the
elements of this matrix, we first decompose the effective Hamiltonian
for the Standard Model into a QCD+QED part, $H_\mathrm{QCD+QED}$, and
a weak part, $H_W$, and work to second order in the weak
interaction. Then, 
\be
H_{ij}=M_{K^0}\,\delta_{ij}+\frac{\langle i\vert H_W\vert
  j\rangle}{2M_{K^0}} +
\frac1{2M_{K^0}}\sum_n\hspace{-0.5cm}\int\hspace{0.15cm}
\frac{\langle i\vert H_W\vert n\rangle\langle n\vert H_W\vert
  j\rangle}{M_{K^0}-E_n+i\epsilon}
+\cdots\ ,
\labell{eq:Hij}\ee
where $M_{K^0}$ is the mass common to $K^0$ and $\bar K^0$, as given
in QCD and QED, and $E_n$ is the energy of the intermediate state
$\vert n\rangle$.

Now, Cauchy's theorem implies (with $P$ the principal part)
\be
\frac1{\omega-E+i\epsilon}=P\left(\frac{1}{\omega-E}\right)-
i\pi\delta(E-\omega)
\ ,\ee
where the first term on the RHS of this equation will yield the
dispersive contribution to the Hamiltonian of \eq{eq:Hijdef} (i.e. the mass
term) and the second term, the absorptive part (i.e. the width
term). Then, the off-diagonal element of the mass matrix is
\be
M_{12}=\frac{\langle K^0\vert H_{\Delta S=2}\vert
  \bar K^0\rangle}{2M_{K^0}} +
\frac1{2M_{K^0}}P\sum_n\hspace{-0.5cm}\int\hspace{0.15cm}
\frac{\langle K^0 \vert  H_{\Delta S=1}\vert n\rangle\langle n\vert  
H_{\Delta S=1}\vert
  \bar K^0\rangle}{M_{K^0}-E_n}
+\cdots\ ,
\labell{eq:M12}\ee
where the term with the double insertion of the $\Delta S=1$
Hamiltonian gives rise to long distance contributions, since the states
$\vert n\rangle$ which can contribute are light. For instance, $\vert
n\rangle$ can be a $\pi^+\pi^-$ state.

The off-diagonal element of the width matrix is given by the
absorptive part of the integral:
\be
\Gamma_{12}=\frac1{2M_{K^0}}\sum_n\hspace{-0.5cm}\int\hspace{0.15cm}
{\langle K^0 \vert  H_{\Delta S=1}\vert n\rangle\langle n\vert  
H_{\Delta S=1}\vert
  \bar K^0\rangle}(2\pi)\delta(E_n-M_{K^0})
\labell{eq:G12}\ee

Now, the physical states $K_L$ and $K_S$ are the eigenstates of
$H_{ij}$ with eigenvalues $M_L-\frac{i}2\Gamma_L$ and
$M_S-\frac{i}2\Gamma_S$, respectively. It is straightforward to express
these quantities in terms of the $M_{ij}$ and $\Gamma_{ij}$. Defining
\be
\Delta M_K\equiv M_{K_L}-M_{K_S}\qquad\mbox{and}\qquad \Delta \Gamma_K
\equiv \Gamma_{K_S}-\Gamma_{K_L}
\ ,\ee
one obtains:
\be
\frac{1+\tilde\epsilon}{1-\tilde\epsilon}=2\frac{M_{12}-\frac{i}2\Gamma_{12}}
{\Delta M_K+\frac{i}2\Delta \Gamma_K}
\ .\ee
Solving for $\tilde\epsilon$ gives us explicit expressions for
computing the relationship of the physical eigenstates $K_L$ and $K_S$ to
the flavor eigenstates $K^0$ and $\bar K^0$ through
\eq{eq:KSLintermsK0K0bar}. It is worth noting that
\be
\Delta M_K\simeq 2M_{12}\qquad\mbox{and}\qquad \Delta \Gamma_K\simeq
-2\Gamma_{12}
\ ,
\labell{eq:DMKDGKapprox}\ee
to first nontrivial order in the weak phases.

Having implicitly worked out the relation between physical and flavor
eigenstates, we return to $K\to\pi\pi$ decays. What is actually
measured are the amplitude ratios
\be
\eta_{00}\equiv\frac{T[K_L\to\pi^0\pi^0]}
{T[K_S\to\pi^0\pi^0]}\qquad\mbox{and}\qquad
\eta_{+-}\equiv\frac{T[K_L\to\pi^+\pi^-]}
{T[K_S\to\pi^+\pi^-]}
\ .\ee
Experimentally, $|\eta_{00}|\simeq2\times 10^{-3}$, and
$|\eta_{00}/\eta_{+-}|\simeq 1$
\shortcite{Nakamura:2010zzi}. These ratios are clearly CP violating
since $K_L$ does not decay into two pions if CP is conserved. The CP
violating decays $K_L\to\pi\pi$ can occur in two ways. As seen in
\eq{eq:KSLintermsK12}, $\vert K_L\rangle$ can acquire a small CP even
component proportional to $\tilde\epsilon\vert K_+\rangle$ through
$K^0$-$\bar K^0$ mixing. This component can then decay into two pions
without violating CP. However, the CP odd component of $\vert
K_L\rangle$, proportional to $\vert K_-\rangle$, can directly decay
into two pions if CP is violated in the decay. These two decay modes
are illustrated in
\begin{center}
\includegraphics[width=0.9\textwidth]{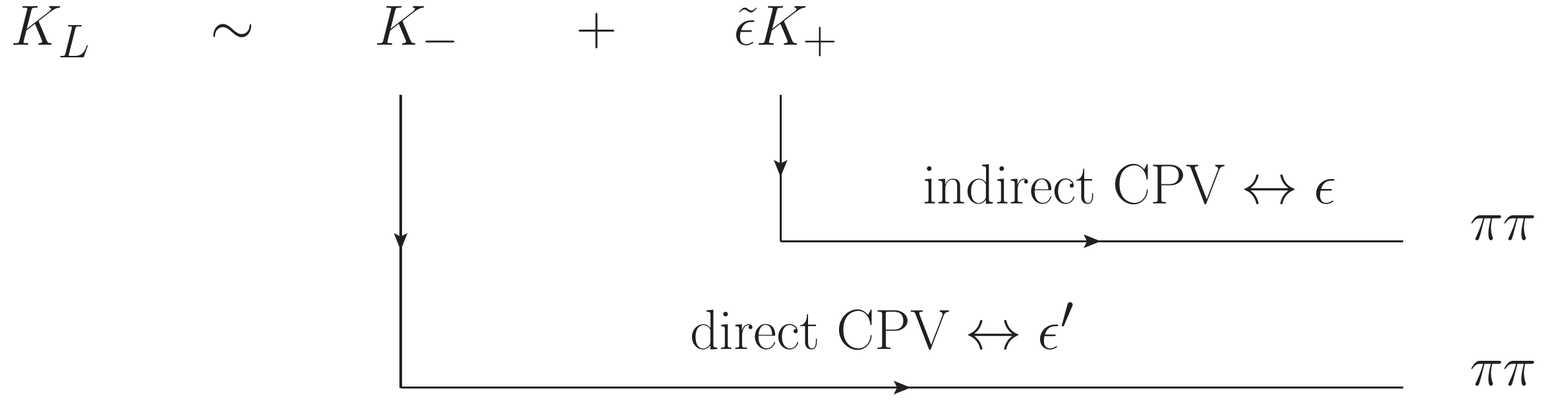}
\end{center}
The first mode of decay is called indirect CP violation and is
parametrized by a small number $\epsilon$ whose relation to
$\tilde\epsilon$ will be given shortly. The second mode is call
direct CP violation and is parametrized by an even smaller number
$\epsilon'$. These parameters are defined through
\be
\epsilon=\frac{T[K_L\to(\pi\pi)_0]}{T[K_S\to(\pi\pi)_0]}
\labell{eq:epsilon}\ee
and
\be
\epsilon'=\frac1{\sqrt2}\Biggl(\frac{T[K_L\to(\pi\pi)_2]}{T[K_S\to(\pi\pi)_2]}
-\epsilon\underbrace{\frac{T[K_S\to(\pi\pi)_2]}{T[K_S\to(\pi\pi)_0]}}_{\dot{\equiv}\omega}\Biggr)
\ .\ee
A little algebra allows us to relate these CP violating parameters to
the measured quantities $\eta_{00}$ and $\eta_{+-}$:
\bea
\eta_{00}&=&\epsilon - 2\frac{\epsilon'}{1-\sqrt2 \omega}\\
\eta_{+-}&=&\epsilon + \frac{\epsilon'}{1+\omega/\sqrt2}
\ .\eea
Moreover, under the assumption that $(\pi\pi)_0$ dominates the
sum for $\Gamma_{12}$ in \eq{eq:G12}, the relationship of $\epsilon$ to
$\tilde\epsilon$ can be calculated to be
\be
\epsilon \simeq \frac{\tilde\epsilon+i\frac{\im A_0}{\re A_0}}
{1+i\tilde\epsilon\frac{\im A_0}{\re A_0}}\simeq\tilde\epsilon+
i\frac{\im A_0}{\re A_0}
\ee
to first non-trivial order in the small CP violating quantities
$\frac{\im A_0}{\re A_0}$ and $\tilde\epsilon$. A bit more algebra and
some simplifications allow us to express $\epsilon$, $\epsilon'$ and
$\omega$ in terms of $A_0$, $A_2$ and $\im
M_{12}$~\shortcite{deRafael:1995zv} which, in principle, can be
calculated using lattice QCD:
\bea
\omega&\simeq& \frac{\re A_2}{\re A_0}e^{i(\delta_2-\delta_0)}\\
\epsilon&\simeq& e^{i\phi_\epsilon}\sin\phi_\epsilon\left\{\frac{\im
  M_{12}}{\Delta M_K}+\frac{\im A_0}{\re A_0}\right\}\labell{eq:eps}\\
\epsilon'&\simeq& \frac{e^{i(\delta_2-\delta_0)}}{\sqrt2}\im\frac{A_2}{A_0}
\ ,\eea
where the phase of $\epsilon$ is approximately given by
$\phi_\epsilon\simeq (2\phi_{+-}+\phi_{00})/3\sim\pi/4$, with
$\phi_{+-}$ and $\phi_{00}$ the phases of $\eta_{+-}$ and $\eta_{00}$,
respectively.~\footnote{\labell{fn:ldcontrib}In many phenomenological studies,
$\phi_\epsilon$ is fixed to $\pi/4$ and $\frac{\im A_0}{\re A_0}$ is
neglected. However, at the levels of accuracy currently reached in the
computation of the local contributions to $\im M_{12}$, these
approximations are becoming too crude. This has been emphasized
in \shortcite{Buras:2008nn}, where the implications of an estimate of
$\frac{\im A_0}{\re A_0}$ and of the deviation of $\phi_\epsilon$ from
$\pi/4$ has been investigated. Moreover, as explained
in \shortcite{Buras:2010pza}, if $\frac{\im A_0}{\re A_0}$, which
approximates $-\frac{\im\Gamma_{12}}{2\re\Gamma_{12}}$, is included
in \protect\eq{eq:eps}, consistency requires that one also account for
long distance contributions to $\im M_{12}$.} In the expression for
$\epsilon'$, the imaginary part of $A_2/A_0$ measures the relative
reality of $A_2$ and $A_0$, which is what we need for direct CP
violation since the latter must arise from the interference between
the two available decay channels. Note that with the approximations
used here, \eq{eq:DMKDGKapprox} gave $\Delta M_K\simeq 2\re M_{12}$,
where $\re M_{12}$ also could, in principle, be computed on the
lattice.

Experimentally, the various quantities which describe the $K^0$-$\bar
K^0$ system and $K\to\pi\pi$ decays are well 
measured~\shortcite{Nakamura:2010zzi}:
\bea
\Delta M_K &=& (3.483\pm 0.006)\times 10^{-12}\,\mev\qquad [0.2\%]\\
\frac1{|\omega|} &\simeq& \left\vert\frac{A_0}{A_2}\right\vert\simeq
22.4\qquad\mbox{($\Delta I=1/2$ rule)}\\
|\epsilon| &\simeq& (2|\eta_{+-}|+|\eta_{00}|)/3=(2.228\pm 0.011)\times
10^{-3}\qquad [0.5\%]\\
\phi_\epsilon &\simeq& (2\phi_{+-}+\phi_{00})/3=43.51\pm0.05\qquad
    [0.1\%]\\
\re\frac{\epsilon'}{\epsilon} &\simeq&
\left(1-\left\vert\frac{\eta_{00}}{\eta_{+-}}\right\vert\right)=(1.65\pm
0.26)\times 10^{-3}\qquad
    [16\%]
\ .\eea

Using lattice QCD, the weak matrix element relevant for the
short-distance, Standard Model contribution of $|\epsilon|$ has been
calculated with a precision of less than 3\% (see
~\shortcite{Lellouch:2009fg,Lubicz:2010nx,Sachrajda:2011tg,Colangelo:2010et}
for recent reviews). We are just beginning to provide
phenomenologically relevant information for the $\Delta I =1/2$ rule
$|A_0/A_2|$~\shortcite{Sachrajda:2011tg,Liu:LGT10,Christ:LGT10}, but
$\re(\epsilon'/\epsilon)$ is still out of reach for the
moment~\shortcite{Sachrajda:2011tg}. Moreover, as already mentioned
and discussed further below, $\Delta M_K$ has long distance
contributions which make its determination on the lattice
difficult. However, for that also, progress has been
made \shortcite{Christ:2010gi}.

\subsectiondum[$K^0$-$\bar K^0$ mixing in the Standard Model]{{\large $K^0$-$\bar K^0$} mixing in the Standard Model}

As we have just seen, $\bar K^0$-$K^0$ mixing arises from the $\Delta
S=2$, $s\bar d\to\bar s d$ FCNC. This mixing is responsible for
the $K_L$-$K_S$ mass difference, $\Delta M_K$, and indirect CP
violation in $K\to\pi\pi$ decays.

In the Standard Model, it occurs at one loop through diagrams such 
as~\cite{Glashow:1970gm,Gaillard:1974hs}
\be
\labell{eq:KKbarboxes}
\parbox{0.4\textwidth}{\includegraphics[width=0.35\textwidth]{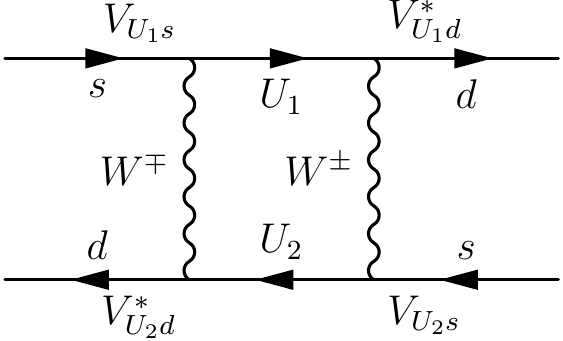}}
\ee
Setting the external four-momenta to zero, the amplitude associated
with this diagram is
\be
-i\cM = \left(\frac{-ig_2}{\sqrt2}\right)^4\int
\frac{d^4k}{(2\pi)^4}iD^W_{\mu\nu}(k)iD^W_{\rho\sigma}(k)
\left(\bar v_{dL}\gamma_\mu iS(k)\gamma_\sigma v_{sL}\right)
\left(\bar u_{dL}\gamma_\nu iS(k)\gamma_\rho u_{sL}\right)
\ ,\ee
where the $u$'s and $v$'s are the usual particle and antiparticle
spinor wavefunctions, and we have defined the amplitude with a minus
sign for convenience.  The $W$ boson propagator in Feynman gauge is
\be
D^W_{\mu\nu}(k)=\frac{-g_{\mu\nu}}{k^2-M_W^2+i\epsilon}
\ee
and the sum of the up quark propagators,
\be
S(k)=\sum_{U=u,c,t}\frac{\lambda_U}{\slash k-m_U+i\epsilon}
\ ,
\ee
with $\lambda_U\equiv V_{Us}V_{Ud}^*$. 

The unitarity of the CKM matrix implies that
$\sum_{U=u,c,t}\lambda_U=0$. Thus, we have (dropping $i\epsilon$ for
the moment)
\be
S(k)=\sum_{U=c,t}\lambda_U\left(\frac1{\slash k-m_U}-\frac1{\slash k-m_u}\right)
\ .\ee
From this we see the GIM mechanism~\shortcite{Glashow:1970gm} in
action: if $m_u=m_c=m_t$ there would be no $K^0$-$\bar K^0$ mixing. In
fact, this process was used to estimate the charm quark mass before
it was actually discovered~\shortcite{Gaillard:1974hs}.

After performing some Dirac algebra, we find:
\be
\cM = \frac{G_F^2M_W^2}{2\pi^2}
(\lambda_t^2 T_{tt} + \lambda_c^2 T_{cc} + 2 \lambda_c\lambda_t
T_{ct})\times (\bar v_{dL}\gamma_\mu v_{sL})(\bar u_{dL}\gamma^\mu u_{sL})
\ ,
\labell{eq:Mint}\ee
where we have used $G_F=g_2^2/(4\sqrt2 M_W^2)$. Setting $m_u=0$,
\be
T_{U_1U_2}=\frac{4i}{\pi^2 M_W^2}
\int d^4k\frac{1}{k^2(1-k^2/M_W^2)^2}\frac{m_{U_1}^2}{k^2-m_{U_1}^2}
\frac{m_{U_2}^2}{k^2-m_{U_2}^2}
\ .
\labell{eq:TU1U2}\ee

With the spinor wavefunction factors appropriately replaced by quark
field operators, $\cM$ can be interpreted as an effective Hamiltonian
whose matrix element between a $K^0$ and a $\bar K^0$ state yields
the off diagonal matrix element (\reff{eq:M12}) of the mass matrix of
\eq{eq:Hijdef}. Of course, to obtain the full effective
  $\Delta S=2$ Hamiltonian, one must include the contributions of all
  of the diagrams which contribute to the
  process~\shortcite{Inami:1980fz}.~\footnote{\protect\eq{eq:Mint}
  and the amplitudes associated with the other contributing diagrams
  must be multiplied by $1/2$ if their spinors factors are replaced by
  operators to yield the effective $\Delta S=2$ Hamiltonian. Indeed,
  the operator $(\bar d_L\gamma_\mu s_L)(\bar d_L\gamma^\mu s_L)$ has
  twice as many contractions with the four external quark states as
  there really are. These extra contractions correspond to a doubling
  of the individual box diagrams.}

Now, the values of the CKM matrix elements as well as of the masses
$m_u$, $m_c$ and $m_t$ imply that $\re M_{12}\gg\im M_{12}$ and that
$\re M_{12}$ is dominated by the $cc$ term. Naively,
\bea
\frac{G_F^2M_W^2}{4\pi^2}T_{cc}&=& \frac{iG_F^2}{\pi^4}\int
d^4k\frac{m_c^4}{(k^2+ i\epsilon)[k^2-m_c^2 + i\epsilon]^2}
+O\left(\frac1{M_W^6}\right)\nonumber\\
&=&\frac{G_F^2m_c^2}{\pi^2}+O\left(\frac1{M_W^6}\right)\ .
\eea
However, a closer look at this loop integral indicates that it is
dominated by momenta in the range between 0 and $m_c$ and we should
not forget that all sorts of gluons with or without quark loops can be
exchanged between the quarks in the diagram. Since these momenta
include scales of $O(\lqcd)$ or below, it is clear that
$\alpha_s$ corrections are out of control and $\re M_{12}$ cannot be
calculated in this way.

Said differently, box diagrams, such as the one of \eq{eq:KKbarboxes},
cannot be viewed like a point interaction, but rather receive
long-distance contributions from intermediate $c\bar c$ states. In the
language of \eq{eq:M12}, these diagrams contribute to the second term
through $(\bar d_L\gamma_\mu s_L)(\bar c_L\gamma_\mu c_L)$ effective
operators in $\cH_{\Delta S=1}$.

The situation is very different for the calculation of the CP
violating parameter $\epsilon\sim \im M_{12}/\re
M_{12}$~\footnote{For the sake of clarity, we
  neglect here the small contribution to $\epsilon$ from $\frac{\im
    A_0}{\re A_0}$ (see \eq{eq:eps}).}. Indeed, indirect CP violation
in $K\to\pi\pi$ decays comes from the interference between the
following types of contributions

\bigskip

\centerline{
\parbox{0.40\textwidth}{\includegraphics[width=0.40\textwidth]{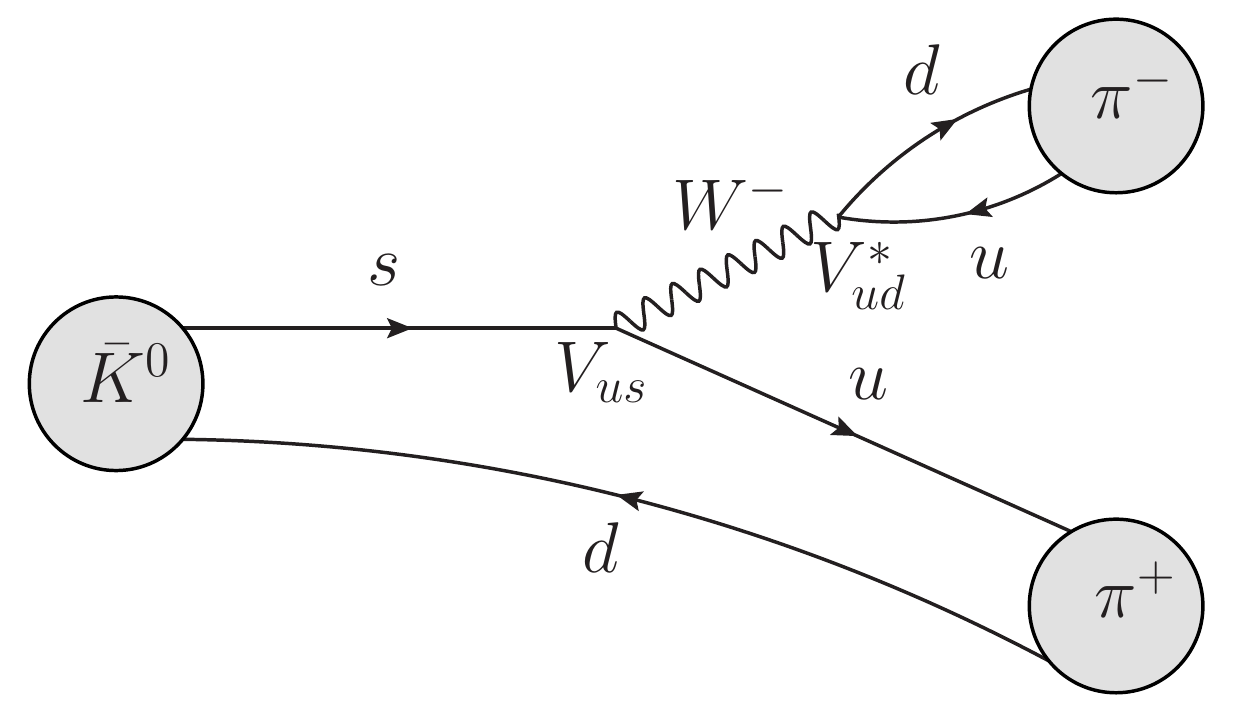}}}

\medskip

\noindent
and

\medskip

\centerline{
\parbox{0.40\textwidth}{\includegraphics[width=0.40\textwidth]{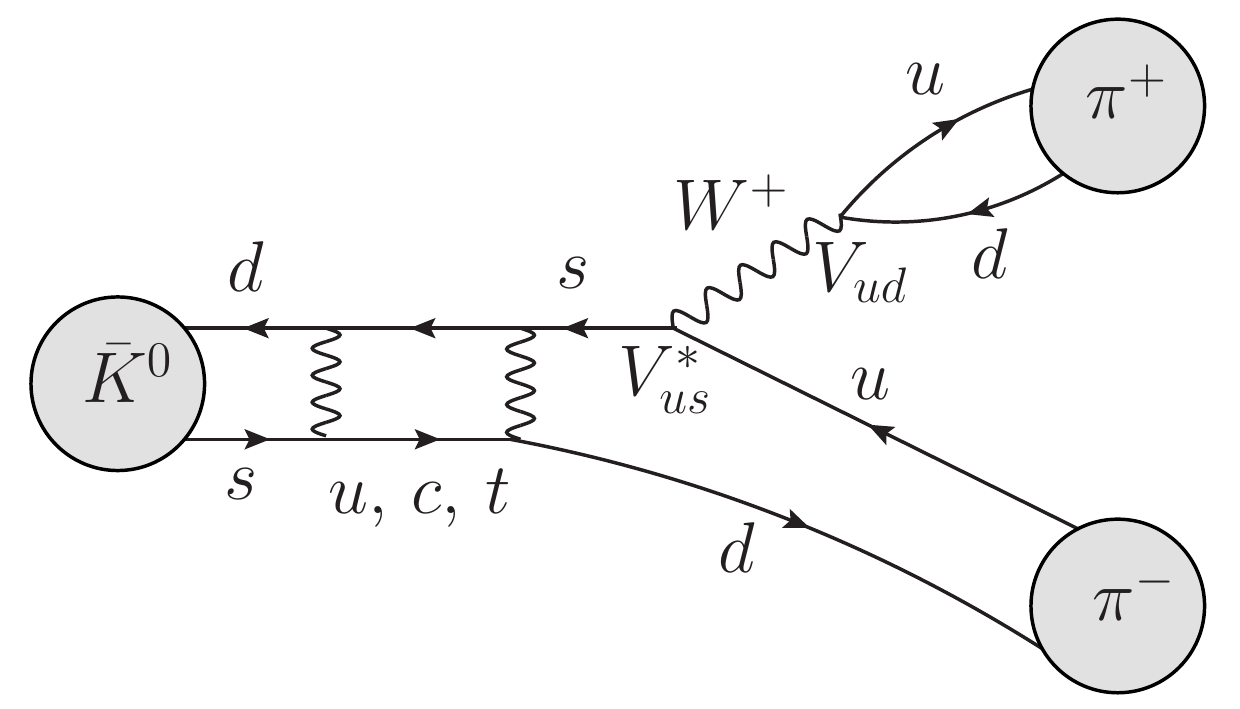}}}

\bigskip

\noindent
So, the relative weak phase between the two types of contributions is
$\arg\{(\lambda_u^*)^2 M_{12}\}$ and thus, what we really have is
\be
\epsilon\sim \frac{\im\left[(\lambda_u^*)^2 M_{12}\right]}
{\re\left[(\lambda_u^*)^2 M_{12}\right]}
\ ,\ee
where the $(\lambda_u^*)^2$ had been canceled in earlier expressions,
using the fact that $\lambda_u$ is real in our conventions. 

Now, using \eq{eq:Mint} with the replacement of wavefunctions by
operators discussed after \eq{eq:TU1U2}, we have:
\bea
(\lambda_u^*)^2
M_{12}&\sim&
\frac{G_F^2M_W^2}{16\pi^2}\Bigl((\underbrace{\lambda_u^*\lambda_t}_{I^{(4)}_{udts}})^2 
T_{tt} + 
(\underbrace{\lambda_u^*\lambda_c}_{I^{(4)}_{udcs}})^2 T_{cc} + 
2 (\lambda_u^*\lambda_c)(\lambda_u^*\lambda_t)
T_{ct}\Bigr)\nonumber\\
&&\times \la K^0|(\bar ds)_{V-A}(\bar ds)_{V-A}|\bar K^0\ra
\ ,
\eea
where
\be
\labell{eq:VmAdef}
(\bar ds)_{V-A}=\bar
d\gamma_\mu(1-\gamma_5) s
\ee
and where the
$I^{(4)}$ are the quartic rephasing invariants defined in
\eq{eq:I4}. As shown in \sec{sec:CKMquartinv},
$\im I^{(4)}_{udcs}=J=-\im I^{(4)}_{udts}$, where $J$ is the Jarlskog
invariant. Therefore,
\bea
\im[(\lambda_u^*)^2
  M_{12}]&\sim&\frac{g_2^4}{8M_W^4}J\left\{\re(\lambda_u^*\lambda_t)(T_{tt}-T_{ct})-
\re(\lambda_u^*\lambda_c)(T_{cc}-T_{ct})\right\}\nonumber\\
&&
\labell{eq:epsOPE}\times 
\la K^0|(\bar ds)_{V-A}(\bar ds)_{V-A}|\bar K^0\ra
\ ,\eea
and $J$ appears as it should in a CP violating quantity.

Moreover, the integrals in $(T_{tt}-T_{ct})$ and $(T_{cc}-T_{ct})$ are
dominated by momenta between $m_c$ and $m_t$. The same is true of the
integrals which appear in the other diagrams~\shortcite{Inami:1980fz}
that contribute to this $\Delta S=2$ process. That means that the QCD
corrections are calculable to the extent that $m_c$ can be considered
a perturbative scale. Thus, under this assumption, and using the
experimental value of $\Delta M_K$ in lieu of
$\re\left[(\lambda_u^*)^2 M_{12}\right]$, we can reliably calculate
$\epsilon$ with the replacement

$$
\im\left\{(\lambda_u^*)^2\parbox{0.35\textwidth}{\includegraphics[width=0.3\textwidth]{figs/kkbar1}}\qquad+\qquad\cdots\right\}
$$

$$\longrightarrow\qquad
\im\left\{(\lambda_u^*)^2\parbox{0.35\textwidth}{\includegraphics[width=0.3\textwidth]{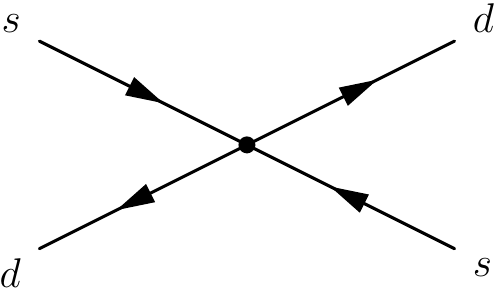}}\right\}
$$

\noindent
i.e. by replacing the box diagram with the local, four-quark operator
of \eq{eq:epsOPE}, and the appropriate short-distance QCD corrections,
omitted here. As mentioned in footnote~\reff{fn:ldcontrib}, corrections to this approximation have been examined in \shortcite{Buras:2010pza}.

\subsectiondum[The theory of $K^0$-$\bar K^0$ mixing]{The theory of {\large $K^0$-$\bar K^0$} mixing}

The calculation of $M_{12}$ of the previous section has actually been
performed to leading-log
(LL)~\shortcite{Vainshtein:1975xw,Witten:1976kx,Vainshtein:1976eu,Gilman:1982ap}
and next-to-leading-log order (NLL)~\shortcite{Buras:1990fn} in QCD
(for a review see~\shortcite{Buchalla:1995vs}). 
The resulting NLL, $\Delta S=2$ effective Hamiltonian is given by
\bea
\labell{eq:HDS2def}
\cH_{\Delta S=2}&=&\frac{G_F^2}{16\pi^2}M_W^2\left[\lambda_t^2
  \eta_{tt}S_{tt} + 
\lambda_c^2 \eta_{cc}S_{cc} + 2 \lambda_c\lambda_t\eta_{ct}
S_{ct}\right]\nonumber\\
&&\times C(\mu)\times (\bar ds)_{V-A}(\bar ds)_{V-A}
\ ,
\eea
where the running of matrix elements of the four quark operator is
canceled by the coefficient $C(\mu)$, the $S_{qq'}$ are the Inami-Lim
functions~\shortcite{Inami:1980fz} which correspond to the electroweak
box contributions in the absence of the strong interaction, and the
$\eta_{qq'}$ are short-distance QCD corrections. In \eq{eq:HDS2def}
all Standard Model degrees of freedom with masses down to, and
including, that of the charm quark are integrated out. Thus, in
\eq{eq:HDS2def}, all QCD quantities must be evaluated with {\em three}
active quark flavors.

It is useful to consider the case of a general number of active quark
flavors, $N_f$, and of colors, $N_c$, when
discussing the running of this $\Delta S=2$ operator. Indeed, the
running is the same for other Standard Model, $\Delta F=2$ operators,
such as $\Delta B=2$ and $\Delta C=2$, except that $N_f$ must be
chosen appropriately.  Moreover, working with general $N_c$ allows one
to consider the large-$N_c$ limit, to be discussed shortly. With
$a_s\equiv\alpha_s/4\pi$, the running of $\alpha_s$ and
of $C(\mu)$ is given by:
\bea
\frac{d\ln a_s}{d\ln\mu^2} & = & -\beta(a_s) = -\beta_0 a_s - \beta_1
a_s^2 + O(a_s^3)\ ,\labell{eq:asrun}\\
\frac{d\ln C}{d\ln\mu^2} & = & \gamma(a_s) = \gamma_0 a_s + \gamma_1
a_s^2 + O(a_s^3)\ ,\labell{eq:Crun}
\eea
where $\beta(a_s)$ is the QCD $\beta$ function and $\gamma(a_s)$ is
the anomalous dimension of the $\Delta S=2$ operator. It is well known
that $\beta_0$, $\beta_1$, and that LO anomalous dimensions are
renormalization scheme independent. At two loops, we
have~\shortcite{Nakamura:2010zzi}
\be
\beta_0=\frac{11N_c-2N_f}3\,,\qquad \beta_1=\frac{34}3 N_c^2-\frac{10}3
N_cN_f-2 C_F N_f\,,\qquad C_F=\frac{N_c^2-1}{2N_c}
\ .
\labell{eq:betafncoeffs}\ee
and, in the $\msbar$-NDR scheme for $\gamma_1$,
\be
\gamma_0=3\frac{N_c-1}{N_c}\,,\qquad
\gamma_1=\frac{N_c-1}{4N_c}\left[-21+\frac{57}{N_c} - \frac{19}{3} N_c
+\frac43 N_f\right]
\ .
\labell{eq:anomdimcoeffs}\ee
It is straightforward to integrate \eq{eq:asrun}:
\be
\frac{1}{a_s(\mu)}-\beta_0\ln\left(\frac{\mu}{\mu_0}\right)^2+\frac{\beta_1}{\beta_0}\ln\left[\frac{a_s(\mu)}{1+(\beta_1/\beta_0)a_s(\mu)}\right]
\ee
$$
=\frac{1}{a_s(\mu_0)}+\frac{\beta_1}{\beta_0}\ln\left[\frac{a_s(\mu_0)}{1+(\beta_1/\beta_0)a_s(\mu_0)}\right]
$$
For reasonably small $a_s(\mu_0)\ln(\mu/\mu_0)^2$, the coupling at $\mu$ can be related to the one at $\mu_0$ through:
\bea
a_s(\mu)=a_s(\mu_0)&&\left\{1-\beta_0a_s(\mu_0)\ln\left(\frac{\mu}{\mu_0}\right)^2
-a_s(\mu_0)^2\ln\left(\frac{\mu}{\mu_0}\right)^2\right.\\
&&\times
\left.\left[\beta_1-\beta_0^2\ln\left(\frac{\mu}{\mu_0}\right)^2\right]\right\}
+O(a_s^3)\ .
\eea
Alternatively, 
we can define $\lqcd$ as the value of $\mu_0$ at which $a_s(\mu_0)$ is infinite, yielding:
\be
a_s(\mu)=\frac1{\beta_0\ln\frac{\mu^2}{\lqcd^2}}\left[1+\frac{\beta_1\ln\ln\frac{\mu^2}{\lqcd^2}}{\beta_0^2\ln\frac{\mu^2}{\lqcd^2}}+\cdots\right]
\ .\ee

It
is also straightforward to integrate \eq{eq:Crun},
\be
\labell{eq:Cmurunning}
\frac{C(\mu)}{C(\mu_0)}=\exp\left\{-\int_{a_s(\mu_0)}^{a_s(\mu)}\frac{da_s}{a_s}
\frac{\gamma(a_s)}{\beta(a_s)}\right\}
\ .\ee
The coefficient $C(\mu)$ is only defined up to an integration
constant. For consistency with \eq{eq:HDS2def}, I consider
\be
C(\mu) = [4\pi a_s(\mu)]^{-\gamma_0/\beta_0} \exp\left\{-\int_{0}^{a_s(\mu)} \frac{da_s}{a_s}\left[\frac{\gamma(a_s)}{\beta(a_s)}-\frac{\gamma_0}
{\beta_0}\right]\right\}
\ .
\labell{eq:Cmu}\ee
Thus, at NLO
\be
C(\mu)=[4\pi a_s(\mu)]^{-\gamma_0/\beta_0}\left[1+
\frac{\beta_1}{\beta_0}a_s+O(a_s^2)\right]^{\left(\frac{\gamma_0}{\beta_0}-
\frac{\gamma_1}{\beta_1}\right)}
\ .
\labell{eq:CmuNLONDR}\ee

For the Standard Model $\Delta F=2$ operator relevant for $K^0$-$\bar
K^0$ mixing, we use the anomalous dimension coefficients
of \eq{eq:anomdimcoeffs}, with $N_f=3$ and, of course, $N_c=3$. Thus,
at NLO,
\be
C(\mu)=\alpha_s(\mu)^{-2/9}\left[1+
\frac{307}{162}\frac{\alpha_s(\mu)}{4\pi}\right]
\ .\ee
Then, in \eq{eq:HDS2def} the QCD corrections $\eta_{qq'}$ are of the form
\be
\eta_{qq'}\propto \frac{\left[1+O(\alpha_s)\right]}{C(m_c)}
\ ,\ee
and where $x_q\equiv (m_q/M_W)^2$.  For details, please see \shortcite{Buchalla:1995vs}.

To calculate $\epsilon$, we must compute the matrix element $\langle
K^0|(\bar ds)_{V-A}(\bar ds)_{V-A}(\mu)|\bar K^0\rangle$, where the
kaons are at rest. This is clearly a nonperturbative QCD quantity
because the typical energies of the quarks within the kaons are on the
order of $100\,\mev$, a regime where perturbation theory fails and
confinement effects must be taken into account. This is where lattice
QCD enters.~\footnote{Here again we choose to neglect the small
  contribution to $\epsilon$ from $\frac{\im A_0}{\re A_0}$. Computing
  it on the lattice is a whole other project which is related to the
  computation of $\epsilon'$.  That computation goes beyond the
  presentation I wish to make here.}

For historical reasons, and because it is very convenient in 
lattice computations, we define a normalized matrix element
\be
B_K(\mu)\equiv\frac{\langle\bar K^0|(\bar sd)_{V-A}(\bar
  sd)_{V-A}(\mu) |K^0\rangle}{\frac83\langle\bar K^0|\bar
  s\gamma_\mu\gamma_5 d|0\rangle\langle 0|\bar
  s\gamma_\mu\gamma_5 d|K^0\rangle}
\ ,
\labell{eq:BKdef}\ee
where we have considered the $\Delta S=-2$ matrix element to conform
with convention. The benefit of this normalization on the lattice, is
that the resulting $B_K$ parameter is dimensionless. Therefore it does
not suffer from ambiguities due to scale setting, something which was
particularly bad in old quenched-calculation days. Moreover, the
numerator and denominator are very similar, and both
statistical and systematic uncertainties cancel in the ratio. Of
course, the convenience of this normalization would be limited if the
denominator were an unknown, nonperturbative quantity. However, the
matrix elements in the denominator define the leptonic decay constant
of the kaon, $f_K$,~\footnote{Note that the $K^0$ does not actually
  decay leptonically: the $K^{\pm}$ do. However, in the isospin limit,
  the decay constant defined in \eq{eq:FKdef} is equal to the physical
  decay constant $f_K$ of the $K^\pm$.}
\be
\labell{eq:FKdef}
\langle 0|\bar s\gamma_\mu\gamma_5 d(x)|K^0(p)\rangle=i f_K
p_\mu e^{-ip\cdot x}
\ ,\ee
which is well measured experimentally or straightforward
to compute on the lattice. (In the convention
used here, $f_K\simeq 156\,\mev$.) Thus, once
the $B$-parameter has been computed, the normalizing factor is a
known quantity and the desired matrix element is easily obtained, from
\be
\langle\bar K^0|\underbrace{(\bar sd)_{V-A}(\bar
  sd)_{V-A}(\mu)}_{O_{\Delta S=-2}^\mathrm{SM}}
|K^0\rangle=\frac{8}{3} f_K^2 M_K^2 B_K(\mu)
\ ,\ee
where $M_K^2$ is best taken from experiment.

Historically, the denominator in \eq{eq:BKdef} was an approximation
used to estimate the matrix element of $O_{\Delta
  S=-2}^\mathrm{SM}$. It is called the vacuum saturation
approximation, or VSA for short. It is obtained by inserting the
vacuum in all possible ways between all possible quark-antiquark field
pairs formed from the fields of the four-quark operators, using Fierz
transformations if necessary to bring the fields together. For the
case of interest here,
\be
\labell{eq:FKovNc}
\langle 0|\bar s_b\gamma_\mu(1-\gamma_5)
d^a(x)|K^0(p)\rangle=-i\frac{\delta_b^a}{N_c} f_K
p_\mu e^{-ip\cdot x}
\ ,\ee
where $a$ and $b$ are color indices and the dependence on the number
of colors $N_c=3$ is made explicit. Therefore
\be
\langle\bar K^0|O_{\Delta S=-2}^\mathrm{SM}(\mu) |K^0\rangle_{VSA} =
2\Bigl(\langle\bar K^0|\bar s_a\gamma^\mu(1-\gamma_5)
d^a|0\rangle\langle 0|\bar s_b\gamma_\mu(1-\gamma_5)
d^b|K^0\rangle\nonumber
\ee
\be
+\langle\bar K^0|\bar s_a\gamma^\mu(1-\gamma_5)
d^b|0\rangle\langle 0|\bar s_b\gamma_\mu(1-\gamma_5)
d^a(x)|K^0\rangle\Bigr)
\ ,
\labell{eq:BKVSA}\ee
where repeated color indices are summed over and where the factor of 2
comes from the fact that the two factors of $(\bar sd)_{V-A}$ in 
$O_{\Delta S=-2}^\mathrm{SM}$ are
interchangeable. Plugging (\reff{eq:FKovNc}) in (\reff{eq:BKVSA})
yields
\bea
\langle\bar K^0|O_{\Delta S=-2}^\mathrm{SM}(\mu) |K^0\rangle_{VSA} &=&
\frac{2}{N_c^2}f_K^2
p^2\l[\delta_a^a\delta_b^b+\delta_a^b\delta_b^a\r]\nonumber
\\
&=& 2\frac{N_c+1}{N_c}M_K^2f_K^2
\ .
\eea
This is clearly a rather crude approximation, as the LHS of
\eq{eq:BKVSA} is $\mu$ dependent while the RHS is not. This
approximation introduces a renormalization scale dependence which is
unphysical.

A more modern approximation to the matrix element is obtained by
keeping the leading term in a large-$N_c$ expansion. The
large-$N_c$, or 't Hooft limit~\shortcite{'tHooft:1973jz}, is defined by taking
$N_c\to\infty$ while holding $\alpha_sN_c$ fixed. By counting the
number of $\alpha_s$ and loop factors of $N_c$ in the various
contributions to the relevant correlation functions (see below), it is
straightforward to convince oneself that in the large-$N_c$ limit,
$B_K(\mu)=3/4$. This corresponds to dropping the second term in
\eq{eq:BKVSA}, which is clearly suppressed by a factor of $1/N_c$
compared to the first. As in the VSA approximation $B_K(\mu)$ is $\mu$
independent, but here the $\mu$ dependence is also absent in the
short-distance, Wilson coefficient, as can be
seen by taking the large-$N_c$ limit in \eqs{eq:betafncoeffs}{eq:Cmu}: the
large-$N_c$ approximation is a well-defined and self-consistent
approximation scheme.

Before closing this section, it is worth mentioning that one can
define a renormalization scheme and scale independent $B$-parameter,
$B_K^\RGI$, by multiplying $B_K(\mu)$ by $C(\mu)$ of \eq{eq:Cmu} (with
$N_c=3$ and $N_f=3$):
\be
B_K^\RGI=C(\mu)\times B_K(\mu)
\ .
\labell{eq:BKRGI}\ee

\subsectiondum[Computation of bare $B_K$]{Computation of bare {\large $B_K$}}

On the lattice, the numerator of $B_K$ is obtained from three-point
functions. Quark propagators are given by:
\be
S_q[\vec{x},t;\eta,t_s;U] = \sum_{\vec x_s} D^{-1}[\vec{x},t;\vec x_s,t_s;m_q;U]
\eta(\vec x_s)
\ ,
\labell{eq:ptprop}\ee
where $D$ is the lattice Dirac operator associated with the chosen
fermion action, $m_q$ the quark $q$'s mass, $U$ the gauge field
configuration on which the propagator is computed, $\eta(\vec x_s)$ is
a three dimensional source which may be a delta function or may
have some spatial extent and $t_s$ is the timeslice at which the
source is placed. If only propagators from a point source at, for
instance, $t_s=0$ and $\vec x_s=\vec 0$ (i.e. $\eta(\vec x_s)=\delta_{\vec
  x_s,\vec 0}$), are available, then we can consider the following 
three-point function, in Euclidean spacetime of course:
\be
C_3(t_i,t_f)=\sum_{\vec x_i,\vec x_f}
\langle\bar d\gamma_5 s(\vec x_f,t_f)O_{\Delta S=-2}^\mathrm{SM}(0) 
\bar d\gamma_5 s(\vec x_i,t_i)\rangle
\ ,\ee
where the argument of $O_{\Delta S=-2}^\mathrm{SM}$ is its spacetime
position, not the renormalization scale. For $T/2\gg -t_i\gg 1/\Delta
E_K$ and $T/2\gg t_f\gg 1/\Delta E_K$, $d\gamma_5 s(\vec
x_i,t_i)$ creates a $K^0$ at $t=t_i$, this kaon then propagates to
$t=0$ where $O_{\Delta S=-2}^\mathrm{SM}(0)$ transforms it into a $\bar K^0$
and $\bar d\gamma_5 s(\vec x_f,t_f)$ destroys the resulting $\bar K^0$
at $t=t_f$. $\Delta E_K$ is the energy of the first excited state in
the neutral kaon channel minus $M_K$. The sums over $\vec x_i$ and $\vec
x_f$ put the initial and final kaons at rest. Thus, in this
limit
\be
C_3(t_i,t_f)\stackrel{T/2\gg -t_i\gg 
1/\Delta E_K,\,T/2\gg t_f/\gg
1/\Delta E_K}{\longrightarrow} \frac{e^{-M_K(t_f-t_i)}}{4M_K^2}\langle 0|d\gamma_5
s(0)|\bar K^0(\vec 0)\rangle
\labell{eq:3ptleading}\ee
\be
\times\underbrace{\langle \bar K^0(\vec 0)|O_{\Delta
    S=-2}(0)|K^0(\vec 0)\rangle}_{\mbox{desired mat. elt.}}
\langle K^0(\vec 0)|\bar d\gamma_5 s(0)|0\rangle\nonumber
\ .
\ee
This result is obtained by inserting a complete set of hadron states
between $O_{\Delta S=-2}^\mathrm{SM}(0)$ and $\bar d\gamma_5 s(\vec x_i,t_i)$,
between $\bar d\gamma_5 s(\vec x_f,t_f)$ and $O_{\Delta S=-2}^\mathrm{SM}(0)$
and between $\bar d\gamma_5 s(\vec x_i,t_i)$ and $\bar d\gamma_5
s(\vec x_f,t_f)$, keeping the sequence of states which gives rises to
the smallest exponential suppression and which has the appropriate
quantum numbers to give nonvanishing matrix elements. If $T/2\gg
t_f,-t_i$ is not realized, then contributions which are not
significantly exponentially suppressed compared to the one in
\eq{eq:3ptleading} will have to be added.
 
Similarly, to obtain the matrix elements required to construct the
denominator of \eq{eq:BKdef}, we can consider the following two-point
function
\be
C_{2,\mu}(t)=\sum_{\vec{x}}\langle\bar d\gamma_5 s(\vec x,t)\bar
s\gamma_\mu\gamma_5 d(0)
\rangle\stackrel{t\gg\Delta_{E_K}}{\longrightarrow}
\ee
\be
\frac{(e^{-M_Kt}-e^{-M_K(T-t)})}{2M_K}\langle 0|d\gamma_5
s(0)|\bar K^0(\vec 0)\rangle\underbrace{\langle \bar K^0(\vec 0)|\bar 
s\gamma_5\gamma_\mu d(0)|0\rangle}_{\mbox{denom. mat. elt.}}
\ ,\ee
where I have not assumed here that $T/2\gg t$ to illustrate the
additional contributions which arise in that case, and where I have used the
properties of the correlation function under time reversal.

Then, $B_K$ is obtained from the ratio
\be
\frac{C_3(t_i,t_f)}{\sum_{\mu}C_{2,\mu}(t_f)C_{2,\mu}(t_i)}
\stackrel{T/2\gg -t_i\gg 
1/\Delta E_K,\,T/2\gg t_f/\gg
1/\Delta E_K}{\longrightarrow} B_K(a)
\ ,
\labell{eq:BKlat}\ee
where I have reinstated $T/2\gg t_f-t_i$ to get rid of ``backward'' 
contributions. In \eq{eq:BKlat}, the argument $a$ of $B_K$ is there to
indicate that this is the value of $B_K$ in the lattice regularized
scheme and that it still requires renormalization.

To actually compute $C_3(t_i,t_f)$, we have to take the propagators of
\eq{eq:ptprop}, contract them in the following way and average these
contractions over the gauge ensemble, i.e.
\bea
C_3(t_i,t_f) &=& \frac2{\cN_U}\sum_{U}\sum_{\vec x_i,\vec x_f}\Bigg\{
\underbrace{\parbox{0.45\textwidth}{\includegraphics[width=0.45\textwidth]{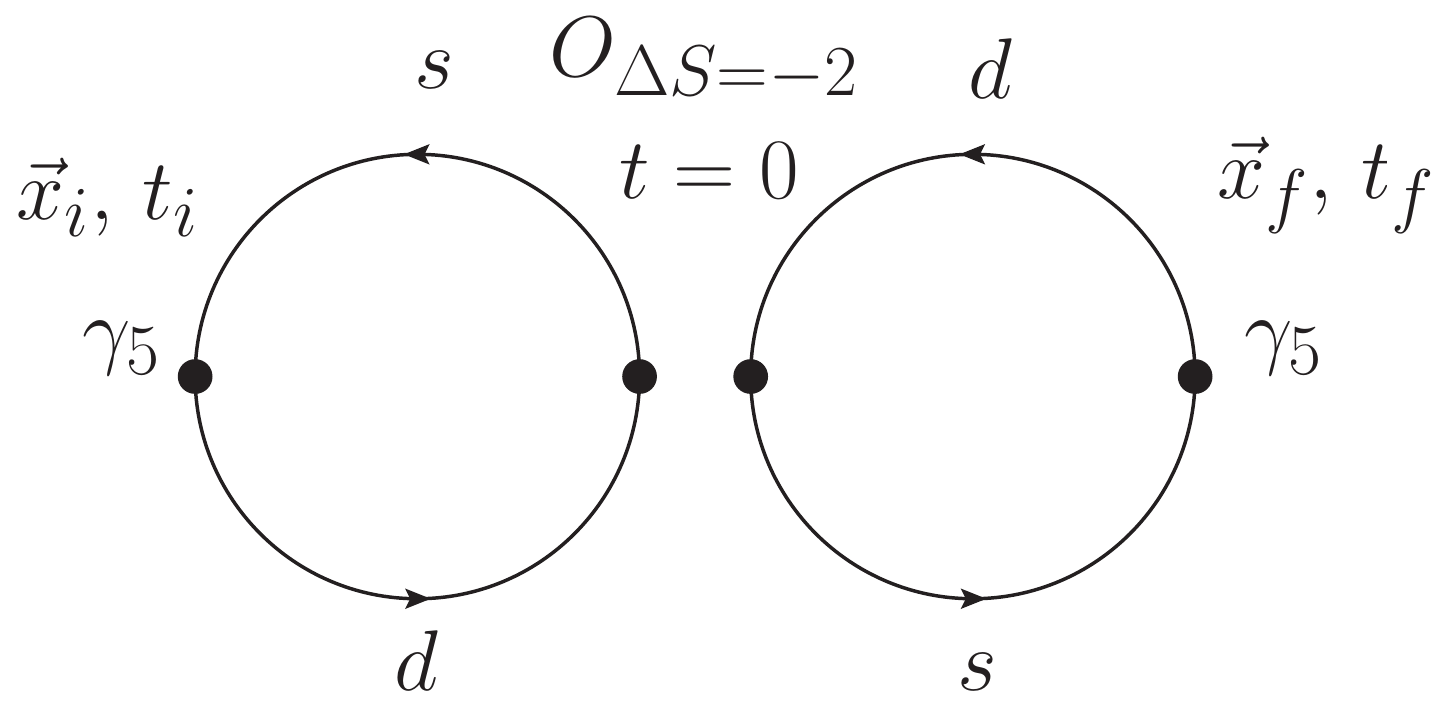}}}_{\mbox{double trace term}\propto N_c^2}\nonumber\\
&&-\underbrace{\parbox{0.45\textwidth}{\includegraphics[width=0.45\textwidth]{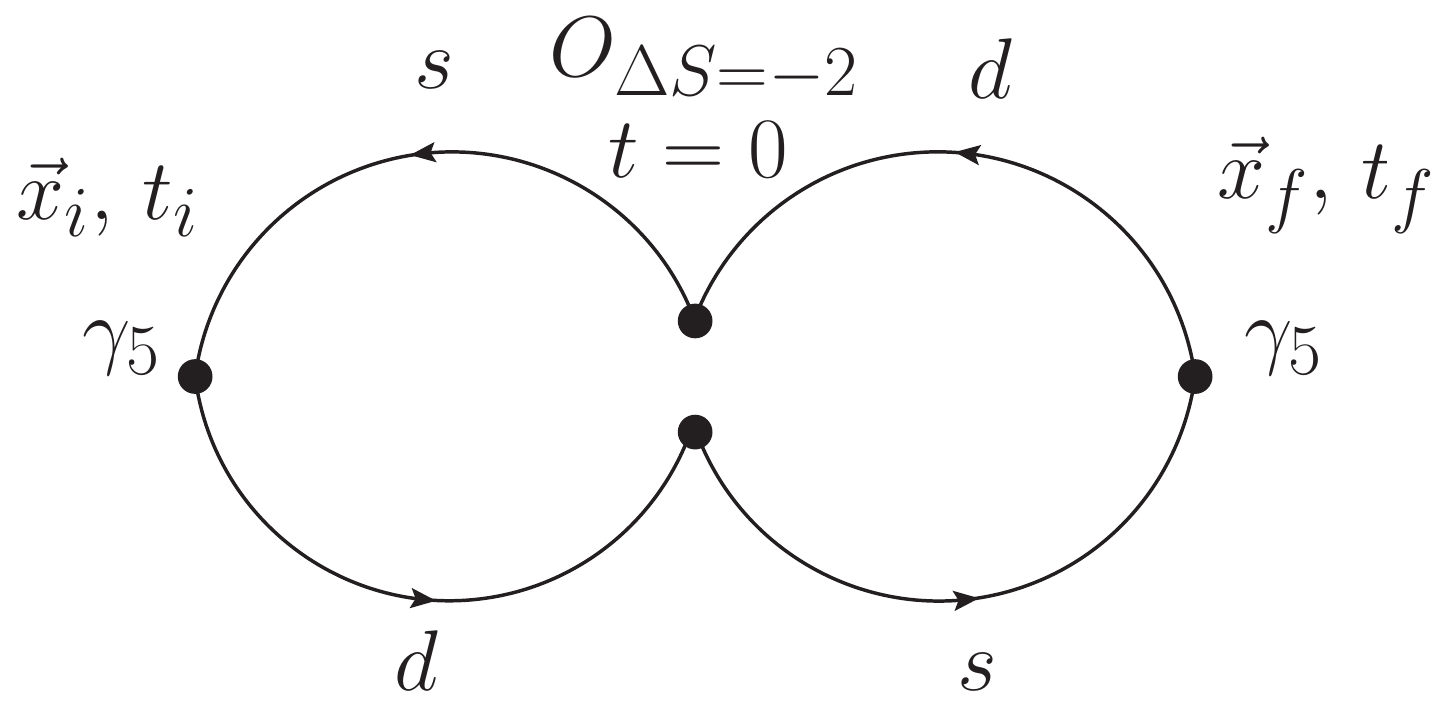}}}_{\mbox{single trace term}\propto N_c}\Bigg\}
\ ,
\labell{eq:C3contract}\eea
where $\cN_U$ is the number of independent gauge configurations. To
get the time-reversed propagator required to construct the correlation 
function, one usually uses the $\gamma_5$ hermiticity
which most lattice Dirac operators have,
\be
S_d[\vec{0},0;\vec{x},t;U]=\gamma_5S_d[\vec{x},t;\vec{0},0;U]^\dagger\gamma_5
\ .\ee

In \eq{eq:C3contract}, one clearly sees how, in the large-$N_c$
limit, only the first contraction survives. The second contraction
provides a $1/N_c$ suppressed correction. In practice, the two
contractions have the same sign, and a cancellation operates in the
calculation of $B_K$.

The method described above is actually a poor way to obtain $B_K$
because the matrix elements of interest, $\langle \bar K^0(\vec
0)|O_{\Delta S=-2}(0)|K^0(\vec 0)\rangle$ and $\langle 0|\bar
\gamma_5\gamma_\mu d(0)|K^0(\vec 0)\rangle$, are sampled at only one
point on the lattice: at the origin. We would gain a factor of roughly
$(LM_\pi)^3$ in statistics if we could sample them over the whole
three dimensional volume of the lattice.~\footnote{This is because
  the longest correlation length in the system is $1/M_\pi$ so that
  regions separated by that distance should be reasonably
  decorrelated.} Thus, a better way to calculate $B_K$ is to have two
zero momentum sources for the quark propagators, one at $t_i$ and the
other at $t_f$. One can consider, for instance, wall sources:
\be
S_q[\vec{x},t;W,t_W;U] = \sum_{\vec x_s} D^{-1}[\vec{x},t;\vec x_s,t_W;m_q;U]
\sum_{\vec y}\delta_{\vec x_s,\vec y}
\ .
\labell{eq:Wprop}
\ee
Then, one constructs the following three-point function
\be
C_3(t)=\frac1{N_U}\sum_U\sum_{\vec x}\gamma_5\parbox{0.5\textwidth}{
\includegraphics[width=0.5\textwidth]{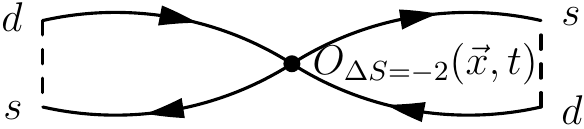}}\gamma_5
\ ,\ee
where $t_i<0$ and $t_f>0$ are chosen so that there is a range of $t$
around $t=0$ such that the correlation function is dominated by the
propagation of a zero momentum $K^0$ state between $t_i$ and $t$ and
the propagation of a zero momentum $\bar K^0$ state between $t$ and
$t_f$. The gauge field configurations are usually gauge-fixed on the
walls because wall sources correspond to meson sources and sinks of the form
$\sum_{\vec{x},\vec{y}} \bar d_a(\vec{x},t_W)\gamma_5
s^a(\vec{y},t_W)$ which are clearly not gauge invariant, except 
for terms along the ``diagonal'' $\vec{x}=\vec{y}$. The gauge is
usually fixed to Coulomb gauge. However, one can also not fix the
gauge, the result being that the sums over the two quark positions in
the sources and sinks reduce to a single diagonal sum over the
positions after the average over gauge configurations is taken.

Using the wall sources, we also construct the two two-point functions
($t_W=t_i,\,t_f$),
\be
C_{2,\mu}(t,t_W)=\frac1{N_U}\sum_U\sum_{\vec x}\gamma_5\parbox{0.15\textwidth}{
\includegraphics[width=0.15\textwidth]{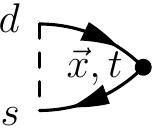}}\gamma_\mu
\gamma_5
\ .\ee
Then we study the ratio of correlators
\be
R(t)\equiv\frac{C_3(t)}{\sum_\mu C_{2,\mu}(t,t_f)C_{2,\mu}(t,t_i)}
\ee
as a function of $t$. For $t_i\ll t\ll t_f$, $R(t)$ develops a plateau
(see \fig{fig:BKplat}) such that
\be
R(t)\stackrel{t_i\ll t\ll t_f}{\longrightarrow} B_K(a)
\ ,\ee
so that $B_K(a)$ is obtained by either averaging $R(t)$ or fitting it
to a constant over the plateau region.

\begin{figure}[t]
\centering
\includegraphics[width=0.65\textwidth]{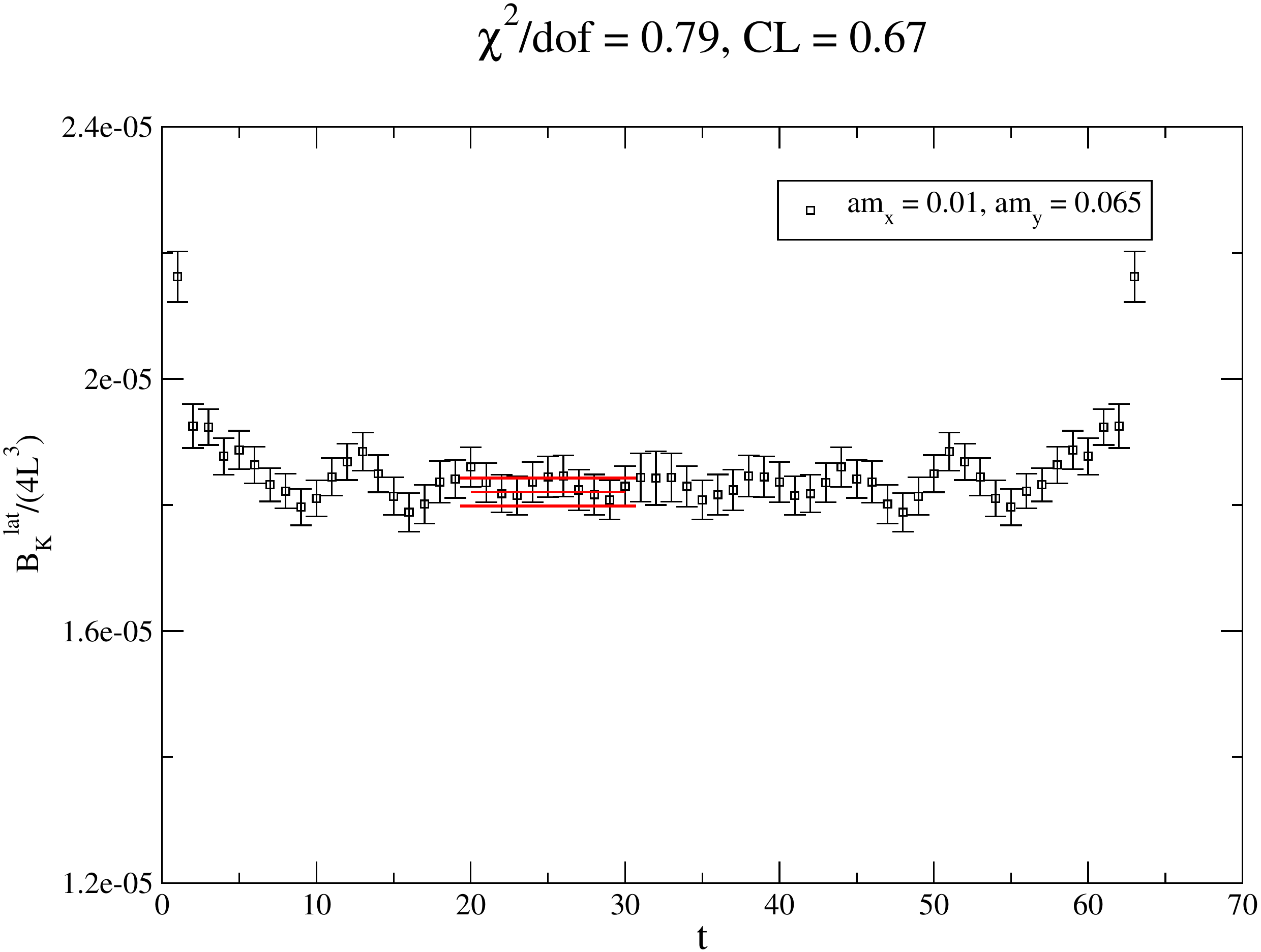}
\caption{Plateau fit to $R(t)/(4L^3)$ on the coarse $N_f=2+1$,
  staggered, MILC, $am_l/am_h = 0.007/0.05$ ensemble. The legend shows
  the nondegenerate pair of domain-wall, valence quark masses making
  up the kaon in the three-point correlation function. The correlated
  $\chi^2/dof$ and confidence level of the fit to a constant (red
  lines) are given in the title. Taken from \protect\shortcite{Aubin:2009jh}.}
\labell{fig:BKplat}
\end{figure}

\subsectiondum[Renormalization of the Standard Model $|\Delta S|=2$ operator]{Renormalization of the Standard Model {\large $|\Delta S|=2$} operator}

We are not done, however. As we already stated, simply inserting the
lattice operator $O_{\Delta S=-2}^\mathrm{SM}$ in $C_3(t)$ and computing $R(t)$
yields the bare $B_K(a)$. This quantity is divergent in the continuum
limit and must be renormalized. And it must be done so in a
renormalization scheme which matches the one used in the perturbative
calculation of the short-distance Wilson coefficients.

It is straightforward to show that the full set of $\Delta S=-2$,
$\Delta D=2$ operators of dimension $d\le 6$ can be written as:
\bea
\labell{eq:O1def}
O_1 &=& O_{\Delta S=-2}^\mathrm{SM}=(\bar sd)_{V-A}(\bar
  sd)_{V-A}\qquad \mbox{(unmix)}\ ,\\
O_{2,3} &=& (\bar sd)_{S-P}(\bar sd)_{S-P}\qquad \mbox{(unmix, mix)}\ ,\\
\labell{eq:O45def}
O_{4,5} &=& (\bar sd)_{S-P}(\bar sd)_{S+P}\qquad \mbox{(unmix, mix)}\ ,
\eea
where he subscripts $(\bar sd)_{S-P}$ and $(\bar sd)_{S+P}$ are
defined in analogy with $(\bar sd)_{V-A}$ in \eq{eq:VmAdef}. In
\eqs{eq:O1def}{eq:O45def} 
``unmix'' and ``mix'' refer to the color indices. In the
``unmix'' case, the color indices of the quark-antiquark pairs within
parentheses are contracted; in the ``mix'' case, the color
index of the quark of one pair is contracted with the color index of
the antiquark of the other pair, and vice versa. Because $O_1$ Fierz 
transforms
into itself, the ``mix'' and ``unmix'' $O_1$ are the same operator.

To understand the renormalization patterns of $O_{\Delta
  S=-2}^\mathrm{SM}$ and the other $\Delta S=-2$ operators, it is
useful to consider their transformation properties under various
symmetry groups. Because we will only work in massless renormalization
schemes, the $SU(3)_L\times SU(3)_R$ chiral group is a symmetry which
is relevant here. Under the action of this group, $(\bar
sd)_{V-A}(\bar sd)_{V-A}$ transforms in the $(27,1)$ representation,
i.e. it is a 27 under $SU(3)_L$ and clearly a singlet under $SU(3)_R$,
since it is composed only of left-handed fields. It is straightforward
to derive the $SU(3)_L\times SU(3)_R$ representations to which the
five $\Delta S=-2$ operators belong:
\bea
O_1\sim (27,1) &\stackrel{SU(3)_V}{\longrightarrow}& 27\otimes 1 =
27\ ,\nonumber\\
\labell{eq:DS2chirep}
O_{2,3}\sim (6,\bar 6) &\stackrel{SU(3)_V}{\longrightarrow}& 6\otimes \bar 6 =
27\oplus 8 \oplus 1\ ,\\
O_{4,5}\sim (8,8) &\stackrel{SU(3)_V}{\longrightarrow}& 8\otimes 8 =
27\oplus 10 \oplus \overline{10} \oplus 8 \oplus 8 \oplus 1\nonumber
\ .
\eea
In \eq{eq:DS2chirep} I have also worked out the 
reduction of these representations to
the diagonal, $V=L+R$, Eightfold Way, $SU(3)_V$ group, for reasons which
will be clear shortly. Note that the ``mix'' versus ``unmix'' feature
of these operators has no bearing on their flavor transformation
properties as these features pertain solely to color.

From this, we see that $O_{\Delta S=-2}^\mathrm{SM}$ is the only
$\Delta S=-2=-\Delta D$ operator of dimension 6 or less, which
transforms as $(27,1)$. Thus, in any regularization which preserves
$SU(3)_L\times SU(3)_R$ symmetry (or at least in the valence sector),
$O_{\Delta S=-2}^\mathrm{SM}$ renormalizes multiplicatively. 
This includes overlap and
domain wall fermions, for sufficiently large fifth dimension.
Similarly, \eq{eq:DS2chirep} indicates that the operator pairs $(O_2,O_3)$
and $(O_4,O_5)$ may mix within each pair
under renormalization, but not with any of the other $\Delta S=-2=
-\Delta D$ operators.

The situation is very different for Wilson fermions. Indeed, the
Wilson-Dirac operator breaks the chiral symmetry of continuum QCD
explicitly, down to the vector flavor symmetry $SU(3)_V$. As
\eq{eq:DS2chirep} shows, $SU(3)_V$ is not sufficient to forbid
$O_{\Delta S=-2}^\mathrm{SM}$ from mixing with the four other 
$\Delta S=-2=-\Delta D$ operators, $O_{2,\cdots,3}$, under
renormalization.

To push things further, we turn to parity. Since parity is preserved by
Wilson fermions, we can consider separately the renormalization of the
parity even and parity odd components of the operators. For $B_K$ we
are clearly interested in the parity even part:
\be
\langle\bar K^0|(\bar sd)_{V-A}(\bar
  sd)_{V-A}|K^0\rangle = \langle\bar K^0|(\bar sd)_{V}(\bar
  sd)_{V}+(\bar sd)_{A}(\bar
  sd)_{A}|K^0\rangle
\ .\ee
So let us begin with the renormalization of the parity even part.

At this point, it is useful to invoke a discrete symmetry
transformation known as CPS \shortcite{Bernard:1985wf}. It consists in
performing a CP transformation, followed by a switching
$s\leftrightarrow d$. Note that this vector flavor symmetry is only
softly broken by the mass terms in the action. Therefore, violations
must appear multiplied by factors of $(m_s-m_d)$.  Under CPS, we have:
\bea
\bar s\gamma^\mu d&\stackrel{CPS}{\longrightarrow}&-\bar s\gamma_{\mu} d\\
\bar s\gamma^\mu\gamma^5 d&\stackrel{CPS}{\longrightarrow}&-\bar s\gamma_{\mu}
\gamma^5 d\\
\bar s d&\stackrel{CPS}{\longrightarrow}&\bar sd\\
\bar s\gamma^5 d&\stackrel{CPS}{\longrightarrow}&-\bar s\gamma^5 d
\ .\eea
Thus, the parity even components of the $\Delta S=-2$ operators
transform under CPS as
\bea
O_1^+  &=& (\bar sd)_{V}(\bar
  sd)_{V}+(\bar sd)_{A}(\bar sd)_{A}\stackrel{CPS}{\longrightarrow} O_1^+\ ,\\
O_{2,3}^+ &=& (\bar sd)_{S}(\bar sd)_{S}+(\bar sd)_{P}(\bar sd)_{P}
\stackrel{CPS}{\longrightarrow} O_{2,3}^+\ ,\\
O_{4,5}^+ &=& (\bar sd)_{S}(\bar sd)_{S}+ (\bar sd)_{P}(\bar sd)_{P}
\stackrel{CPS}{\longrightarrow}O_{4,5}^+\ .
\eea
All of these operators are CPS eigenstates, with eigenvalue
+1. Like $SU(3)_V$ symmetry, CPS does not forbid $O_1^+$ to mix with
$O_{2,\cdots,3}^+$, under renormalization. In fact, there is no
symmetry which forbids $O_1^+$ to mix with the other operators and one
finds, in practice, that they do mix. We have,
\be
O_1^+(\mu)= Z_1^+(a,\mu)\left[O_1^+(a)+\sum_{i=2}^5 
z_{1i}(a)O_i^+(a)\right]
\ ,\ee
where $Z_1^+(a,\mu)$ is
logarithmically divergent in the continuum limit, while the mixing
factors, $z_{1i}(a)$, are finite~\shortcite{Testa:1998ez}. Using the
fact that the values of $B_K^\RGI$ of \eq{eq:BKRGI}, obtained in the
continuum renormalization scheme and in the lattice regularization
scheme, must be identical, one easily obtains a formal expression for
the renormalization constant $Z_1^+(a,\mu)$, to all orders in
perturbation theory:
\be
\labell{eq:z1pallorders}
Z_1^+(a,\mu)=
\exp\left\{\int_{0}^{a_s(\mu)}
\frac{da_s}{a_s}\frac{\gamma(a_s)}{\beta(a_s)}-\int_{0}^{a_s^\lat}
\frac{da_s}{a_s}\frac{\gamma^\lat(a_s)}{\beta^\lat(a_s)}\right\}
\ ,\ee
with $a_s^\lat=g_0^2/(4\pi)^2$, where $g_0$ is the bare lattice coupling. 
Expansions in the bare
lattice coupling are notoriously poorly behaved, and one can usually
improve their convergence by expressing them in
terms of a renormalized continuum coupling, such as $a_s(\mu)$ at $\mu=1/a$
for example. Applying this to \eq{eq:z1pallorders}, we find
\be
\labell{eq:z1pfinal}
Z_1^+(a,\mu)=
\exp\left\{\int_{0}^{a_s(\mu)}
\frac{da_s}{a_s\beta(a_s)}\gamma(a_s)-\int_{0}^{a_s(1/a)}
\frac{da_s}{a_s\beta(a_s)}\bar\gamma^\lat(a_s)\right\}
\ ,\ee
where $\bar\gamma^\lat$ is the anomalous dimension obtained by
rewriting $\gamma^\lat$ in terms of the continuum coupling
$a_s$. Obviously this change of variable would make no difference to
$Z_1^+(a,\mu)$ were it to be computed to all orders in perturbation
theory. However, at finite order this reordering of the expansion may
improve the convergence of the series.

It is interesting to expand \eq{eq:z1pfinal} to one loop--this could
be done for \eq{eq:z1pallorders} instead--to explicitly see the
relationship between the coefficients of the expansion and the anomalous 
and beta function coefficients. We find:
\be
Z_1^+(a,\mu)=1-\frac{\alpha_s}{4\pi}\left(\gamma_0\ln(a\mu)^2+
\frac{\bar\gamma_1^\lat-\gamma_1}{\beta_0}\right)+O(\alpha_s^2)
\ .\ee
One clearly sees that the constant $O(a_s)$ term knows about the
two-loop anomalous dimension (it is a subleading log) whereas the
leading log term is given by $\gamma_0$.
At this same order, the mixing coefficients are finite and given by
\be
z_{1i}(a)=z_{1i}^{(1)}\frac{\alpha_s}{4\pi}+O(\alpha_s^2)\qquad i=2,\cdots,5
\ ,
\ee
where the $z_{1i}^{(1)}$ are constants.

Now, the $SU(3)_L\times SU(3)_R$ properties of the operators indicate
that the physical contributions in $\langle\bar
K^0|O_1^+(a)|K^0\rangle$ are chirally suppressed compared to the
nonphysical ones: $O(p^2)$ vs $O(1)$ in $\chi$PT counting. Thus, even
though the mixing of $O_1$ with $O_2$,\ldots,$O_5$ is $\alpha_s$
suppressed, this suppression can easily be compensated by an
$O(10-20)$ enhancement from the matrix
element~\shortcite{Babich:2006bh}.  This means that the necessary
subtractions are delicate and it is preferable to avoid calculating
$B_K$ with Wilson fermions.

It is also interesting to study the transformation properties of
the parity odd components of $O_1,\cdots,\,O_5$ under CPS. We find
\bea
O_1^-=-2(\bar sd)_V(\bar sd)_A&\stackrel{CPS}{\longrightarrow}&
O_1^-\\
O_{2,3}^-=-2(\bar sd)_S(\bar sd)_P&\stackrel{CPS}{\longrightarrow}&
-O_{2,3}^-\\
O_{4,5}^-=0
\ .\eea
Thus, CPS forbids $O_1^-$ to mix with $O_{2,3}^-$ and $O_{4,5}^-$
vanish anyway. We conclude that $O_1^-$, unlike $O_1^+$, renormalizes
multiplicatively:
\be
O_1^-(\mu)=Z_1^-(a,\mu)O_1^-(a)
\ .
\ee
This turns out to be useful for twisted-mass QCD (tmQCD), which is
described in detail in~\shortcite{tassos}. For instance, twisting
$(u,\,d)$ by an angle $\alpha$ and leaving the strange quark $s$
untwisted, we have
\bea
\left[O_{VV+AA}\right]_\mathrm{QCD}&\equiv& \left[(\bar sd)_V(\bar
  sd)_V+(\bar sd)_A(\bar sd)_A \right]_\mathrm{QCD}\nn\\
&=& \cos\alpha\left[O_{VV+AA}\right]_\mathrm{tmQCD}-i\sin\alpha
\left[O_{VA+AV}\right]_\mathrm{tmQCD}
\ .
\eea
So, picking $\alpha=\pi/2$ (i.e. maximal twist), the above equation
implies the following relationship between the renormalized matrix
elements:
\be
\langle\bar
K^0|O_{VV+AA}(\mu)|K^0\rangle_\mathrm{QCD}=-i\langle\bar
K^0|O_{VA+AV}(\mu)|K^0\rangle_\mathrm{tmQCD}
\ .\ee
Then, since CPS symmetry is only softly broken in tmQCD,
$\langle\bar K^0|O_{VA+AV}(a)|K^0\rangle_\mathrm{tmQCD}$
renormalizes multiplicatively.~\footnote{CPS violating terms are
  proportional to $(m_s-m_d)$, so that $O_{VV+AA}(a)$ can only mix
  with higher dimensional operators which are suppressed by powers of
  the lattice spacing and the latter can only contribute
  discretization errors.} Thus, by working in tmQCD at maximal twist,
one can compute $B_K(\mu)$ performing only multiplicative
renormalization (please see Tassos' course notes~\shortcite{tassos}
for details).

Though we discussed renormalization mostly in terms of perturbation
theory, I greatly encourage you to perform this renormalization
nonperturbatively, with one of the methods explained in
Peter~\shortcite{peter} or Tassos'~\shortcite{tassos} course
notes. You may also want to look into the renormalization procedure
put forward in \shortcite{Durr:2010vn,Durr:2010aw}, which enhances the
RI/MOM method of \shortcite{Martinelli:1994ty} with nonperturbative,
continuum running (see also \shortcite{Arthur:2010ht}). Or if you
choose to renormalize perturbatively, you should at least do so to two
loops to ensure that you have some control over the perturbative
series. Of course, one may argue that the short distance coefficients
in \eq{eq:HDS2def} are only known to NLO, and that there is no point
in doing much better in the lattice to continuum matching. Moreover,
there are other uncertainties in the relation (\reff{eq:eps}) of
$\epsilon$ to $B_K$, such as the one associated with the error on the
determination of $|V_{cb}|$ or with the neglect of $1/m_c^2$ and of
the $\im A_0/\re A_0$ corrections (see
e.g. \shortcite{Lellouch:2009fg}). However, it is admittedly a pity to
perform a careful nonperturbative computation of the bare matrix
elements only to introduce perturbative uncertainties through the
matching procedure.

\bigskip

\subsectiondum[Final words on $K^0$-$\bar K^0$ mixing]{Final words on {\large $K^0$-$\bar K^0$} mixing}

Given a preferably nonperturbatively renormalized $B_K$, it must be
matched to the scheme used in the computation of the Wilson
coefficients which appear in \eq{eq:HDS2def}, and computed for a
variety of lattice spacings and quark masses. Then you must use the
methods described in, for instance,
\shortcite{Durr:2008zz,Lellouch:2009fg,Durr:2010vn,Durr:2010aw} and/or
in Pilar~\shortcite{pilar}, Peter~\shortcite{peter} and
Maarten's~\shortcite{maarten} course notes to extrapolate (preferably
interpolate) to the physical values of $m_{ud}$, $m_s$ and extrapolate
to the continuum limit. Finally, you must perform a complete
systematic error analysis, such as the ones performed in
\shortcite{Durr:2008zz,Durr:2010hr,Durr:2010vn,Durr:2010aw}. 
For a recent, comprehensive review of lattice calculations of $B_K$,
see for instance \shortcite{Colangelo:2010et}.

Before concluding this discussion of $K^0$-$\bar K^0$ mixing, it is
worth mentioning that lattice QCD can also provide information that is
relevant for this process beyond the Standard Model. Quite
generically, when one adds heavy degrees of freedom to those of the
Standard Model, such as in supersymmetric extensions, one finds that
the full set of $\Delta S=-2$ operators of \eqs{eq:O1def}{eq:O45def}
contribute to the low-energy effective Hamiltonian. Of course, the
lattice can also be used to compute the matrix elements of the four
additional operators, between $K^0$ and $\bar K^0$ states. This has
been studied in the quenched approximation in
\shortcite{Donini:1999nn,Babich:2006bh}. Ref.~\shortcite{Babich:2006bh}
finds ratios of non Standard Model to Standard Model matrix elements
which are roughly twice as large as those in
\shortcite{Donini:1999nn}. As explained in \shortcite{Babich:2006bh},
this is most probably due to discretization errors present in
\shortcite{Donini:1999nn}. This picture appears to be confirmed by the
preliminary $N_f=2$ results of \shortcite{Dimopoulos:2010wq}.

\subsectiondum[Phenomenology of the $\Delta I=1/2$ rule]{Phenomenology of the {\large $\Delta I=1/2$} rule}

\labell{sec:kpipiinfv}

The goal here is to compute nonleptonic weak decay amplitudes, such as
those for $K\to\pi\pi$, directly in the Standard Model, without any
extraneous model assumptions, nor potentially uncontrolled
approximations such as LO, $SU(3)$ $\chi$PT. This is critical for
showing that QCD is indeed responsible for surprising phenomena such
as the $\Delta I=1/2$ rule, or deciding whether New Physics is hidden
in the experimental measurement of direct CP violation in $K\to\pi\pi$
(i.e. of $\epsilon'$). In the following I will concentrate on the
$\Delta I=1/2$ rule as the calculation of $\epsilon'$ is significantly
more difficult. For the latter, the renormalization of the relevant
matrix elements is more complicated~\shortcite{Dawson:1997ic}. There
are more matrix elements involved and there appear to be important
cancellations between them (see
e.g.\ \shortcite{Buchalla:1995vs}). Moreover, final-state
interactions, such as the ones discussed in
\shortcite{Pallante:2000hk,Buras:2000kx}, seem to play an important
role. Indeed, recent attempts to exhibit the $\Delta I=1/2$ rule and
to determine $\re(\epsilon'/\epsilon)$ using soft pion theorems appear
to fail due to the large chiral corrections required to translate the
$K\to\pi$ and $K\to 0$ amplitudes computed on the lattice into
physical $K\to\pi\pi$
amplitudes~\shortcite{Li:2008kc,Sachrajda:2011tg,Christ:LGT10}.


Experimentally, the partial widths measured in the different $K\to\pi\pi$
decay channels, together with the corresponding isospin changes
between the final two-pion and initial kaon states, 
are~\cite{Nakamura:2010zzi}:
\bea
\Gamma_{+-}&=&\Gamma(K_S\to\pi^+\pi^-)=5.08\times 10^{-12}\mev\qquad \Delta I=\frac12,\frac32\\
\Gamma_{00}&=&\Gamma(K_S\to\pi^0\pi^0)=2.26\times 10^{-12}\mev\qquad \Delta I=\frac12,\frac32\\
\Gamma_{+0}&=&\Gamma(K^+\to\pi^+\pi^0)=1.10\times 10^{-14}\mev\qquad \Delta I=\frac32
\ .\eea
Using these results, we find
\be
\frac{\Gamma_{+-}}{\Gamma_{+0}}=463.\quad\mbox{and}\quad \frac{\Gamma_{00}}{\Gamma_{+0}}=205.
\ ,\ee
whereas $\Gamma_{+-}/\Gamma_{+0}$ should be $\sim 1$ in the 
electroweak theory and in the absence of the strong interaction. Now,
in terms of the amplitudes, the rates are
\bea
\labell{eq:Gammapm}
\Gamma_{+-}&=&\frac{\gamma}3\l[2|A_0|^2+|A_2|^2+2\sqrt2\re
\l(A_0A_2^*e^{i(\delta_0-\delta_2)}\r)\r]\ ,\\
\labell{eq:Gamma00}
\Gamma_{00}&=&\frac{\gamma}3\l[|A_0|^2+2|A_2|^2-2\sqrt2\re
\l(A_0A_2^*e^{i(\delta_0-\delta_2)}\r)\r]\ ,\\
\labell{eq:Gammap0}
\Gamma_{+0}&=&\frac34\gamma|A_2|^2\ ,
\eea
with $\gamma=\sqrt{M_K^2-4M_\pi^2}/(16\pi M_K^2)$. Considering
$\Gamma_{+-}+\Gamma_{00}$ and $\Gamma_{+0}$, and taking $M_\pi=134.8.\,\mev$ and 
$M_K=494.2\,\mev$ (i.e. isospin limit values), we find

\be
|A_0| = 4.66\times 10^{-4}\,\mev \quad\mbox{and}\quad
|A_2| = 2.08\times 10^{-5}\,\mev
\ee
\be
\rightarrow\frac{|A_0|}{|A_2|}=22.4
\labell{eq:DI12expt}\ ,\ee
whereas the combined chiral and large-$N_c$ limit predicts
$\sqrt2$~\shortcite{Lellouch:2001su4}!  It is the huge enhancement of
over 400 in the rate or 20 in the amplitude which is known as the
$\Delta I=1/2$ enhancement. It was termed a ``rule'' because this
enhancement of $\Delta I=1/2$ over $\Delta I=3/2$ transitions is also
observed in other decays, such as $\Lambda\to N\pi$, $\Sigma\to N\pi$
and $\Xi\to\Lambda\pi$.

\subsectiondum[The $\Delta I=1/2$ rule in the Standard Model]{The {\large $\Delta I=1/2$} rule in the Standard Model}
\labell{sec:DI12inSM}

At leading order in the weak and strong interactions, $\Delta S=-1$,
$\Delta D=1$ transitions occur through the tree level diagram

\begin{center}
\includegraphics[width=0.55\textwidth]{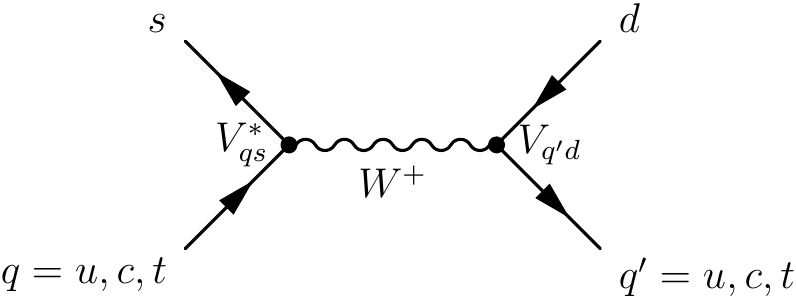}
\end{center}

As usual, we integrate out the heavy $W$ boson and $t$ quark. At
leading order in QCD, this yields the following four-quark, local
vertex
$$
\parbox{0.4\textwidth}{\includegraphics[width=0.4\textwidth]{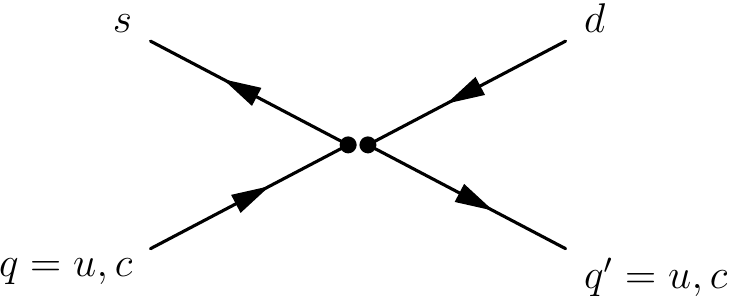}}
\longrightarrow\frac{G_F}{2\sqrt2}\sum_{q,q'=u,c} V_{q'd}V_{qs}^*Q_2^{qq'}
$$
with
\be
Q_2^{qq'}=(\bar sq)_{V-A}(\bar q'd)_{V-A}
\ .
\ee

Now, if we include $\alpha_s$ corrections~\shortcite{Gilman:1979bc},
such as
\begin{center}
\includegraphics[width=0.35\textwidth]{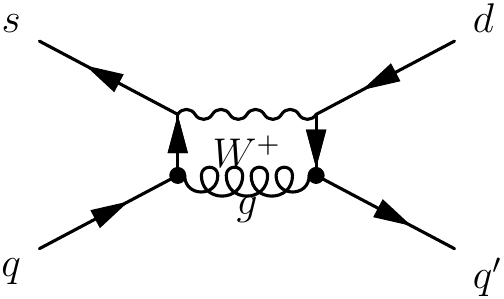}
\end{center}
we generate a new operator:
\be
Q_1^{qq'}=(\bar sd)_{V-A}(\bar q'q)_{V-A}
\ .\ee
These corrections also lead to the penguin diagram
\begin{center}
\includegraphics[width=0.45\textwidth]{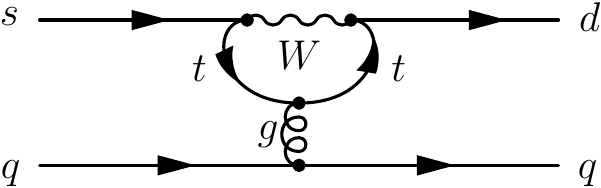}
\end{center}
with $q=u,d,s,c,b$. However, one finds that they yield tiny
contributions to the CP conserving parts of $K\to\pi\pi$ decays. 
Thus, we neglect these here. Moreover, we are only interested in
external states with $u$, $d$ and/or $s$ quarks. So we do not need
operators with a single charm leg. To simplify the operator structure,
we can also use the unitarity of the
CKM matrix:
\be
V_{cd}V_{cs}^*=-V_{ud}V_{us}^*-V_{td}V_{ts}^*
\ ,\ee
where the second term on the RHS of this equation can be neglected
because it is multiply Cabibbo suppressed by a factor of $\lambda^4\sim
0.003$ compared to the first term. Then, for CP conserving $\Delta
S=-1$, $\Delta D=1$ transitions, we have the effective Hamiltonian
\be
\cH^{\Delta
  S=-1}_\mathrm{CPC}=\frac{G_F}{\sqrt2}V_{ud}V_{us}^*\sum_{i=\pm}C_i(\mu)O_i
\labell{eq:DSeq1}\ee
with
\be
O_\pm=\left[(\bar su)_{V-A}(\bar ud)_{V-A}\pm(\bar sd)_{V-A}(\bar uu)_{V-A}\right]
-\left[u\to c\right]
\ ,
\labell{eq:Opm}
\ee
where the second term, in which $u$ is replaced by $c$ with an overall
minus sign, implements the GIM suppression
mechanism~\shortcite{Glashow:1970gm}. It is straightforward to show
that $O_+$ is in the $(84,1)$ representation of the $SU(4)_L\times
SU(4)_R$ chiral group, appropriate for classifying operators composed
of $u$, $d$, $s$ and $c$ quarks for renormalization in massless
schemes: it is a completely symmetric and traceless tensor in two
fundamental and two conjugate $SU(4)_L$ indices. $O_-$, on the other
hand, is in the $(20,1)$ representation of this group (it is a
completely antisymmetric and traceless tensor in two fundamental and
two conjugate $SU(4)_L$ indices). Thus, $O_+$ and $O_-$ do not mix
under renormalization. In \eq{eq:DSeq1}, the short distance Wilson
coefficients are given, at $O(\alpha_s)$, by~\shortcite{Gaillard:1974nj,Altarelli:1974exa,Altarelli:1980te}:
\be
C_\pm(\mu)=1+\frac{\alpha_s}{4\pi}\left(\gamma_\pm^{(0)}\ln
\frac{\mu^2}{M_W^2}+\delta_\pm\right)
\ ,\ee
with, in the $\msbar$-NDR scheme:
\be
\gamma_\pm^{(0)}=\pm3\frac{N_c\mp1}{N_c}\quad\mbox{and}\quad
\delta_\pm = \pm\frac{11}2\frac{N_c\mp1}{N_c}
\ .\ee
One often pushes the short distance calculation further and
integrates out the charm quark. But the GIM mechanism is very useful
for reducing the divergences of relevant weak matrix element on the
lattice. (Unfortunately, we will not have the time to cover this here,
but I refer the interested reader to \shortcite{Dawson:1997ic}, for
instance, for a discussion of this and related issues.)

A straightforward analysis of the isospin structure of the operators
of \eq{eq:Opm} show that $O_-$ is pure $I=\frac12$, while $O_+$ has
both $I=\frac12$, $\frac32$ components.~\footnote{In terms of
  $SU(3)_L\times SU(3)_R$ representations, $O_-$ is pure $(8,1)$ while
  $O_+$ contains both $(8,1)$ and $(27,1)$ representations.} One may
wonder then, whether the $\Delta I=1/2$ enhancement comes from the
running of the Wilson coefficients $C_\pm(\mu)$ from the scale
$\mu\sim M_W$, where the ratio $C_-(\mu)/C_+(\mu)$ is 1 plus small
corrections of order $\alpha_s(M_W)\sim 0.1$, down to a scale $\mu\sim
2\,\gev$. This would mean that the $\Delta I=1/2$ rule could be
understood as a short-distance enhancement. At leading-log order
(LL)~\shortcite{Gaillard:1974nj,Altarelli:1974exa}, using
\eq{eq:Cmurunning}, we find $C_-(2\,\gev)/C_+(2\,\gev)=
\left(\alpha_s(2\,\gev)/\alpha_s(m_b)\right)^{18/25}$
$\left(\alpha_s(m_b)/\alpha_s(M_W)\right)^{18/23}$ $(C_-(M_W)/C_+(M_W))\sim
2$. So there is some short-distance enhancement, but not enough by a
long shot to explain the $\Delta I=1/2$ rule.~\footnote{If you have
  been reading these notes carefully, you will be quick to point out
  that this statement has its limitations. Indeed, beyond LLO, the
  running of the Wilson coefficients is scheme dependent. However, it
  is difficult to justify not going beyond that order at scales
  $\mu\sim 2\,\gev$. So the statement should be be understood as
  applying to commonly used schemes.} In turn, this means that most of
the enhancement in
\be
\left|\frac{A_0}{A_2}\right| = 
\frac{\langle (\pi\pi)_0|C_+O_++C_-O_-|K^0\rangle}
{\langle (\pi\pi)_2|C_+O_+|K^0\rangle}
\ee
must come from long distances.

One may also wonder what role the charm quark plays in these
decays. In particular, one might consider an imaginary world in which
the GIM mechanism is exact, i.e. $m_c=m_u$. Does the $\Delta I=1/2$
enhancement persist in that limit? One way to address this problem is
to work in the $SU(4)$ chiral limit and determine the LECs
corresponding to $O_-$ and $O_+$, i.e. the $(20,1)$ and $(84,1)$
couplings (see comments after
\eq{eq:Opm})~\shortcite{Giusti:2006mh}. Numerically, in the quenched
approximation, it is found that there is an enhancement of $|A_0/A_2|$
over naive expectation (e.g. large-$N_c$), but that this enhancement
is roughly a factor of four too small.

An interesting way to understand this $SU(4)$ chiral limit enhancement
is to consider the three-point function contractions required to
determine the matrix elements $\la\pi^+|O_{\pm}|K^+\ra$. It is
straightforward to see that in the case of $O_+$, with the appropriate
flavor replacements, the contractions are the same as those of
\eq{eq:C3contract}, up to the overall factor of 2. On the other hand,
for the pure $\Delta I=1/2$ operator $O_-$, the contractions are those
\eq{eq:C3contract}, but with a plus sign between the double and single
trace terms. Thus, in the large-$N_c$ limit, the two matrix elements
coincide and we find the value of $|A_0/A_2|=\sqrt{2}$ discussed after
\eq{eq:DI12expt}. However, for finite $N_c$, the single trace term
contributes, and does so with opposite signs to $\la\pi^+|O_+|K^+\ra$
and to $\la\pi^+|O_-|K^+\ra$. Since both trace contractions are
positive, the larger the single trace term, the larger is
$\la\pi^+|O_-|K^+\ra$ and the smaller is $\la\pi^+|O_+|K^+\ra$. This
creates a $\Delta I=1/2$ enhancement in which both the numerator
$|A_0|$ is enhanced and the denominator $|A_2|$ is depressed. The
argument also implies an anticorrelation between $B_K$, the
$B$-parameter of $K^0$-$\bar K^0$ mixing, and the $\Delta I=1/2$
enhancement of $|A_0/A_2|$. Indeed, in the chiral limit, the smaller
$B_K$ is compared to its large-$N_c$ value of $3/4$, the larger
$|A_0/A_2|$ is compared to $\sqrt{2}$, as first noted in
\shortcite{Pich:1995qp}.

\medskip

Given the argument which we made earlier, one would think that LQCD is
well suited to study $K\to\pi\pi$ decays. However, there are many
conceptual problems one encounters when trying to study these decays
on the lattice. Some of these are:
\begin{itemize}

\item The renormalization of the $\Delta S=-1$ effective Hamiltonian
  is difficult on the lattice, even more so for the CP violating part
  (see e.g. \shortcite{Dawson:1997ic});

\item lattices are only a few fermi in size and the final-state
  hadrons cannot be separated into isolated, asymptotic states;

\item only approximately evaluated Euclidean correlation functions
  are available

\end{itemize}
There are also technical challenges. For instance, the study of
these decays requires the calculation of 4-point functions. Moreover,
power divergences must be subtracted, if the charm quark is integrated
out in the case of CP conserving decays, and once the $W$ and $t$ are
integrated out in the CP violating case. Both these points make the
study of $K\to\pi\pi$ decays very demanding numerically.

\subsectiondum{Euclidean correlation functions and the Maiani-Testa theorem}

For well over a decade, it was believed that $K\to\pi\pi$ amplitudes
could not be studied directly on the lattice. Indeed, these amplitudes
have both real and imaginary strong-interaction contributions while,
in the Euclidean, correlation functions are purely real (or
imaginary). Thus, it was difficult to see how such amplitudes could be
extracted from a lattice calculation, necessarily performed in the
Euclidean.

Of course, the Osterwalder-Schrader theorem
\shortcite{Osterwalder:1973dx,Osterwalder:1974tc} guarantees that Euclidean
correlation functions can be continued to Minkowski space, at least in
principle. However, in practice such analytical continuations are
essentially impossible with approximate Euclidean results.

This led Maiani and Testa, in 1990, to investigate what can be
extracted from Euclidean correlation functions without analytical
continuation~\shortcite{Maiani:1990ca}. They considered the following type
of center-of-mass frame Euclidean correlation function,
\be
\langle\pi(\vec{p},t_1)\pi(-\vec{p},t_1) \cH^{\Delta
  S=-1}_\mathrm{eff} K^\dagger(\vec{0},t_i)\rangle
\ ,\ee
and asked the question: what information does this correlation function 
contain regarding physical $K\to\pi\pi$ decay amplitudes
in the usual lattice, asymptotic limit, i.e. $t_1,t_2\gg 1/M_\pi$,
$-t_i\gg 1/M_K$?

To ``simplify'' the problem and disentangle Euclidean from other
possible lattice effects, they chose to work in a large,
quasi-infinite volume. This apparently innocuous assumption has rather
important consequences. For one, in infinite volume, the $\pi\pi$
spectrum is continuous. This means that in the limit $-t_i\gg 1/M_K$
and $t_1,t_2\gg 1/M_\pi$, only the ground state contribution can be
picked out: there is no known numerical technique to isolate an excited
state in a continuous spectrum. In turn, this implies that one can
only extract information about the matrix element
$\langle\pi(\vec{0})\pi(\vec{0})| \cH^{\Delta S=-1}_\mathrm{eff}|
K(\vec{0})\rangle$, in which all mesons are at rest:
the physical decay is not directly accessible on the lattice.

This statement became known as the ``Maiani-Testa theorem.'' It was a
formalization of the general belief that $K\to\pi\pi$ decays could not
be studied directly on the lattice but rather that
approximations~\shortcite{Bernard:1985wf} or
models~\shortcite{Ciuchini:1996mq} were needed to obtain information
about physical $K\to\pi\pi$ decays from lattice calculations.

However, as is often the case with ``no-go'' theorems, the solution is
found by questioning the underlying, apparently innocent
assumptions. Here, it was the infinite volume assumption that brought
in all of the difficulties while, for simulations performed in boxes
with sides of a few fermi at most, it does not even approximately
hold.

\subsectiondum{Two-pion states in finite volume}

In a finite box with sides $L$,~\footnote{$L$ must be large enough so
  that the range of the interaction between the two pions is contained
  within the box.} two pions cannot be isolated
into noninteracting asymptotic states. Rather, the $\pi\pi$
eigenstates are the result of a stationary scattering process. In
addition, boundary conditions generically imply that the particles'
momenta are quantized. For periodic boundary conditions, they come in
discrete multiples of $2\pi/L$: $\vec{k}=\vec{z}(2\pi/L)$, with
$\vec{z}\in\mathbb{Z}^3$. In turn, this means that the spectrum is
discrete and the splitting is actually rather large for box sides of a
few fermi. In such boxes, the typical spacing between momenta is
$\Delta p=2\pi/L=1.2\,\gev/L[\fm]$. This is clearly quite different from
the continuous spectrum found in infinite volume.

In the free theory, in the center-of-mass frame and in the $A_1^+$
(i.e. cubic spin-0) sector, the energy of the $n$-th excited state is
given by:
\be
W_n^{(0)}=2\sqrt{M_\pi^2+n\left(\frac{2\pi}{L}\right)^2}
\ ,\ee
for $n\le 6$--there is no integer three-vector with norm squared, 7,
nor 8 for that matter. This spectrum is shown as the dashed lines in
\fig{fig:2pispect}.

\begin{figure}[t]
\centering
\includegraphics[width=0.75\textwidth]{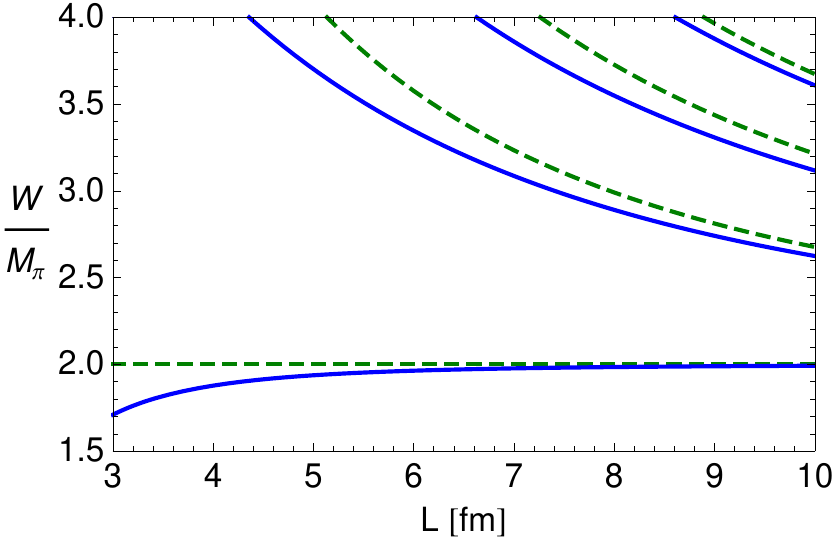}
\caption{The two-pion spectrum in a box of volume $L^3$ as a function
  of $L$ in the $I=J=0$ channel, under the four-pion threshold
  $W/M_\pi=4$. The free spectrum is depicted by the green dashed
  curves. The solid blue curves show the interacting spectrum as
  obtained from \eq{eq:MartinQC}, using the one-loop $I=J=0$ phase
  shift from \protect\shortcite{Gasser:1990ku,Knecht:1995tr}.}
\labell{fig:2pispect}
\end{figure}

In the presence of interactions, the energy $W_n$ of the fully
interacting state was worked out by Martin to all orders in
relativistic quantum field theory
\shortcite{Luscher:1986pf,Luscher:1990ux} for two-pion energies below
the four-pion threshold, up to corrections which fall of exponentially
with the box size (i.e. up to finite-volume, vacuum polarization effects). 

The ground state $n=0$ requires special treatment. In the isospin
$I=0$ and spin $J=0$ channel, we have~\shortcite{Luscher:1986pf}:
\be
W_0=2M_\pi-\frac{4\pi a_I}{M_\pi L^3}\left\{1+c_1\frac{a_I}{L}+
c_2\left(\frac{a_I}{L}\right)^2\right\}+O(L^{-6})
\ee
where
\be
c_1=-2.837297,\qquad c_2=6.375183\ ,
\ee
where the $S$-wave scattering length in the appropriate channel is $a_I$ with
\be
a_I=\lim_{k\to 0}\frac{\delta_I(k)}k
\ ,\ee
and where $k$ is the momentum of the pions in the center-of-mass frame.

For excited states, the results are given in terms of the scattering
phase, $\delta_I$, of the relevant isospin channel. The appearance of
scattering phases should not be surprising. We know from scattering
theory that, under reasonable conditions, potentials can be
reconstructed from these phases. Martin finds
that~\shortcite{Luscher:1990ux}
\be
W_n=2\sqrt{M_\pi^2+k_n^2}\qquad n=1,2,3,\ldots
\ ,
\ee
where $k_n$ is a solution of the quantization equation
\be
\labell{eq:MartinQC}
n\pi-\delta_I(k_n)=\phi(q_n)
\ ,\ee
with $q_n\equiv k_n L/(2\pi)$ and with
\be
\labell{eq:tanphi}
\tan\phi(q)=-\frac{\pi^{3/2}q}{Z_{00}(1;q^2)}
\ .\ee
$\phi(q)$ is defined for $q\ge 0$. It is such that $\phi(0)=0$ and
it depends continuously on $q$. It is given in terms
of the zeta function of the Laplacian
\be
\labell{eq:zeta00}
Z_{00}(s;q^2)=\frac{1}{\sqrt{4\pi}}\sum_{\vec{n}\in\mathbb{Z}^3}\left(\vec{n}^2-
  q^2\right)^{-s}
\ ,\ee
for $\re\ s>3/2$ and by analytic continuation elsewhere. A useful
integral representation for evaluating $Z_{00}(1;q^2)$ numerically is
given in 
\sec{sec:integzeta}.

Solving \eq{eq:MartinQC}, one generically finds
\be
W_n=2\sqrt{M_\pi^2+n\left(\frac{2\pi}{L}\right)^2}+O\left(\frac{1}{L^3}\right)
\ ,\ee
where the equation actually gives the whole tower of $1/L$ corrections
once the scattering phase is specified. As already stated, the
equation holds for $n\le 6$ and there is no integer three-vector whose
squared norm is 7 or 8. Then, beginning at 9, there are integer
vectors, $\vec{z}$, which are not related by cubic rotations but which
have the same squared norm, e.g. $\vec{z_1}=(2,2,1)$ and
$\vec{z_2}=(3,0,0)$. In a treatment where only the $O(3)$ spin-0
component of the $A_1^+$ representation is taken into
account~\footnote{Because the cube is not invariant under generic
  rotations, the irreducible representations of the cubic group are
  resolved into many irreducible representations of $O(3)$. For the
  $A_1^+$ cubic representation, the relevant spin representations are
  spin 0, 4,\ldots.}, the states associated with such three-vectors in
the free case do not feel the interaction in the interacting case
either, and have $k=|\vec{z}|(2\pi/L)$. This is because they combine
into states with $O(3)$ spin-4 and/or higher spins. Such states will
be ignored in the following and we will consider only $n\le 6$, which
is certainly not a limitation in practice.

The full solution (taking for instance $I=J=0$) 
is shown in \fig{fig:2pispect}. The first remark which can be made is
that the two-pion spectrum on lattices which can be considered in the
foreseeable future is far from being continuous. The second is that 
distortions due to interactions are quite small: the volume
suppression of the corrections is effective when
$L\ge3\,\fm$. Finally, it is clear that by studying the
energies $W_n(L)$ as a function of box size $L$, we can turn
\eq{eq:MartinQC} around and reconstruct, at least for a few discrete
momenta, the scattering phase $\delta_I(k)$.

Now, suppose that there is a single resonance $R$ in this $\pi\pi$
channel, with mass $M_R< 4 M_\pi$, i.e. under the inelastic threshold,
and width $\Gamma_R$. To explain how Martin's equation works in that
case, it is useful to turn to quantum mechanics. A quantum mechanics
approach is actually justified because corrections suppressed
exponentially in $L$ are neglected in the derivation of the
quantization formula. Such suppression factors generically correspond
to tunneling phenomena. Here they are associated with features of
relativistic quantum field theories which are absent in quantum
mechanics: the exchange of a virtual particle around the box.

We begin by decomposing the total Hamiltonian $H$ of the two-pion
system into a free part $H_0$ and an interaction $H_{int}$:
\be
H=H_0+H_{int}
\ .\ee
Then we consider the $n$-th free $\pi\pi$ state ($n\le 6$), 
$|n_0\rangle$, in our
chosen channel (here isospin $I=0$ or $2$ and $J^p=0^-$). It is such that
$\langle n_0|n_0\rangle=1$ and
\be
H_0|n_0\rangle = W_n^{(0)}|n_0\rangle
\ .
\ee

To understand what this energy becomes in the presence of interactions
and of the resonance, it is useful to consider a perturbative
expansion in the interaction $H_{int}$, though the final result is
accurate to all orders. Denoting the resulting energy $W_n$, and
$|n\rangle$, the corresponding fully-interacting eigenstate, we have,
to second order in $H_{int}$:
\bea
\labell{eq:WnPT}
W_n&=&\langle n |H|n\rangle\nn \\
&=& W_n^{(0)} + \langle n_0 |H_{int}|n_0\rangle + \sum_\alpha\frac{\langle
  n_0 |H_{int}|\alpha\rangle\langle \alpha |H_{int}|n_0\rangle}{W_n^{(0)}-W_\alpha}
+\cdots\ .\eea
Here $\alpha$ runs over 2, 4, $\ldots$ pion states, as well as any
other state which appears in the given channel, and $W_\alpha$ is the
corresponding energy. Factors of the form $(W_n^{(0)}-W_\alpha)$ also
appear in the higher-order terms of the perturbative expansion
represented by the ellipsis.

As long as $L$ is such that $W_\alpha=M_R$ is far from the
free two-pion energy, $W_n^{(0)}$, all of the terms in the
perturbative series are regular and can therefore be
resummed. Now, if the resonance is narrow, the coupling of the
resonance to the $n_0$, $\pi\pi$ state will be small. In turn, this
means that the matrix element $\la\pi\pi|H_{int}|R\ra$ is small
compared to the mass and typical energies of the system.~\footnote{We
  assume here that the narrowness of the resonance is not only due to
  phase-space suppression.} Moreover, the leading correction which the
resonance brings to the interacting two-pion energy, $W_n$, appears at
second order in the expansion. Thus, it is of order
$|\la\pi\pi|H_{int}|R\ra|^2$ and is therefore small.

When $L=L_R$ such that $W_n^{(0)}(L_R)=M_R$, the effect
of the resonance is radically different. Its second and higher order
contributions to $W_n$ blow up. In such a situation, we have to resort
to degenerate perturbation theory and first diagonalize $H$ in the
two-state subspace $\{|n_0\rangle,\,|R\rangle\}$. This means
diagonalizing the $2\times 2$ matrix
\be
\left(
\begin{array}{cc}
\langle n_0|H|n_0\rangle & \langle n_0|H|R\rangle\\
\langle R|H|n_0\rangle & \langle R|H|R\rangle
\end{array}
\right)
=\left(
\begin{array}{cc}
M_R & M_n\\
M_n^* & M_R 
\end{array}
\right)
\ ,\ee
where $M_n\equiv \langle n_0|H|R\rangle$ is the transition
amplitude between the resonance and the two-pion state
$|n_0\rangle$. A straightforward diagonalization yields
\be
W_n^\pm=M_R\pm |M_n|
\ ,\ee
thereby lifting the degeneracy and giving rise to a typical level
repulsion phenomenon. Thus, in solving Martin's formula
(\reff{eq:MartinQC}), we would find a dependence of the two-pion
energy as a function of $L$, which looks like what is depicted in
\fig{fig:avoidedlc}.

\begin{figure}[t]
\centering
\includegraphics[width=0.75\textwidth]{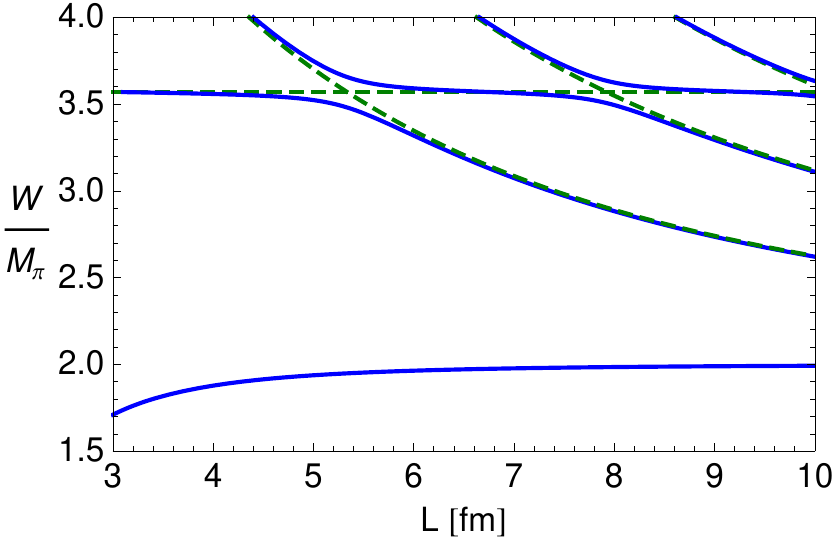}
\caption{The two-pion spectrum in a box of volume $L^3$ as a function
  of $L$ in the $I=J=0$ channel with an additional, fictitious
  resonance of mass $M_R=M_{K_S}$ and width
  $\Gamma_R=10^{12}\times(\Gamma_{+-}+\Gamma_{00}-(4/3)\Gamma_{+0})
  \simeq 7.3\,\mev$ with $\Gamma_{+-}$, $\Gamma_{00}$, and $\Gamma_{+0}$ given in 
\eqs{eq:Gammapm}{eq:Gammap0}. The dashed green
  curves represent the two-pion spectrum interacting through
  $\delta_0$ (they correspond to the solid blue curves in
  \protect\fig{fig:2pispect}). The solid blue curves show the
  interacting spectrum as obtained from \eq{eq:MartinQC}, with the
  contribution of the resonance to the phase shift. At the points at
  which the resonance (i.e. the horizontal green dashed curve at
  $M_{K_S}/M_\pi\sim 3.57$) crosses the excited $I=J=0$ states, one
  clearly sees a level repulsion effect. The effect is rather small
  and limited to a smallish region around the crossing point because
  the resonance is narrow: $\Gamma_R/M_R\simeq 1.5\%$.}
\labell{fig:avoidedlc}
\end{figure}

\bigskip

\subsectiondum[$K\to\pi\pi$ in finite volume]{{\large $K\to\pi\pi$} in finite volume}
\labell{sec:KpipiFV}

What Martin and I realized over ten years ago is that to study
$K\to\pi\pi$ decays in finite volume, we could treat the kaon as an
infinitesimally narrow resonance in the weak interaction contribution
to the scattering of the two pions~\shortcite{Lellouch:2000pv}. In
that case, one considers the free Hamiltonian to be the QCD Hamiltonian,
i.e.
\be
H_0=H_\mathrm{QCD}
\ ,\ee
and the perturbation, $H_{int}$, to be the effective weak Hamiltonian
relevant for $K\to\pi\pi$ decays, i.e.
\be
H_{int}=H_W=\int_{x_0=0}d^3x\;\cH_W(x)
\ .\ee
Then, since the amplitudes for $K\to\pi\pi$ decays, $T(K\to\pi\pi)$,
are computed at $O(G_F)$, we can perform all of our computations to
that order.

Following what was done in the preceding section for the resonance, we
tune the size of the box to $L=L_K$, such that for some level $n$,
\be
\labell{eq:WnLK}
W_n(L_K)=M_K
\ .\ee
In that case, the corresponding pion momentum is the momentum,
$k_\pi$, which the pions would have in the physical kaon decay, i.e.
\be
\labell{eq:knLK}
k_n(L_K)=k_\pi\equiv\sqrt{\frac{M_K^2}{4}-M_\pi^2}
\ .\ee
Then, the transition matrix element in the finite volume $V=L_K^3$,
\be
\labell{eq:MnIdef}
M_n^I\equiv\langleV(\pi\pi)_In|H_W|K\rangleV
\ ,\ee
is an energy conserving matrix element. Since finite-volume
corrections to a single, stable particle state are exponentially small
in $L$ and since we neglect such corrections here, $|K\ra_V$ is
identical to $|K\ra$, up to a purely kinematic normalization factor.

However, we are not interested in the finite-volume matrix element of
\eq{eq:MnIdef}. What we want is the infinite-volume transition
amplitude,
\be
T_I\equiv \langle(\pi\pi)_In,out|\cH_W|K\rangle
\ ,\ee
where the corresponding $A_I$ of \eq{eq:isodecomp} can be made real in the
CP conserving case.

Because it is important here, let us pause to say a few words about
the normalization of states used. In finite volume we use the usual 
quantum mechanical normalization of states to unity. Thus, for a spinless
particle of mass $m$ and momentum $\vec{p}$:
\be
\labell{eq:FVnorm}
\langleV \vec{p}|\vec{p}'\rangleV=\delta_{\vec{p},\vec{p}'}
\ .\ee
In infinite volume, it is the
standard relativistic normalization of states, i.e.
\be
\labell{eq:IVnorm}
\langle p|p'\rangle=2p^0(2\pi)^3\delta^{(3)}(\vec{p}-\vec{p}')
\ ,\ee
with $p^0=\sqrt{m^2+\vec{p}^2}$, which is implemented.

To obtain the relationship between these two amplitudes, we compute
the shift in energy brought about by the presence of the weak
interactions in two different ways, and require the two results to agree.

We begin by repeating the degenerate perturbation theory of the
preceding section. We obtain
\be
\labell{eq:Wnpm}
W_n^\pm=M_K\pm |M_n^I|
\ ,\ee
where $M_n^I$ is clearly $O(G_F)$. Then, we turn to Martin's
finite-volume 
quantization formula (\reff{eq:MartinQC}). In the presence of
the weak interaction, the phase shift receives a weak
contribution, $\delta_W$. Thus, in that formula,
we have to perform the replacement
\be
\delta_I\to \bar\delta_I = \delta_I+\delta_W
\ .\ee

\begin{figure}[t]
\centering
\includegraphics[width=0.4\textwidth]{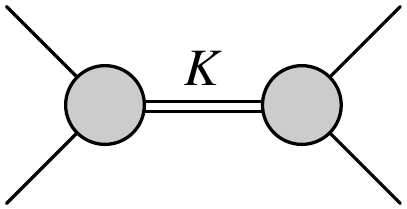}
\caption{Kaon contribution to the elastic weak
  scattering of two pions in the $S$-channel. The shaded vertices
  represent the $K\to\pi\pi$ transition amplitudes at $O(G_F)$.}
\labell{fig:pipiKpipi}
\end{figure}

We are interested in this phase shift at the values of momenta
$k=k_n^\pm$, corresponding to the perturbed energies $W_n^\pm$ of
\eq{eq:Wnpm}:
\be
k_n^\pm=k_n\pm\Delta k=k_n\pm\frac{W_n|M_n^I|}{4k_n}+O(G_F^2)
\ .\ee
Indeed, we know that those values must come out of the
quantization formula because the two methods of
determining the energy shifts must give the same result. 
Because $W_n^\pm$ are
``infinitesimally'' close to $M_K$ (in our LO counting in $G_F$),
$\delta_W(k_n^\pm)$ is dominated by the $s$-channel
kaon exchange depicted in \fig{fig:pipiKpipi}. Any other contribution
will be at least $O(G_F^2)$. In the $s$-channel, however, the factor
of $G_F^2$ coming from the two $K$-$\pi\pi$ vertices is compensated by
a propagator enhancement, due to the fact that we are sitting only
$O(G_F)$ away from the peak of the resonance. Indeed, the scattering amplitude 
corresponding to \fig{fig:pipiKpipi} is
\bea
\labell{eq:Spipiscat}
S[\mbox{\protect\fig{fig:pipiKpipi}}]&=&-\frac{\overbrace{T_I^*
  T_I}^{O(G_F^2)}}{\underbrace{(W_n^\pm)^2-M_K^2}_{O(G_F)}-iM_K
\underbrace{\Gamma_K}_{O(G_F^2)}}+O(G_F^2)\\
&=&\mp \frac{|A_I|^2}{2W_n^{(0)}|M_n^I|}+O(G_F^2)
\ ,\eea
where we have used the fact that the vertices in \fig{fig:pipiKpipi}
are the on-shell transition amplitudes up to higher order corrections
in $G_F$. To translate this amplitude into a scattering phase, we use
the partial wave decomposition of a $\pi\pi$ scattering amplitude $S$
in the center of mass frame:~\footnote{Note that $\delta_l^I$'s
  subscript in \protect\eq{eq:Spartialwave} corresponds to angular
  momentum $l$, while the superscript is the isospin $I$. This
  notation will be used in this equation only. Elsewhere, $\delta$'s
  subscript will be the isospin, except for $\delta_W$ where $W$
  stands for weak.}
\be
\labell{eq:Spartialwave}
S=16\pi\,W\sum_{l=0}^\infty (2l+1)P_l(\mbox{cos}\;\theta)\,e^{i\delta_l^I(k)}
\,\frac{\mbox{sin}\;\delta_l^I(k)}{k}
\ ,\ee
where $W=2\sqrt{M_\pi^2+k^2}$ is the energy of the
pion pair. Since the amplitude of \eq{eq:Spipiscat} has no angular
dependence, it only leads to a zero angular momentum phase shift,
which is the weak phase shift of interest. Thus, we find
\be
\delta_W(k_n^\pm)=\mp \frac{k_n^\pm|A_I|^2}{32\pi
  (W_n^{(0)})^2|M_n^I|}+O(G_F^2)
\ .\ee
We can now include this contribution into the total phase shift
$\bar\delta_I$ and write down the resulting quantization equation:
\be
n\pi-\delta_I(k_n\pm\Delta k)\pm\frac{k_n^\pm|A_I|^2}{32\pi
  (W_n^{(0)})^2|M_n^I|}=\phi(q_n\pm\Delta q)+O(G_F^2)
\ ,\ee
with $\Delta q=L\Delta k/(2\pi)$. Expanding this equation to $O(G_F)$,
we finally find the relationship between the desired, infinite-volume
amplitude $A_I$ and the finite-volume matrix elements $M_n^I$,
computed on the lattice~\shortcite{Lellouch:2000pv}:
\be
\labell{eq:LLformula}
|A_I|^2=8\pi\left\{q\frac{\partial}{\partial
  q}\phi(q)+k\frac{\partial}{\partial
  k}\delta_I(k)\right\}_{k=k_n}\frac{(W_n^{(0)})^2M_K}{k_n^3}L^6|\cM_n^I|^2
\ ,\ee
were we have used the definition
\be
\cM_n^I\equiv\langleV(\pi\pi)_In|\cH_W(0)|K\rangleV=M_n^I/L^3
\ .\ee
In our derivation we have tuned the size of the box to $L_K$ such that
$W_n^{(0)}=M_K$ and thus, $k_n=k_\pi$ (\eqs{eq:WnLK}{eq:knLK}).  However,
\eq{eq:LLformula} is also valid for $W_n^{(0)}\ne M_K$, as can be seen
in \sec{sec:kpipirqft} and as was derived using the fact that the
matching factor is related to the density of interacting two-pion
states in finite volume~\shortcite{Lin:2001ek}. An interesting
discussion of this and other ways of looking at this formula is given
in \shortcite{Testa:2005zm}.  In \shortcite{Kim:2005gf}, 
\eq{eq:LLformula} was further generalized
to moving frames, i.e. frames in which the
center of mass has a nonvanishing momentum. And in
\shortcite{Kim:2010sd}, it is shown how partially-twisted boundary
conditions can be used to obtain the phase shift $\delta_2(k)$
and its derivative in the isospin-2 channel. With twisted boundary
conditions, one allows some of the quark flavors to be periodic
only up to
a phase. This phase forces the flavors concerned to carry a momentum
which is proportional to the phase. The boundary conditions are called
partially-twisted when it is only the valence flavors which are given
a twist.

The proportionality factor (\reff{eq:LLformula}) is to a large extent
kinematic, as it accounts for the difference in normalization of
states in finite and infinite volumes, given in
\eqs{eq:FVnorm}{eq:IVnorm}. This can easily be seen in the absence of
interactions. To reach the $n^\mathrm{th}$ two-pion with energy $W_n^{(0)}$
and pion momentum $k_n^{(0)}$, the cube must have sides
\be
\labell{eq:Lnfree}
L_n=\frac{2\pi}{k_n^{(0)}}\sqrt{n}
\ .\ee
Then, \eq{eq:LLformula} assumes the form
\be
\labell{eq:LLfree}
|A_I|^2=\frac{4}{\nu_n}(W_n^{(0)})^2M_KL^3|M_n^I|^2
\ee
where
\be
\labell{eq:nundef}
\nu_n\equiv\mbox{number of $\vec{z}\in\mathbb{Z}^3\;\ni\;\vec{z}^2=n$}
\ .\ee
The proportionality constant is precisely the relative normalization
of free kaon and two-pion states in finite and infinite volume
projected onto the $A_1^+$ and spin-0 sectors, respectively (see also
\sec{sec:kpipirqft}). The constant is the product of the ratio
of squared norms of the kaon state, $2M_KL^3$, and of the two-pion
state, $2(W_nL^3)^2$,~\footnote{The factor of 2 is required because
  the two pions in the final state are identical particles in the
  isospin limit which we consider here.} 
times the square of the factor relating
$H_W$ and $\cH_W(0)$, i.e. $1/L^6$, times $1/\nu_n$ since the
finite-volume, $A_1^+$ state is obtained by summing over the $\nu_n$
pion momentum directions.


\subsectiondum[$K\to\pi\pi$ in finite volume: a simple relativistic quantum 
field theory example]{{\large $K\to\pi\pi$} in finite volume: a simple relativistic quantum 
field theory example}
\labell{sec:kpipirqft}

To understand how the finite-volume effects predicted by
\eqs{eq:MartinQC}{eq:LLformula} show up in correlation functions
similar to those one would use in numerical simulations, it is useful
to consider them in the context of a relativistic field theory in
which all quantities of interest can be computed analytically. Because
the form of the finite-volume formulae (\reff{eq:MartinQC}) and
(\reff{eq:LLformula}) does not depend on the details of the dynamics,
we choose to work in a world in which this dynamics is simplified, so
as not to obscure the discussion of finite-volume effects with
superfluous technical details. The calculations below were summarized in
\shortcite{Lellouch:2000pv}.


\subsubsectiondum{Specification of the model}

We consider a theory of a single, neutral, spinless pion field
$\pi(x)$, of mass $M_\pi$. In the notation of
\sec{sec:KpipiFV}, the unperturbed Hamiltonian density is
\be
\cH_0=\cH_\mathrm{kin}+\cH_S
\ ,
\ee
where $\cH_\mathrm{kin}$ is the usual kinetic Hamiltonian of a scalar
field and the ``strong'' interaction between the pions is given by
\be
\cH_S=\frac{\lambda}{4!}\pi^4
\ .\labell{lst}
\ee
We assume here that the theory is perturbative in $\lambda$ and we
will work to first nontrivial order in $\lambda$,
i.e. $O(\lambda)$. Because the lattice calculations are performed in
Euclidean spacetime, we rotate this theory into the Euclidean and
consider Euclidean correlation functions. Moreover, our
calculations will be performed in a three-volume $L^3$ with periodic
boundary conditions, but the time direction will be considered
of infinite extent.

To make the perturbation theory completely well-defined, we introduce
a Pauli-Villars cutoff $\Lambda$.  At tree level the Euclidean pion
propagator is then given by
\be
\labell{eq:piprop}
S_\pi(x)=\int_x\,e^{ik\cdot
  x}\langle\pi(x)\pi(0)\rangle=\frac1{p^2+M_\pi^2}-
\frac1{p^2+\Lambda^2}
\ .\ee
The cutoff should be large enough so that ghost particles cannot
be produced at energies below the four-pion threshold, but in view of
the universality of \eq{eq:MartinQC} and (\reff{eq:LLformula}) there is no need
to take $\Lambda$ to infinity at the end of the calculation.

Since we are going to be computing correlation functions in the
time-momentum representation, it is useful to have the pion propagator
in this same representation. We have
\be
S_\pi(t;\vec{k})\equiv\int_{\vec{x}}e^{-i\vec{k}\cdot \vec{x}}
\la\pi(t,\vec{x})\pi(0)\ra
\nonumber
\ee
\be
\labell{piprop}
=\frac{1}{2E_k}\;e^{-E_k\,|t|}-\frac{1}{2\cE_k}\;e^{-\cE_k\,|t|}
\ ,
\ee
where $E_k=\sqrt{\vec{k}^2+M_\pi^2}$ and $\cE_k=\sqrt{\vec{k}^2+\Lambda^2}$.

As far as the kaon and its decays into two pions are concerned, the
least complicated possibility is to describe it by a free hermitian
field $K(x)$ with mass $M_K$ and to take 
\be
\labell{lwk}
\cH_W=\frac{g}{2}K\pi^2
\ee
as a weak Hamiltonian density. We will only work here to leading order
in the weak coupling, also.


\subsubsectiondum{Determination of the phase shift}

Let us first determine the phase shift, $\delta(k)$, in the model
of \eq{lst}. This is an infinite-volume, Minkowski space
calculation, though at the order at which we work this fact makes very
little difference. The partial wave decomposition of the invariant,
scattering amplitude $S$, in the center-of-mass frame, is given in
\eq{eq:Spartialwave}. At
$\ord{\lambda}$, $S=-\lambda$, and therefore
\be
\delta(k)=-\frac{\lambda}{16\pi}\frac{k}{W}+O(\lambda^2)
\ ,\labell{phase1}
\ee
where $W=2E_k$ is the free two-pion energy.


\subsubsectiondum{Two-pion energies using \protect\eq{eq:MartinQC}}

In the absence of interactions, i.e. for $\delta(k)\equiv 0$, the
solutions of \eq{eq:MartinQC}, for $n=1,2,\ldots,6$, are the free,
finite-volume momentum magnitudes,
$\kLO{n}\equiv\sqrt{n}(2\pi/L)$.~\footnote{Here and in the following,
  quantities with the superscript $(0)$ are computed at $O(\lambda^0)$,
  while those without a superscript are the same quantities in the
  presence of the ``strong'' interaction of \protect\eq{lst}.} For weakly
interacting pions, the solutions are small perturbations about these
values. Thus, the rescaled momenta are
\be
\qNLO{n}=\frac{\kNLO{n} L}{2\pi}=\qLO{n}+\Delta
q_n=\sqrt{n}+\Delta q_n
\ ,\labell{qNLOdef}
\ee
where $\Delta q_n$ is the small perturbation. 

To first order in $\Delta q_n$ and $\lambda$
\be
\mathrm{tan}\,\phi(\qNLO{n})=\frac{4\pi^2}{\nu_n}n\Delta
q_n+\ord{\Delta q_n^2}
\ ,
\labell{tanphifirst}
\ee
and \eq{eq:MartinQC} yields
\be
\Delta q_n=\lambda\frac{\nu_n}{32\pi^2\sqrt{n}}
\frac{1}{\WLO{n} L}+\ord{\lambda^2}
\ ,\labell{deltaq}
\ee
where $\WLO{n}=2\sqrt{M_\pi^2+(\kLO{n})^2}$ is the energy of two free pions with
opposite momenta of magnitude $\kLO{n}$.  Thus, the energy of the
corresponding two-pion state, in the presence of interactions, is
\be
\WNLO{n}=\WLO{n}\l(1+\lambda\frac{\nu_n}{2}\frac{1}{(\WLO{n} L)^3}
+ \ord{\lambda^2}\r)
\ .\labell{2pien}
\ee
%


\subsubsectiondum{Two-pion energies from perturbation theory}

In perturbation theory, the two-pion energy corresponding to pions
whose momenta would have magnitude $\kLO{n}$, $n=1,\ldots,6$, in the
absence of interactions, can be extracted from the $\pi\pi\to\pi\pi$
correlation function
\be
\Cpipi(t)=\la \cO_n(t)\cO_n(0)\ra_\mathrm{conn}
\ ,
\labell{pipicf}\ee
where
\be
\cO_n(t)=\frac{1}{\nu_n}
\sum_{\{\vec{k}_n\}}\int_{\vec{x_1}\vec{x_2}}\,e^{i\vec{k}_n
\cdot(\vec{x_2}-\vec{x_1})}\pi(t,\vec{x_2})
\pi(t,\vec{x_1})
\labell{pipiop}
\ee
is an operator which has overlap with zero-momentum, cubically
invariant, two-pion states and $\nu_n$ is the number of momenta
$\vec{k}_n$ such that $|\vec{k}_n|=\kLO{n}$ (see \eq{eq:nundef}).  The sum in
\eq{pipiop} is over these momenta, all related by cubic
transformations. The operator $\pi(x)$ has overlap with single pion
states. In the limit of large $t$, the contribution of the two-pion states, 
$|\pi\pi\,l\rangleV$, to the correlation function of \eq{pipicf}, is
\be
\Cpipi(t)\longrightarrow \sum_{l=0}^6\l|\la0|\cO_n(0)|\pi\pi\,l\rangleV\r|^2
\,e^{-\WNLO{l} t}+\cdots
\ ,\labell{cpipik}
\ee
where the ellipsis stands for terms which decay more rapidly. The
states $|\pi\pi\,l\rangleV$ are normalized to one.

\begin{figure}
\centering
\includegraphics[width=0.8\textwidth]{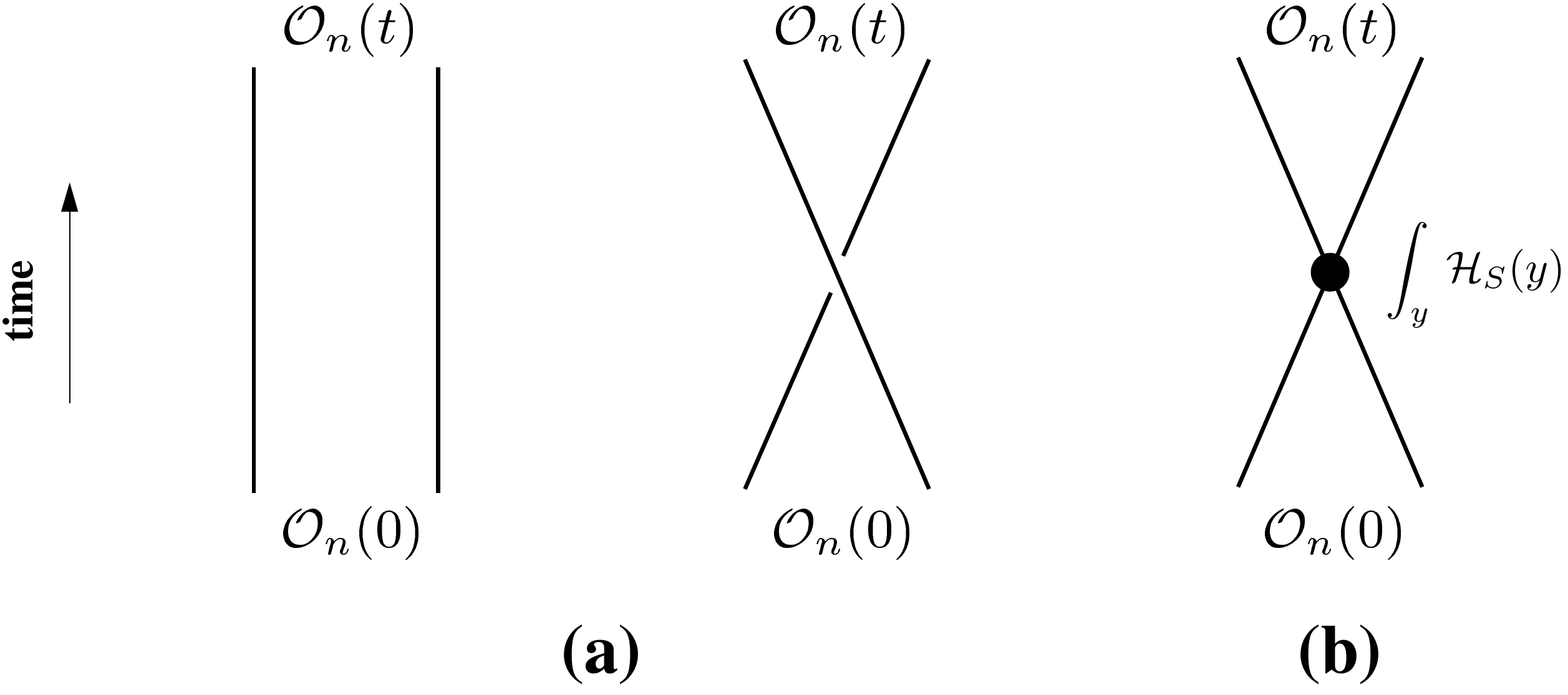}
\caption{Diagrams which contribute to $\Cpipi(t)$ at $\ord{\lambda^0}$
(a) and $\ord{\lambda}$ (b).}
\labell{fig:pipi}
\end{figure}
A straightforward calculation of the diagrams of \fig{fig:pipi}, using
the propagator of \eq{piprop}, gives for the correlation function of
\eq{pipicf}, at $\ord{\lambda}$,
$$
\left.\Cpipi(t)\right\vert_n= \frac{2}{\nu_n}\l(\frac{L^3}{\WLO{n}}\r)^2e^{-\WLO{n} t}
\Biggl\{1-\lambda\frac{\nu_n}{2L^3}
\Biggl(\frac{t}{(\WLO{n})^2}-\frac{1}{(\WLO{n})^3}
$$
\be
\labell{cpipires}
-
\frac{\Rlam{(\kLO{n})^2}{(\kLO{n})^2}}{2}\Biggr)+\ord{\lambda^2}\Biggr\}
\ee
$$
=\frac{2}{\nu_n}\l(\frac{L^3}{\WNLO{n}}\r)^2e^{-\WNLO{n} t}
\l\{1+\lambda\frac{\nu_n}{2L^3}\l(
\frac{1}{(\WLO{n})^3}+\frac{\Rlam{(\kLO{n})^2}{(\kLO{n})^2}}{2}\r)+\ord{\lambda^2}\r\}
\ ,
$$
where we have only retained the contribution which decays
exponentially with the rate corresponding that of the two-pion state,
$|\pi\pi\,n\rangleV$.  In \eq{cpipires}, the two-pion energy, $\WNLO{n}$, is
the same as that obtained from the finite-volume formula of
\shortcite{Luscher:1990ux} (see \eq{2pien}) and the regulator
contribution is given by
\be
\Rlam{\vec{p}^2}{\vec{k}^2}=
\frac{4(E_p+\cE_p)}{E_p\cE_p\l[(E_p+\cE_p)^2-4E_k^2\r]}
-\frac{1}{\cE_p(E^2_p-\cE^2_p)}
\ ,\labell{regcontr}
\ee
with self-explanatory notation.

Comparison of this result with \eq{cpipik} further gives the matrix
element of $\cO_n$ between the vacuum and the two-pion state,
$|\pi\pi\,n\rangleV$:
$$
\l|\la 0|\cO_n(0)|\pi\pi\,n\rangleV\r|=\sqrt{\frac{2}{\nu_n}}
\l(\frac{L^3}{\WNLO{n}}\r)\Biggl\{1+
\lambda\frac{\nu_n}{4L^3}\Biggl[\frac{1}{(\WLO{n})^3}+\frac{\Rlam{(\kLO{n})^2}{(\kLO{n})^2}}{2}
\Biggr]
$$
\be
+\ord{\lambda^2}\Biggr\}
\ .\labell{pipime}
\ee
\begin{figure}
\centering
\includegraphics[width=0.45\textwidth]{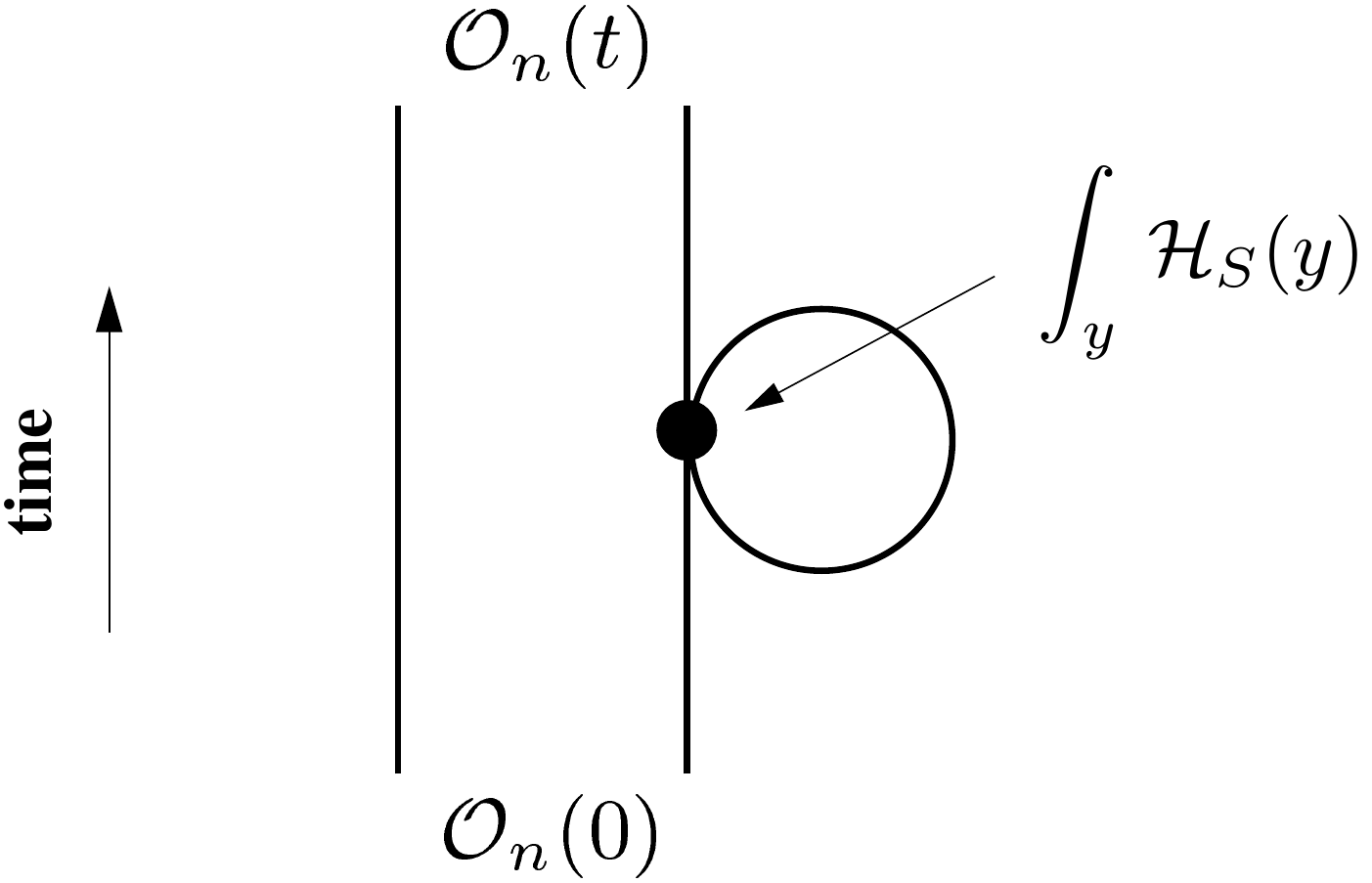}
\caption{Example of a tadpole contribution to $\Cpipi(t)$ at
$\ord{\lambda}$.}
\labell{fig:tad}
\end{figure}

The observant reader will have noticed that we have not taken into
account contributions from diagrams such as the one of \fig{fig:tad},
which also appear at $\ord{\lambda}$. As can be verified explicitly,
these diagrams amount to a shift of the pion mass by terms which are
independent of $L$, up to exponentially small corrections.  Since such
corrections are neglected here, these contributions will affect none
of our finite-volume results, once the mass has been appropriately
renormalized in infinite volume. The details of this renormalization
are irrelevant here and we assume that the renormalization has been
adequately performed. Furthermore, the coupling $\lambda$ and the
field $\pi(x)$ only get renormalized at $O(\lambda^2)$, which is
beyond the order at which we are working.


\subsubsectiondum{Matching of finite to infinite matrix elements using \protect\eq{eq:LLformula}}

Here we consider a weak transition
between the state of a kaon at rest and a two-pion state,
$|\pi\pi\,n\rangleV$, in finite volume. The amplitude for this transition is
\be
M=\int_{\vec{x}}\langleV\pi\pi\,n|\cH_W(0,\vec{x})|K\rangleV
\ .\labell{mdef}\ee
The corresponding infinite-volume transition amplitude, $T$, is given
by
\be
T=\la\pi(\vec{p})\pi(-\vec{p}),out|\cH_W(0)|K(\vec{0})\ra
\ .\ee
Again, infinite-volume states are relativistically normalized here. 

To compare with the perturbative results obtained below, we must 
compute the factor relating $|T|$ and $|M|$ in
\eq{eq:LLformula} to $O(\lambda)$. Using the expressions in
\eqs{qNLOdef}{deltaq} for $\qNLO{n}$, and after some algebra, we find
\be
\qNLO{n}\phi'(\qNLO{n})=\frac{4\pi^2
n^{3/2}}{\nu_n}\l\{1+\frac{\lambda}{8\pi^2}\frac{1}{\WLO{n} L}\l[
\frac{\nu_n}{n}+z_n\r]+O(\lambda^2)\r\}
\ ,\labell{qphip}
\ee
where $z_n$ is the constant given by
\be
z_n=\lim_{q^2\to n}\l\{\sqrt{4\pi}\cZ_{00}(1;q^2)+\frac{\nu_n}{q^2-n}\r\}
\ .\labell{zndef}
\ee

Now, using the result of \eq{phase1} for $\delta(k)$ and
\eqs{qNLOdef}{deltaq} for $\kNLO{n}$, we find
\be
\kNLO{n}\delta'(\kNLO{n})=-\frac{\lambda}{8}
\frac{\sqrt{n}}{\WNLO{n} L}\l[1-n\l(\frac{4\pi}{\WLO{n} L}\r)^2+O(\lambda^2)\r]
\ .\labell{kdelp}
\ee

Combining \eqs{qphip}{kdelp}, we find for the factor which relates
$|T|$ and $|M|$ in \eq{eq:LLformula},
\be
8\pi\,\l\{q\frac{\partial\phi}{\partial
q}+k\frac{\partial\delta_0}{\partial
k}\r\}_{\kNLO{n}}\frac{M_K\WNLO{n}^2}{\kNLO{n}^3}
\labell{relfact1}\ee
$$
=
\frac{4M_K\WNLO{n}^2L^3}{\nu_n}\l\{1+\frac{\lambda}{2}\frac{1}{\WLO{n} L}\l[
\frac{z_n}{4\pi^2}+\frac{\nu_n}{(\WLO{n} L)^2}\r]+O(\lambda^2)\r\}
\ ,$$
where we have used the expression for $\kNLO{n}$ obtained in
\eq{qNLOdef} and (\reff{deltaq}).  It should be remarked that, unless the pions
interact strongly, the size of this factor is essentially
determined by mismatches in the definitions of $T$ and $M$ and in the
normalization of states in finite and infinite volume (see \eq{eq:LLfree} and
subsequent discussion).

\subsubsectiondum{Matching of finite to infinite matrix elements from perturbation
theory}

\begin{figure}
\centering
\includegraphics[width=0.8\textwidth]{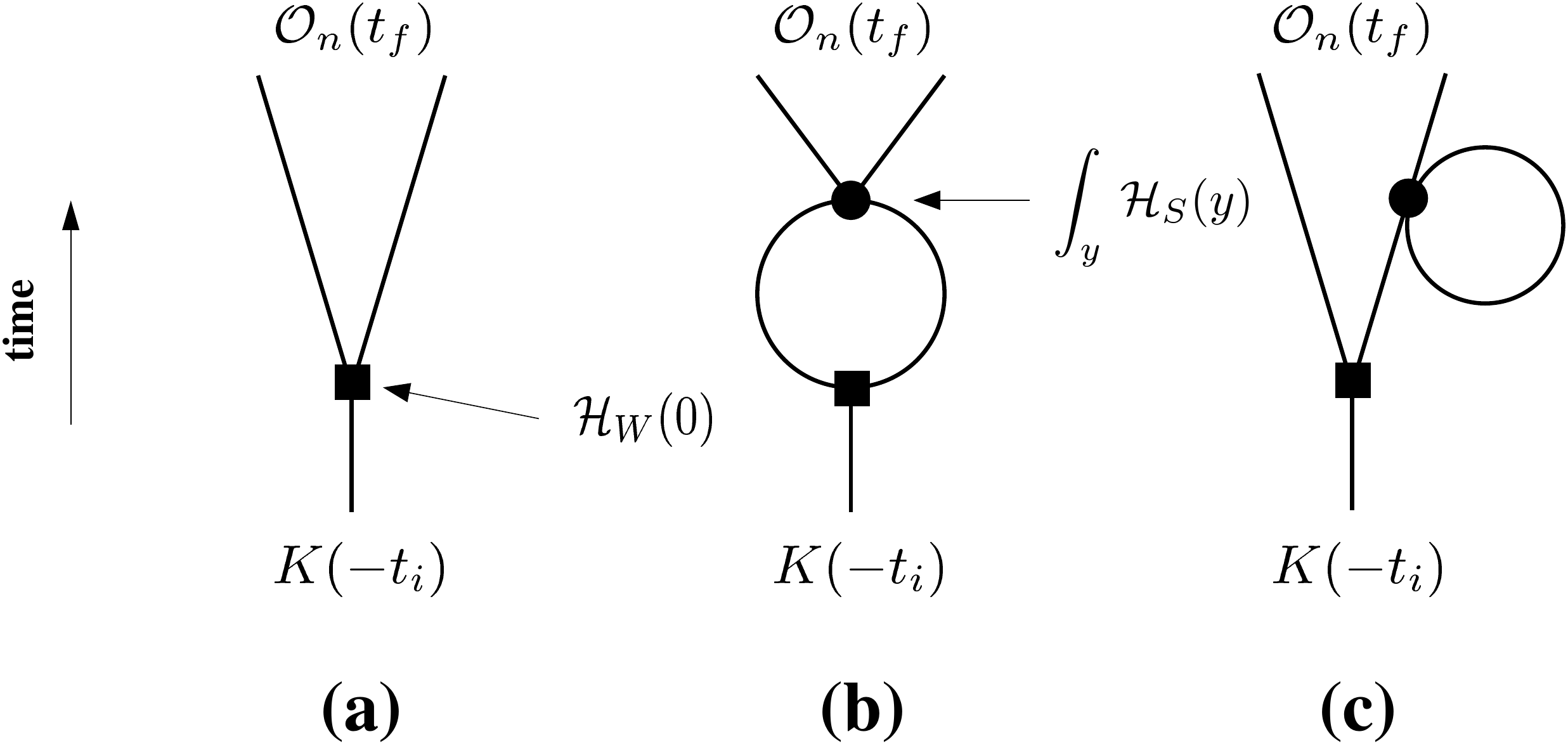}
\caption{Diagrams which contribute to $\Ckpipi(t_i,t_f)$ 
at $O(\lambda^0)$ (a) and
$O(\lambda)$ (b,c). }
\labell{fig:kpipi}
\end{figure}

The relevant correlation function here is
\be
\Ckpipi(t_i,t_f)=\int_{\vec{x}}\,\la\cO_n(t_f)
\cH_W(0)K(-t_i,\vec{x})\ra
\ .\labell{kpipicf}
\ee
At $O(\lambda^0)$, it is given by the diagram
in \fig{fig:kpipi}.a. One trivially obtains,
\be
\left.\Ckpipi^{\mbox{\small(\reff{fig:kpipi}.a)}}(t_i,t_f)\right\vert_n=-g
\frac{e^{-M_Kt_i}}{2M_K}\frac{e^{-\WLO{n}t_f}}{(\WLO{n})^2}
\ .\labell{ckpipi0}
\ee
At $O(\lambda)$, the contribution from the diagram in \fig{fig:kpipi}.b
gives, using the pion propagators defined in \eq{piprop},
$$
\left.\Ckpipi^{\mbox{\small(\reff{fig:kpipi}.b)}}(t_i,t_f)\right\vert_n=
\lambda\frac{g}{2}\,\frac{e^{-M_Kt_i}}{2M_K}
\int_{t_y}\frac{1}{L^3}\sum_{\vec{p}}\{S_\pi(t_y;\vec{p})\}^2
\times \{S_\pi(t_f-t_y;\vec{k}_n)\}^2
$$
$$
\longrightarrow 
-\left.\Ckpipi^{\mbox{\small(\reff{fig:kpipi}.a)}}(t_i,t_f)\right\vert_n\times\lambda
\Biggl\{\frac{\nu_n}{2L^3}\l(
\frac{t}{(\WLO{n})^2}+\frac{1}{(\WLO{n})^3}-
\frac{\Rlam{(\kLO{n})^2}{(\kLO{n})^2}}{2}
\r)
$$
\be
\labell{ckpipi1}
+\frac{1}{8}\cA(\kLO{n})\Biggr\}
\ ,\ee
where
\be
\cA(\kLO{n})=\frac{1}{L^3}\sum_{\vec{p}}
\sumprime\l[\frac{1}{E_p(\vec{p}^2-(\kLO{n})^2)}-
\Rlam{\vec{p}^2}{(\kLO{n})^2}\r]
\ ,\labell{akndef}
\ee
and where we have only kept, in the second line of \eq{ckpipi1}, the
terms in the momentum sum of the first line which fall off as
$e^{-\WLO{n}t_f}$. 

The sum in \eq{akndef} is restricted to momenta $\vec{p}$ such that
$|\vec{p}|\ne \kLO{n}$. To evaluate it, we use the asymptotic,
large-volume expansion of \shortcite{Luscher:1986pf}. Up to terms that
vanish more rapidly than any power of $1/L$, we find
\be
\cA(\kLO{n})\sim
I_1(\kLO{n})+\frac{z_n}{2\pi^2\WLO{n}L}
+\frac{\nu_n}{L^3}\l[\frac{4}{(\WLO{n})^3}+\Rlam{(\kLO{n})^2}{(\kLO{n})^2}\r]
\ ,\labell{aexp}
\ee
where $z_n$ is defined in \eq{zndef}. In \eq{aexp},
$I_1(\kLO{n})$ is the infinite-volume contribution:
\be
I_1(\kLO{n})=\int_{\vec{p}}\l\{\frac{1}{E_p}
\re\l[\frac{1}{\vec{p}^2-(\kLO{n})^2-i\epsilon}
\r]-\Rlam{\vec{p}^2}{(\kLO{n})^2}\r\}
\ .
\labell{i1def}
\ee

As discussed following \eq{pipime}, tadpole diagrams such as the one of
\fig{fig:kpipi}.c solely contribute to the renormalization of the mass
of the corresponding leg and do not affect our finite-volume
expressions. 

Combining \eqss{ckpipi0}{ckpipi1}{aexp}, we find, at
$O(\lambda)$,
\be
\Ckpipi(t_i,t_f;\kLO{n})\longrightarrow
-g\l\{1-\frac{\lambda}{8}I_1(\kLO{n})\r\}
\,\frac{e^{-M_Kt_i}}{2M_K}\frac{e^{-\WNLO{n} t_f}}{\WNLO{n}^2}
\times
\labell{ckpipires}
\ee
$$
\times
\l\{1-\frac{\lambda}{8}\l(\frac{z_n}{2\pi^2\WLO{n} L}-
\frac{\nu_n\Rlam{(\kLO{n})^2}{(\kLO{n})^2}}{L^3}
\r)+O(\lambda^2)\r\}
\ ,
$$
where, again, $\WNLO{n}$, is given by \eq{2pien}.

\subsubsectiondum{Putting it all together}

The contributions of the two-pion states, $|\pi\pi\,l\rangleV$, to
$\Ckpipi(t_i,t_f)$ are, in the limit of large $t_i$ and $t_f$,
$$
\Ckpipi(t_i,t_f)\longrightarrow
\sum_{l=0}^6 e^{-M_Kt_i-\WNLO{l} t_f}\,\la 0|\cO_n(0)|\pi\pi\,l\rangleV
\langleV \pi\pi\,l|\int_{\vec{x}}\cH_W(0,\vec{x})|K\rangleV
$$
\be
\times\langleV
K|K(0)|0\ra+\cdots
\ ,\labell{ckpipinp}
\ee
where the ellipsis stands for terms which decay more rapidly. In
\eq{ckpipinp}, $|K\rangleV$ is a zero-momentum state and, again, all states
are normalized to 1. With these normalizations,
\be
\labell{kmenorm}
\langleV K|K(0)|0\ra=\sqrt{\frac{1}{2M_KL^3}}
\ .
\ee
Combining this matrix element with \eqs{ckpipires}{ckpipinp} and 
the result of \eq{pipime} for $|\la
0|\cO_n(0)|\pi\pi\,n\rangleV|$, we find
$$
|M|
=g\l\{1-\frac{\lambda}{8}I_1(\kLO{n})\r\}\frac{1}{2}\sqrt{
\frac{\nu_n}{M_K\WNLO{n}^2L^3}}
\Biggl\{1-\frac{\lambda}{4}\frac{1}{\WLO{n} L}\l[
\frac{z_n}{4\pi^2}+\frac{\nu_n}{(\WLO{n}L)^2}\r]
$$
\be
\labell{mres}
+O(\lambda^2)\Biggr\}
\ ,
\ee

Now, a straightforward evaluation of the infinite-volume analogs of the
diagrams of \fig{fig:kpipi} yields:
\be
T=-g\l\{1-\frac{\lambda}{8}\int_{\vec{p}}\l[
\frac{1}{E_p(\vec{p}^2-(\kLO{n})^2-i\epsilon)}
-\Rlam{\vec{p}^2}{(\kLO{n})^2}\r]\r\}
\ .
\labell{ivampres}\ee
Therefore,
\be
\frac{|T|}{|M|}
=2\sqrt{\frac{M_K\WNLO{n}^2L^3}{\nu_n}}
\l\{1+\frac{\lambda}{4}\frac{1}{\WLO{n} L}\l[
\frac{z_n}{4\pi^2}+\frac{\nu_n}{(\WLO{n}L)^2}\r]+O(\lambda^2)\r\}
\ ,\labell{finpertres}\ee
which is in perfect agreement with the result of \eq{relfact1},
predicted by the finite-volume formula of \eq{eq:LLformula}. In obtaining
\eq{finpertres}, we have used the fact that
$|T|=|\re\,T|+O(\lambda^2)$.

\subsubsectiondum{$K\to\pi\pi$ in finite volume: physical kaon decays}

For illustration, let us suppose that the $S$-wave scattering phases
$\delta_I$, $I=0,2$, are accurately described by the one-loop formulae
of chiral perturbation theory \shortcite{Gasser:1990ku,Knecht:1995tr}.
The two-pion energy spectrum can then be calculated in the isospin $I$
channel and in a box of size $L$ where level $n=1$ coincides with the
kaon mass.  With this input, the proportionality factor in \eq{eq:LLformula}
is easily evaluated and one ends up with (cf. \tab{tab:LLeval})
\bea
  |A_0|&=&44.9\times|M_0|,\\
  |A_2|&=&48.7\times|M_2|,\\
  |A_0/A_2|&=&0.92\times|M_0/M_2|.
\eea
As these results show, the large difference
between the scattering phases in the two isospin channels 
(about $45^{\circ}$ at $k=k_\pi$) does not lead to 
a big variation in the proportionality factors.
In fact, if we set the scattering phases to zero altogether,
\eqs{eq:Lnfree}{eq:LLfree} give
$|A_I|=47.7\times|M_I|$ for $n=1$, which is not far from 
the results quoted above. This may be surprising at first sight, since
the interactions of the pions in the spin and isospin 0 state are quite
strong. However, one should take into account the fact that the
comparison is made for box sizes $L$ which are greater than
5~fm. Hence, it is quite plausible that the finite-volume matrix
elements already include most of the final-state interaction
effects. Apart from a purely kinematic factor, only a small
correction is then required to obtain the infinite-volume matrix
elements, from the finite-volume ones.

\begin{table}
\begin{center}
\begin{tabular}{ccccc}
\hline\hline
$I$ & $L$ [fm] & $q$ & $q\partial\phi/\partial q$ &
$k\partial\delta_I/\partial k$\\
\hline
0 & 5.34 & 0.89 & 4.70 & 1.12\\
2 & 6.09 & 1.02 & 6.93 & -0.09\\
\hline\hline\\
\end{tabular}
\caption{\labell{tab:LLeval}Calculation of the proportionality factor in
  \protect\eq{eq:LLformula} at the first level crossing}
\end{center}
\end{table}

The proportionality factor in \eq{eq:LLformula} thus appears to be 
only weakly dependent on the final-state interactions.
In particular, if the theory is to reproduce the $\Delta I=1/2$
enhancement, the large factor has to come from the ratio of the 
finite-volume matrix elements $M_I$.  In fact, if you carry out
this calculation, you should see an 8\% enhanced $\Delta I=1/2$
enhancement!

To carry out this calculation you will have to address a couple of
issues which have not yet been discussed here. The first is that, in
the absence of twisted boundary conditions, at least the first excited
$\pi\pi$ energy will have to be extracted from the lattice
calculation, This requires cross-correlator techniques constructed
from operators such as the one given in \eq{pipiop} and solving the
resulting generalized eigenvalue problem (GEVP), as described in
\shortcite{Luscher:1990ck}. The second issue is the one of the
renormalization of the lattice matrix elements. This is an important,
but fairly technical problem, which depends sensitively on the fermion
discretization used. It is usually referred to as the {\em ultraviolet}
problem, as opposed to the {\em infrared} problem which we dealt with
here, which is associated with the continuation of the theory to
Euclidean spacetime and the use of a finite volume in numerical
simulations. Unfortunately I will not have the time to cover the
ultraviolet problem here. This problem has been studied quite
extensively, and I refer you to the original literature, as well as to
the lectures of Peter \shortcite{peter} and Tassos \shortcite{tassos}
in this volume. For Wilson fermions, which explicitly break chiral
symmetry, but retain a full flavor symmetry, the problem has been
studied in the following series of papers
\shortcite{Bochicchio:1985xa,Maiani:1986db,Bernard:1987pr,Dawson:1997ic}. For
twisted-mass QCD, the reference is \shortcite{Frezzotti:2004wz}. When
considering domain-wall fermions with a finite fifth dimension, one
should follow the renormalization set forth for Wilson
fermions. However, for domain-wall fermions the required subtractions
should be significantly smaller, since the chiral symmetry breaking
should be significantly suppressed compared to what it is for Wilson
fermions. Regarding discretizations which have the full, continuum
chiral-flavor symmetry at finite lattice spacing (e.g. overlap
fermions, or domain-wall fermions with a practically infinite fifth
dimension), the renormalization is much simplified and will proceed as
in the continuum. Finally, for staggered fermions, these issues are
discussed in \shortcite{Sharpe:1986xu,Sharpe:1993ur}.

To conclude, if you wish to be the first particle theorist to
unambiguously see the $\Delta I=1/2$ rule in $K\to\pi\pi$ decays and
determine $\epsilon'$ with controlled errors, I hope that you begun
working on the problem immediately after the course was given. If not,
you would do better to hurry because the RBC-UKQCD collaboration is
making quick progress on these problems
\shortcite{Christ:LGT10,Liu:LGT10,Sachrajda:LGT10}.

\sectiondum[Appendix: integral representation for $Z_{00}(1;q^2)$]{Appendix: integral representation for {\huge $Z_{00}(1;q^2)$}}
\labell{sec:integzeta}

Here we derive an integral representation for the zeta function
$Z_{00}(1;q^2)$ of \eq{eq:zeta00}, which is a meromorphic function of
$q^2$, with poles at $q^2=\vec{n}^2$, $\vec{n}\in\mathbb{Z}^3$. This
representation is particularly effective for evaluating numerically
the kinematic function $\phi(q)$ that appears in Martin's
two-particle momentum quantization formula \eq{eq:MartinQC}. It
differs from the one given in Appendix C of
\shortcite{Luscher:1990ux}.

The definition of $Z_{00}(s;q^2)$ is given by \eq{eq:zeta00}:
\be
Z_{00}(s;q^2)=\frac{1}{\sqrt{4\pi}}\sum_{\vec{n}\in\mathbb{Z}^3}\left(\vec{n}^2-
  q^2\right)^{-s}
\ .\nonumber\ee
We define
\be
Z_{00}^{>\Lambda}(s;q^2)= \frac{1}{\sqrt{4\pi}}\sumngtq\left(\vec{n}^2-
  q^2\right)^{-s}
\ , \ee
where the sum runs over all $\vec{n}\in\mathbb{Z}^3$ such that
$\vec{n}^2>\Lambda$ with $\Lambda\ge \re\,q^2$. For $\re\, s>0$,
\bea
Z_{00}^{>\Lambda}(s;q^2)&=&\frac1{\Gamma(s)}\sumngtq\int_0^\infty
dt\,t^{s-1}
\,e^{-t(\vec{n}^2-q^2)}\nonumber\\
\labell{eq:Z00gtqDelta}
&=& \frac1{\Gamma(s)}\sumn\int_0^1
dt\,t^{s-1}
\,e^{-t(\vec{n}^2-q^2)}+\Delta(s;q^2)
\ ,\eea
with
\be
\Delta(s;q^2)=\frac1{\Gamma(s)}\left\{\sumngtq\int_1^\infty dt-\sumnleq\int_0^1
dt\right\}\,t^{s-1}
\,e^{-t(\vec{n}^2-q^2)}
\ ,\ee
where the second sum runs over all $\vec{n}\in\mathbb{Z}^3$ such that
$\vec{n}^2\le\Lambda$.

To evaluate the sum in \eq{eq:Z00gtqDelta}, we use a Dirac comb:
\bea
\sumn f(\vec{n})&=&
\sumn\int d^3x\,f(\vec{x})\,\delta^{(3)}(\vec{x}-\vec{n})\nonumber\\
\labell{eq:sumncomb}
&=&\sumn\int d^3x\,f(\vec{x})\,e^{i2\pi\vec{n}\cdot\vec{x}}
\ . \eea
With $f(\vec{n})=e^{-t\vec{n}^2}$ and
\be
\int_{-\infty}^{+\infty} dx\, e^{-tx^2+i2\pi n x}=\sqrt{\frac{\pi}{t}}
e^{-\frac{\pi^2n^2}{t}}
\ ,\ee
\eq{eq:sumncomb} yields
\be
\sumn e^{-t\vec{n}^2}=\sumn\left(\frac{\pi}{t}\right)^{3/2}\,
e^{-\frac{\pi^2\vec{n}^2}{t}}
\ .\ee
Thus
\be
\labell{eq:Z00gtqIsn}
Z_{00}^{>\Lambda}(s;q^2)=\frac1{\Gamma(s)}\sumn I(s;\vec{n})+
\Delta(s;q^2)
\ ,\ee
with
\be
I(s;\vec{n})=\pi^{3/2}\int_0^1 dt\, t^{s-5/2}e^{tq^2-\frac{\pi^2\vec{n}^2}{t}}
\ .\ee

Viewed as a function of $\vec{n}$, $I(s;\vec{n})$ is singular only for
$\vec{n}=\vec{0}$ when $\re\,s\le 3/2$. In that case, $I$'s integrand
goes like $t^{s-1-3/2}(1+tq^2+O(t^2)$ when $t\to 0$. Thus,
$\re\,s>3/2$ yields the half-plane in complex $s$ for which the
expression of \eq{eq:Z00gtqIsn} gives a finite result. For such $s$ we can write
\be
I(s;\vec0)=\pi^{3/2}\int_0^1 dt\, t^{s-5/2}\left(e^{tq^2}-1\right)
+\frac{\pi^{3/2}}{s-3/2}
\ ,\ee
which is actually well defined for $\re\,s>1/2$ and $s\ne3/2$. Thus,
for all $s$ in the half plane $\re\,s>1/2$, we obtain
\bea
\labell{eq:zoointreps}
\sqrt{4\pi}Z_{00}(s;q^2)&=&
\sumnleq(\vec{n}^2-q^2)^{-s}+\frac{\pi^{3/2}}{\Gamma(s)}
\left\{\frac1{s-3/2}+\int_0^1dt\,t^{s-5/2}(e^{tq^2}-1)\right.\nonumber\\
&&\left.+\sum_{\vec{n}^2\ne
  0}\int_0^1dt\,t^{s-5/2} e^{tq^2-\frac{\pi^2\vec{n}^2}{t}}\right\}+
\Delta(s;q^2)\ ,
\eea
where the last runs over all $\vec{n}\in\mathbb{Z}^3$ such that
$\vec{n}^2\ne 0$.

Now, for the case $s=1$ which is of interest to us here, it is
straightforward to compute $\Delta(s;q^2)$:
\bea
\Delta(1;q^2)&=& \left\{\sumngtq\int_1^\infty dt-\sumnleq\int_0^1
dt\right\}\,e^{-t(\vec{n}^2-q^2)}\nonumber\\
&=& \sumngtq \frac{e^{-(\vec{n}^2-q^2)}}{\vec{n}^2-q^2}+
\sumnleq \frac{e^{-(\vec{n}^2-q^2)}-1}{\vec{n}^2-q^2}
\ .\eea
Using this result in the expression of \eq{eq:zoointreps} for $Z_{00}(s;q^2)$
with $s=1$, we obtain:
\bea
Z_{00}(1;q^2)&=&-\pi+\frac1{\sqrt{4\pi}}\sumn
\frac{e^{-(\vec{n}-q^2)}}{\vec{n}-q^2} + \frac{\pi}{2}
\int_0^1\frac{dt}{t^{3/2}}(e^{tq^2}-1)\nonumber\\ 
&&+ \frac{\pi}{2}\sum_{\vec{n}\ne \vec{0}}
\int_0^1\frac{dt}{t^{3/2}} e^{tq^2-\frac{\pi^2\vec{n}}{t}}\nonumber\\
&=&
-\pi+\frac1{\sqrt{4\pi}}\sum_{m=0}^\infty
\nu_m\,\frac{e^{-(m-q^2)}}{m-q^2} + \frac{\pi}{2}
\int_0^1\frac{dt}{t^{3/2}}(e^{tq^2}-1)\nonumber\\ 
\labell{eq:zoointrep1}
&&+ \frac{\pi}{2}\sum_{m=1}^\infty\nu_m\,
\int_0^1\frac{dt}{t^{3/2}} e^{tq^2-\frac{\pi^2m}{t}}
\ ,\eea
where $\nu_m$ counts the $\vec{n}\in\mathbb{Z}^3$ such that
$\vec{n}^2=m$ (see \eq{eq:nundef}). \eq{eq:zoointrep1} is the integral
representation that we were after. Using this representation, it is
straightforward to calculate $Z_{00}(1;q^2)$ numerically with good
efficiency and high precision, using standard integration routines.

\bibliographystyle{OUPnamed_notitle}
\bibliography{./lellouch}

\end{document}